\documentclass[floatfix,twocolumn,showpacs,aps,tightenlines,superscriptaddress,nofootinbib]{revtex4-1}
\usepackage{color}
\usepackage{xspace}
\usepackage{graphicx}
\usepackage{amsmath,amssymb,amsfonts}
\usepackage{array}
\usepackage{enumitem}
\usepackage{epsf}
\usepackage{subfig}

\newcommand{\KMS}{\rm km\ s^{-1}}

\newcommand{\bbh}{black hole--black hole binary\xspace}
\newcommand{\bbhs}{black hole--black hole binaries\xspace}
\newcommand{\bns}{neutron star--neutron star binary\xspace}
\newcommand{\bnss}{neutron star--neutron star binaries\xspace}
\newcommand{\bhns}{black hole--neutron star binary\xspace}
\newcommand{\bhnss}{black hole--neutron star binaries\xspace}

\newcommand{\headon}{head-on\xspace}

\def\mnras{MNRAS}
\def\apj{ApJ}

\def\apjl{ApJL}
\def\aap{AAp}
\def\nat{Nature}

\def\prd{Phys. Rev. D}

\begin{document}

\title{Numerical Relativity of Compact Binaries in the 21st Century}
\author{Matthew D. Duez}
\affiliation{Department of Physics and Astronomy, Washington State
University, Pullman, Washington 99164, USA}
\author{Yosef Zlochower}
\affiliation{Center for Computational Relativity and Gravitation
  and School of Mathematical Sciences, Rochester Institute of
Technology, 85 Lomb Memorial Drive, Rochester, New York 14623, USA}

\begin{abstract}

We review the dramatic progress in the simulations of compact objects
and compact-object binaries that has taken place in the first two
decades of the twenty-first century.  This includes simulations of the
inspirals and violent mergers of binaries containing black holes and
neutron stars, as well as simulations of black-hole formation through
failed supernovae and high-mass neutron star--neutron star mergers.  Modeling
such events  requires numerical integration of the field equations of
general relativity in three spatial dimensions, coupled, in the case of
neutron-star containing  binaries, with increasingly sophisticated
treatment of fluids, electromagnetic fields, and neutrino radiation.
However, it was not until 2005 that accurate long-term evolutions of
binaries containing black holes were even
possible~\cite{Pretorius:2005gq, Campanelli:2005dd, Baker:2005vv}.  Since then, there
has been an explosion of new results and insights into the physics of
strongly-gravitating system. Particular emphasis has been placed on
understanding the gravitational wave and electromagnetic signatures
from these extreme events. And with the recent dramatic discoveries
of gravitational waves from merging black holes by the Laser
Interferometric Gravitational Wave Observatory and Virgo, and the
subsequent discovery of both electromagnetic and gravitational wave
signals from a merging \bns,  numerical relativity
became an indispensable tool for the new field of multimessenger
astronomy.
\end{abstract}

\maketitle

\section{Introduction}
\label{sec:introduction}
With the discovery of gravitational waves from merging \bbhs
by the Laser Interferometric Gravitational Wave Observatory
(LIGO) in 2015~\cite{Abbott:2016blz, TheLIGOScientific:2016wfe,
Abbott:2016nmj}, the subsequent observations of other black hole
mergers by LIGO~\cite{Abbott:2017vtc, Abbott:2017gyy} and later by LIGO
  and Virgo~\cite{Abbott:2017oio}, and the simultaneous
observation of gravitational waves and electromagnetic spectra from
the merger of a \bns by LIGO, Virgo, and a large team
of astronomers in 2017~\cite{TheLIGOScientific:2017qsa, GBM:2017lvd,
Monitor:2017mdv}, the new field of gravitational wave and
multimessenger astronomy was born.

Fundamental to this new science is the ability to infer the
dynamics of the sources based on the observed signals, something that
can only be accomplished using detailed theoretical predictions based
on numerical simulations of the nonlinear Einstein field equations of general
relativity both in vacuum and coupled to the equations of
magnetohydrodynamics. Indeed, the body of techniques that emerged based on efforts
to solve this system, known as numerical relativity, was designed largely with
such gravitational wave source modeling in mind, but it has also been
turned to other astrophysical phenomena involving strongly-curved,
dynamical spacetime. 

Only rather exotic phenomena involve sufficiently strong spacetime
curvature to require numerical relativity.  Newtonian gravity clearly works
quite well for main sequence stars, planets, and the like.  As is well-known,
relativity becomes important when speeds approach the speed of light $c$, so
a reasonable guess would be to expect important general relativistic effects
as the escape velocity approaches $c$.  Then an object of mass $M$ and
radius $R$ will require relativistic treatment if $R$ is close to the
gravitational radius $r_G\equiv 2GM/c^2$, the radius of a nonspinning
black hole of mass $M$.  The same condition can be stated in terms of
the dimensionless {\it compaction} $\mathcal{C}\equiv \frac{GM}{Rc^2}$. 
Strong-gravity objects have high compaction (order unity being the standard of
``high'').  Black holes ($\mathcal{C}\sim 1$) and neutron stars
($\mathcal{C}\sim 0.1$) are compact objects
by this definition.  White dwarfs ($\mathcal{C}\sim 10^{-4}$) are a marginal
case--relativity plays a large role in their stability condition but not
their equilibrium structure--and are usually also classified as compact.

Formulating the integration of the Einstein equations so that
evolutions
are stable and the coordinates evolve sensibly turned out to be a difficult
task.  There was some worry that numerical relativity might not be ready
when the advanced gravitational wave detectors needed it.  Finally, in
2005~\cite{Pretorius:2005gq, Campanelli:2005dd, Baker:2005vv}, the first
stable \bbh merger simulations were carried
out.  There followed a race to produce accurate waveforms for
gravitational wave observation efforts, which were already underway.

In this review, we describe how numerical relativity has come to be
a robust tool for studying strong-gravity systems.  We also review
some of the major accomplishments of numerical relativity to date. 
To provide an appropriate scope, we focus on applications to compact
binaries and black hole formation, processes where general relativity
is essential and whose astrophysical importance is clear.

The article is organized as follows.  In the rest of the introduction,
we provide background on general relativity, black holes, and relativistic
stars.  In Section~\ref{sec:NR}, we cover methods for evolving the Einstein
field equations and coupled matter sources.  Particular attention is given to
the historical ``breakthrough'' discoveries that enabled stable evolutions
of multiple-black-hole spacetimes.  The next sections review simulation results,
covering the period before the breakthroughs and after.  Section~\ref{sec:bbh}
is devoted to \bbh mergers.  The following three
sections describe simulations of phenomena with matter, especially neutron
stars.  Finally, in Section~\ref{ligo-compare}, we return to
our original motivation and consider what has been learned by the
confrontation of numerical relativity predictions with actual LIGO-Virgo
observations.

\subsection{The field equations of general relativity}
\label{sec:field_eq}

The theory of special relativity introduced the notion of spacetime.
In that theory, spacetime is a geometrically flat 4-dimensional
manifold. General relativity extends this notion to non-flat
manifolds. In general relativity, the Newtonian notion of a
gravitational force is replaced by geodesic motion in a curved
spacetime. Unless acted on by other (non-gravitational) forces,
objects whose size is much smaller than the local spacetime radius of curvature
travel along geodesics. 

Throughout this paper, we will use geometric units. In these units,
the speed of light, $c$, and Newton's constant $G$, are taken to be 1.
A consequence of this is that distances, time intervals, masses, and
energy all have the same units. By convention, the unit of each of
these is denoted by an arbitrary mass $M$.\footnote{For example, solar
mass intervals of distance and time are about 1.5\,km and $5\times 10^{-6}$s,
respectively.}

In the section below, we will provide an extremely brief overview of
the field equations of general relativity. For a comprehensive
overview, we suggests consulting~\cite{2003gieg.book.....H} for a very accessible
introduction to general relativity, and~\cite{2009fcgr.book.....S},
\cite{2004sgig.book.....C}, and \cite{1984ucp..book.....W} for a more advanced treatment.
The material below was synthesized from these references.

The geometry of a spacetime can be entirely described by a line
element $ds^2$.
In Minkowski spacetime, in Cartesian coordinates, the line element
takes the form
\begin{equation}
  ds^2 = -dt^2 + dx^2 + dy^2 + dz^2.
\label{flat:ds2}
\end{equation}
Note that $ds^2$ is not necessarily positive. For timelike paths, the
proper time along the path is given by the integral of 
$d\tau = \sqrt{-ds^2}$. Equation~(\ref{flat:ds2}) can be written as
\begin{equation}
  ds^2 = \eta_{\mu \nu} dx^{\mu} dx^{\nu},
\end{equation}
where $x^{\mu} = (t, x, y, z)$ and the components of the symmetric tensor $\eta_{\mu
\nu}$ are given by ${\rm diag}(-1,1,1,1)$. Here we used two standard
conventions, the timelike coordinate is listed first and repeated
Greek indices are summed over. By convention, the index of the
timelike coordinate is 0, while the spatial coordinates have indices
1, 2, 3.
In arbitrary coordinates, the line element becomes
\begin{equation} 
ds^2 = g_{\mu \nu} dy^{\mu} dy^{\nu},
\end{equation}
where $y^{\mu}$ is some new set of coordinates and
\begin{equation}
  g_{\mu \nu} = \left(\frac{\partial{x^\alpha}}{\partial y^\mu}\right)
\left(\frac{\partial x^\beta}{\partial y^\nu}\right) \eta_{\alpha
\beta}.
\end{equation}

Here, we will  not make a
distinction between the components of a tensor  and a tensor itself.
For our purposes a tensor $L$ of type $(p,q)$ is a set of $p\times q$
functions, denoted by 
\begin{equation}
  L^{\alpha_1 \alpha_2 \cdots \alpha_p}_{\beta_1 \beta_2 \cdots
  \beta_q},
\end{equation}
which, under a change of coordinates from some coordinate system ${\bf
x}$ to another ${\bf y}$, transform as 
\begin{eqnarray}
  L'&&^{\alpha_1 \alpha_2 \cdots \alpha_p}_{\beta_1 \beta_2 \cdots
  \beta_q}({\bf y}) = 
  \left(\frac{\partial y^{\alpha_1}}{\partial x^{\mu_1}}\right)
  \left(\frac{\partial y^{\alpha_2}}{\partial x^{\mu_2}}\right)
  \cdots
  \left(\frac{\partial y^{\alpha_p}}{\partial x^{\mu_p}}\right)\nonumber\\
  &&\times
  \left(\frac{\partial x^{\nu_1}}{\partial y^{\beta_1}}\right)
  \left(\frac{\partial x^{\nu_2}}{\partial y^{\beta_2}}\right)
  \cdots
  \left(\frac{\partial x^{\nu_q}}{\partial y^{\beta_q}}\right)
   L^{\mu_1 \mu_2\cdots \mu_p}_{\nu_1 \nu_2\cdots \nu_q}({\bf x}).
   \label{eq:tensor_transform}
\end{eqnarray}
Thus the metric $g_{\mu \nu}$ is a tensor. Associated to $g_{\mu \nu}$
is its matrix inverse $g^{\mu \nu}$ (i.e., $g^{\mu \sigma} g_{\sigma
\nu} = \delta^\mu_\nu$, where $\delta$ indicates the usual Kronecker
delta function).

The metric, $g_{\mu \nu}$, in special relativity is intrinsically flat.
By this, we mean any of several equivalent statements: there is a
coordinate transformation
such that the new metric is everywhere identical to $\eta_{\mu \nu}$, as
discussed above, parallel
geodesics will remain parallel,  and the
intrinsic curvature of the metric, as measured by the Riemann
curvature tensor,
vanishes everywhere. We briefly describe how geodesics and the Riemann
curvature tensor are calculated.

A geodesic is the
generalization of a straight line in Euclidean space. In Cartesian
coordinates, the tangent vector to a straight line is a constant. This
can be expressed as
\begin{eqnarray}
  t^{\mu} = \frac{d}{d \lambda} x^\mu(\lambda), \\
  \frac {d}{d \lambda} t^\mu(\lambda)= 0,\label{eq:cartgeo}
\end{eqnarray}
where $t^\nu$ is the tangent vector to the line and $x^\mu(\lambda)$
are the Cartesian coordinates of each point of the line.
Equation~(\ref{eq:cartgeo}) can be re-written as
\begin{equation}
  t^\nu \frac{\partial t^\mu}{\partial x^{\nu}} = 0.
\end{equation}
However, even in flat space, this equation does not hold true in
arbitrary coordinates. To fix this, we replace the ordinary derivative
$\partial/\partial x^\nu$ with the covariant derivative $\nabla_\nu$. 
The equation for a geodesic in arbitrary
coordinates, as well as flat and non-flat metrics, is
\begin{eqnarray}
  \frac{d}{d\lambda} x^\mu(\lambda)= t^{\mu}(\lambda),\\
  t^{\mu}\nabla_\mu t^\nu = 0.
\end{eqnarray}
Finally, $\nabla_\alpha$, the covariant derivative associated with
the metric $g_{\mu \nu}$, is defined by its action on arbitrary
tensors
$L^{\mu_1 \mu_2\cdots}_{\mu_1 \mu_2\cdots}$, which  is given by
\begin{eqnarray}
  \nabla_{\alpha}  L^{\mu_1 \mu_2\cdots}_{\nu_1 \nu_2\cdots} && =
  \partial_\alpha L^{\mu_1
  \mu_2\cdots}_{\nu_1 \nu_2\cdots} \nonumber \\
  &&+ {\Gamma^{\mu_1}}_{\alpha \beta} L^{\beta
  \mu_2\cdots}_{\nu_1 \nu_2\cdots}
  + {\Gamma^{\mu_2}}_{\alpha \beta} L^{\mu_1 \beta
\cdots}_{\nu_1 \nu_2\cdots} + \cdots \nonumber \\&& -
    {\Gamma^{\beta}}_{\alpha
    \nu_1} L^{\mu_1 \mu_2
  \cdots}_{\beta \nu_2 \cdots} - {\Gamma^{\beta}}_{\alpha \nu_2}
  L^{\mu_1 \mu_2
      \cdots}_{\mu_1 \beta \cdots} - \cdots,
\end{eqnarray}
where
\begin{equation}
  {\Gamma^\sigma}_{\mu \rho} = \frac{1}{2} g^{\sigma
  \alpha}\left( \partial_\mu g_{\alpha \rho} + \partial_\rho g_{\alpha
  \mu} - \partial_\alpha g_{\mu\rho}\right),
\label{eq:connection}
\end{equation}
and $\partial_\alpha$ is shorthand for $\partial/\partial
x^{\alpha}$~\footnote{Readers familiar with differential geometry
will notice that in Eq.~\ref{eq:connection} we are limiting ourselves to
coordinate (holonomic) bases, which are sufficient for numerical relativity.}.
The components of ${\Gamma^\sigma}_{\mu \rho}$ are collectively known
as the Christoffel symbols. Unlike the metric and the tensors
constructed from the Christoffel symbols below, the components of the
Christoffel symbols do not transform according to
Eq.~(\ref{eq:tensor_transform}).

The Riemann curvature tensor is constructed from the metric
and has the following form \footnote{
Note that while the components of ${R_{\mu \nu \rho}}^\sigma$
transform according to Eq.~(\ref{eq:tensor_transform}),
the components of ${\Gamma^\sigma}_{\mu \rho}$ do not.}
\begin{equation}
  {R_{\mu \nu \rho}}^{\sigma} = \partial_\nu {\Gamma^\sigma}_{\mu
  \rho} - \partial_\mu {\Gamma^\sigma}_{\nu \rho} +
  {\Gamma^\alpha}_{\mu \rho} {\Gamma^\sigma}_{\alpha \nu}-
  {\Gamma^\alpha}_{\nu \rho} {\Gamma^\sigma}_{\alpha \mu}.
\end{equation}
The Riemann curvature tensor can be further split into a trace-free part, known
as the Weyl tensor $C_{\mu \nu \rho \sigma}$, and the Ricci tensor
\begin{equation}
  R_{\mu \nu} = {R_{\mu \rho \nu}}^\rho
\end{equation}
Finally, the Einstein tensor $G_{\mu \nu}$, is given by
\begin{equation}
  G_{\mu \nu} = R_{\mu \nu} - \frac{1}{2} g_{\mu \nu} g^{\alpha
  \beta}R_{\alpha \beta}.
\end{equation}

Regardless of coordinates, the Riemann curvature tensor is identically
zero in special relativity. General relativity extends the notion of spacetime to
include non-flat metrics, where 
\begin{equation}
G_{\mu \nu} =  8 \pi T_{\mu \nu},\label{EE}\end{equation}
and $T_{\mu \nu}$ is the stress energy tensor, a measure of the total
energy and momentum flux from matter and non-gravitational interactions
and radiation.
 Because the Einstein equations do not constrain
the components of $C_{\mu \nu \rho \sigma}$, even in vacuum there can
be non-trivial curvature. Note that Eq.~(\ref{EE}) is the standard
covariant form
of the Einstein equations.

\subsection{Evolution of matter sources}
\label{sec:matterintro}
In general relativity, the curvature of spacetime possess its own dynamics,
and indeed one of the most important phenomena treated by numerical relativity,
the merger of two black holes, is a {\it vacuum} problem.  That is,
$T_{\mu\nu}=0$.  Numerical relativity is also used to study phenomena
involving matter flows in strongly-curved dynamical spacetimes.  Two
problems where such relativistic effects should be particularly important
are compact object
mergers involving neutron stars and the formation
of black holes by stellar collapse. 

Matter and energy constitute the stress-energy tensor $T_{\mu\nu}$ that
is the source term in Einstein equations.  In general, $T_{\mu\nu}$
will be a sum of stress tensors for the gas, electromagnetic, and
neutrino fields.  The energy and momentum conservation equations
\begin{equation}
\label{divT}
\nabla_\mu T^{\mu\nu} = 0
\end{equation}
provide evolution equations for the matter.

In the non-vacuum systems we will consider, the particles--including
nucleons, nuclei, electrons, positrons, and photons, but not necessarily
neutrinos--form a nearly perfect fluid, meaning the mean free path is very
small compared to the system's scale, making the collection a fluid, and
viscosity and heat transport are small enough to be ignored.  
This fluid will have a stress tensor
\begin{equation}
\label{Tgas}
T^{\rm gas}_{\mu\nu} = (\rho_0 + u + P)u_{\mu}u_{\nu} + P g_{\mu\nu}\ ,
\end{equation}
where $\rho_0$, $u$, $P$, and $u_{\mu}$ are the rest mass density,
internal energy, pressure, and 4-velocity.  (Note $\rho_0$ must be
distinguished from the total energy density $\rho=\rho_0 + u$.)
The 4-velocity has only three independent components, with the
Lorentz factor $W\equiv \alpha u^t$ given by the normalization
condition $u\cdot u=-1$:
\begin{equation}
  W^2 = 1 + \gamma^{ij}u_iu_j
\end{equation}
  
Equations (\ref{divT}) and (\ref{Tgas}) must be supplemented by
the rest mass conservation equation
\begin{equation}
\nabla_\mu (\rho_0u^{\mu}) = 0
\end{equation}
and also an equation of state (EoS)
\begin{eqnarray}
P &=& P(\rho_0,T,X_i) \\
u &=& u(\rho_0,T,X_i)\ ,
\end{eqnarray}
where $T$ is the temperature and $X_i$ are composition variables.

For a detailed exposition of relativistic hydrodynamics, including its
numerical treatment, see the book by Rezzolla and
Zanotti~\cite{2013rehy.book.....R}.

\subsection{Black holes}

Perhaps one of the most interesting predictions of general relativity
is the existence of black holes. Black holes are regions in spacetime
where the curvature is sufficiently strong that light, and therefore
any physical signal, cannot escape. Astrophysically, black holes form
as stellar objects collapse. Despite all the microphysics that
goes into the dynamics of stellar objects, once equilibrated, a black
hole can be completely described by two parameters: its mass and spin\footnote{More correctly, an equilibrated
  black hole is completely described by its mass, spin, and charge.
  However, astrophysical black holes are expected to have effectively zero
  charge because accretion from the  interstellar medium should rapidly
discharge them.}.
In geometric units the magnitude of the spin angular momentum $S$ is bounded by the
mass $m$, where $S < m^2$. Typically one defines a specific spin $a$,
where $a=S/m$ and a dimensionless spin $\chi$, where $\chi = S/m^2$.

If the black hole is non-spinning, it is known as a Schwarzschild
black hole~\cite{Schwarzschild16a,
Schwarzschild16b}, and if $S$ is non-zero, it is known as a Kerr black
hole~\cite{Kerr63}. In both cases the black hole spacetimes are named after
their discoverers. 

A black hole has no material surface, of course, but there is a boundary
separating the region from which it is impossible ever to escape (the black
hole interior) to the outside universe.  This boundary is called the {\it event
horizon}.  The event horizon is a null surface, i.e. one must move at the speed of
light to stay on it. In order to determine if a true event horizon
exists, one needs to know the entire future of the spacetime
(otherwise there is always the possibility  that an observer can escape the supposed
black hole at a later time). 
In practice, numerical relativists find event horizons by evolving a cluster of null
geodesics or a null surface backwards in time starting at the very end
of their simulations. (For a review, see~\cite{Thornburg:2006zb}.) 
We will
see that some methods of numerically handling black hole interiors (excision methods)
require some knowledge of the horizon location during the simulation.  For these
purposes, numerical relativists use the {\it apparent horizon}.
Apparent horizons are two-dimensional surfaces that may exist at each
time in a numerical simulation. They are defined to be surfaces 
from which outward-pointing null rays do not expand. This very unusual
situation can only occur in the vicinity of a black hole, but finding
such surfaces only requires information about the metric and extrinsic
curvature at a given time.
For a
stationary black hole, the apparent horizon will coincide with the event horizon; in a dynamical spacetime, it will
be inside the event horizon.

Black holes can form binaries, as demonstrated by LIGO's recent
detection of gravitational waves~\cite{Abbott:2016blz,
TheLIGOScientific:2016wfe}. In such a case, the black
holes are not truly in equilibrium, but each can still be described
reasonably well as Kerr or Schwarzschild black holes, at least when
the binary components are well separated. Such a binary can then be described by
several intrinsic parameters, such as the mass ratio of the two black holes,
the spin magnitudes and orientations of the two black holes,
and the orbital eccentricity. 

\subsection{Relativistic stars}
\label{sec:stars}

Much astrophysical thinking is guided by idealized equilibria,
such as the spherically symmetric star and the thin accretion disk, and
this remains true in the study of compact object systems.  Neutron stars
are extremely compact objects, containing a little over a solar mass
($r_G\approx 4$km) within a radius of
$\sim 10$\,km, and so must be studied using general relativity; they will
be our primary type of relativistic star. 

\begin{figure}
  \includegraphics[width=\columnwidth]{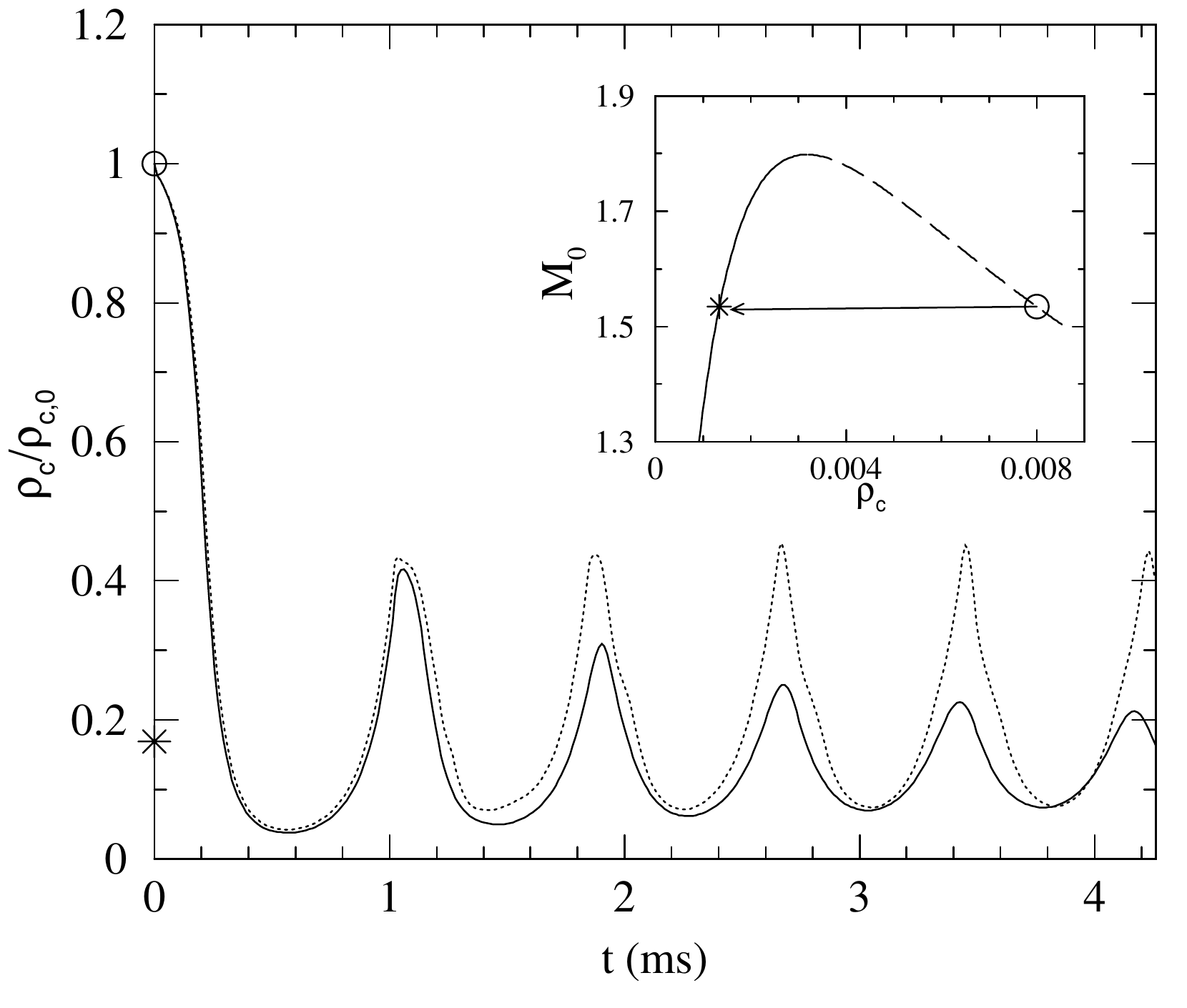}
  \caption{The evolution of an unstable non-rotating star in
numerical relativity.  The inset shows the TOV equilibrium sequence. 
For each central density, there is a unique equilibrium, and the inset
plots the baryonic mass $M_0$ against the central density $\rho_c$. 
Equilibria to the left of the turning point (solid curve) are stable,
to the right (dashed) are unstable.  A star on the unstable branch (open circle)
will (under tiny perturbations) migrate to lower density and oscillate about
the stable equilibrium of the given mass (asterisk).  This evolution is shown in
the main plot.  It illustrates the type of experiments that can be done
in numerical relativity.  The initial state would probably not occur in
any astrophysically realistic scenario.  For the evolution, various pieces
of physics can be turned on or off to study their effect.  The solid line
shows a simulation that allows shock heating, while the dotted line shows
a simulation where this has been artificially turned off, forcing the star
to evolve adiabatically.  Reproduced with permission from~\cite{Font:2001ew}. 
}\label{fig:tov}
\end{figure}

For spherical neutron stars of an assumed barotropic equation of state
$P=P(\rho_0)$, the Tolman--Oppenheimer--Volkoff (TOV)~\cite{Tolman:1939jz,
Oppenheimer:1939ne} equations of hydrostatic equilibrium yield a
sequence of equilibria, one
for each central density $\rho_c$.  At a critical central density
$\rho_{\rm crit}$, the mass reaches a maximum value $M^{\rm
  max}{}_{\rm TOV} = M(\rho_{\rm crit})$, and only the configurations on
the ascending ($dM/d\rho_c>0$) side, which generally turns out to be
$\rho_c<\rho_{\rm crit}$, are stable.  A star on the unstable side
will either collapse to a black hole or undergo large radial
oscillations about the lower density configuration of the same mass.
(See Fig.~\ref{fig:tov}.)

Can neutron stars exist with $M > M^{\rm max}{}_{\rm TOV}$?  If the
neutron star is spinning, this provides some additional support
against gravity.  Codes exist for generating the resulting 2D
(axisymmetric) equilibria (e.g.~\cite{cook94a,Stergioulas:1994ea}). 
The matter is usually be taken
to be a perfect fluid with purely azimuthal flows
[$u^r=u^{\theta}=0$, $u^{\phi}=u^t\Omega$], and a
rotation law for $\Omega$ must be specified.  Rotation might
be uniform ($\Omega=$constant) or differential ($\Omega$ varies
through the star).  Viscosity (or similar angular momentum transport
mechanisms) will tend to produce uniform rotation, but differentially
rotating equilibria can persist on timescales shorter than that of
viscosity.

Rotation in equilibrium stars
is constrained, though, by the mass-shedding limit, at which fluid on the
equator is in geodesic (``Keplerian'') orbit.  Faster rotation at the
equator would centrifugally eject mass.  This restricts the degree of
possible uniform rotation particularly severely, so that the maximum mass
only increases by around 20\%, the most massive configurations being
found close to (but not exactly on) the mass-shedding
limit~\cite{cook94a,cook94c}.  Stars with masses between $M^{\rm
  max}{}_{\rm TOV}$ and this higher limit are called {\it supramassive}.
A sufficient condition for instability of uniformly rotating stars (on
the secular timescale on
which uniform rotation is maintained) can be determined via locating
the turning point in the constant angular momentum
sequence~\cite{1988ApJ...325..722F},
similar to TOV sequences.  (The actual instability onset occurs slightly
on the ``stable'' side~\cite{Takami:2011zc}.)  The point of onset of
dynamical instability can be determined by numerical simulations.

Stars with mass above the supramassive limit are called {\it hypermassive}. 
Such equilibria can exist with the help of differential rotation, providing
rotational support while evading the mass-shedding limit by keeping the
rotation rate sub-Keplerian near the equator.  Numerical relativity
confirms~\cite{Baumgarte:1999cq}
that such stars can persist stably for multiple dynamical timescales,
as we discuss in Section~\ref{sec:stars-stability}.

One might also look to thermal support--hot nuclear matter--to
increase the maximum mass, effectively changing the equation of state
to give more pressure support.  Effects of thermal support have been
studied for uniformly rotating neutron stars by Goussard~{\it et
  al.}~\cite{Goussard:1996dp} and for uniformly and differentially
rotating neutron stars by Kaplan~{\it et al.}~\cite{Kaplan:2013wra}.
The latter suggest an approximate turning point method for assessing
stability which has been numerically confirmed by
Weih~{\it et al.}~\cite{Weih:2017mcw} and used by Bauswein and Stergioulas to
explain some numerical relativity findings on the threshold mass for prompt collapse of a
\bns merger remnant to a black
hole~\cite{Bauswein:2017aur}.

The distinctions introduced above between normal, supramassive, and
hypermassive neutron stars have played a large role in the interpretation
of \bns merger simulations.  When two neutron stars merge,
the resulting object will either collapse to a black hole or settle to a
dynamical equilibrium state in roughly a dynamical timescale
$\sim 10^{-1}$\,ms.  An equilibrium remnant could be described as a type
of relativistic star.  If two 1.4 $M_{\odot}$ neutron stars merge, this
remnant could easily have a mass in excess of $M^{\rm max}{}_{\rm TOV}$.
However, the remnant will also be spinning rapidly and differentially,
and it will have acquired a great deal of heat from the merger shock.
Thus, normal, supramassive, and hypermassive remnants are all possible,
depending on the stars' masses and the unknown value of
$M^{\rm max}{}_{\rm TOV}$.  However, while differentially rotating stars are
in equilibrium on a {\it dynamical}
timescale, they evolve on the secular timescale of effects that transport
angular momentum ($\sim 10$\,ms).  Similarly, the equilibrium will be
adjusted by loss of thermal support on the neutrino cooling timescale
($\sim $sec).  Except in the unlikely event that it sheds enough mass to
drop below the supramassive limit, a hypermassive remnant will ultimately
collapse on one of these timescales.  Thus, the main outline of the
post-merger evolution seems
to depend on one parameter, the mass of the binary, and one EoS-related
number, the neutron star maximum mass.

\subsection{Posing the problem: recasting the field equations as an
initial value problem} 
\begin{figure}
  \includegraphics[width=\columnwidth]{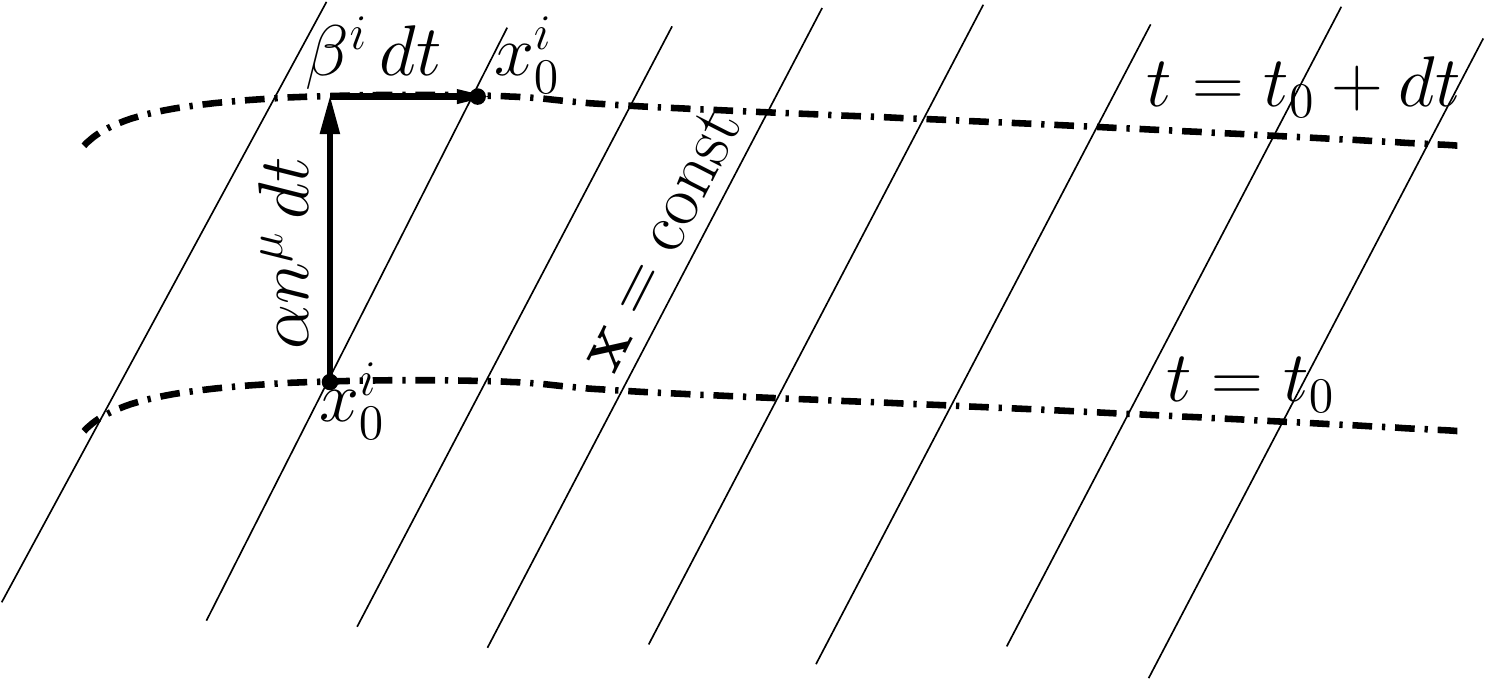}
  \caption{The standard 3+1 coordinate system. Here each point on a
    given spatial slice is specified with a spatial coordinate $x^i$.
    The points labeled with the same value of $x^i$ on two different
    hypersurfaces are connected by a curve (denoted by $x={\bf
    const}$) that is offset from the curve normal to the spatial
    slices (these would be vertical lines in the plot).The shift vector $\beta^i$
  measures how skewed the curves of constant $x^i$ are from the curves
normal to the spatial surfaces and the lapse function $\alpha$ 
measures how far in proper time one slice is from another for
observers traveling along the normal directions. In general
relativity, there is complete freedom in specifying both $\alpha$ and
$\beta^i$ as functions of both space and time.}\label{fig:3+1}
\end{figure}

Returning to the Einstein equations themselves,
similar to how the 4-vector $A^\mu$ in
electromagnetism is not unique due to gauge freedom, the metric that
satisfies Eq.~(\ref{EE}) (and any relevant boundary conditions) is not
unique. In general relativity, the gauge freedom comes in the form of
the freedom to choose coordinates arbitrarily. In order to get a
unique solution, we need to impose gauge conditions.  Many, but not
all, formulations of the Einstein equations for numerical simulations
use what is known as a 3+1 decomposition~\cite{Arnowitt62} (see also recent
texts on numerical relativity~\cite{BaumgarteBook2010,
AlcubierreBook2008, 2016nure.book.....S}). In a 3+1 decomposition, the
coordinates are constructed by using a family of non-intersecting,
spatial hypersurfaces~\footnote{In ordinary Euclidean geometry, a
surface can be obtained by considering the level sets of some function
of space $f(x,y,z)$ (i.e., the points where $f(x,y,z) = {\rm const}$).
A hypersurface is the generalization of this to higher dimensions. A
spatial hypersurface is one where all possible curves on the
hypersurface are spacelike.} The basic setup is illustrated in
Fig.~\ref{fig:3+1}.  The spacetime is split into spatial hypersurfaces
labeled by a coordinate $t$. On each spatial slice, coordinates $x^i$
are specified (we use the convention that Latin indices take on the
values 1, 2, or 3 of the spacelike dimensions) . A point labeled $x^{i}_0$ on one spatial slice and
another with the same label on a different slice may be skewed with
respect to the unit normal direction $n^\mu$ (which must be timelike).
Here, the two points are shifted with respect to each other by a
spatial vector $\beta^i$. If a particle moves from spatial slice $t_0$
to $t_0 + dt$ along the normal, it will experience a proper time
interval of $\alpha\, dt$, where $\alpha$ is known as the lapse
function. The lapse function and shift vector $\beta^i$ are freely specifiable as
a functions of both the spatial and time coordinates. The choice of
lapse function determines how {\it far} neighboring spatial
hypersurfaces are from each other.

On each $t={\rm const}$ hypersurface, there is an induced metric
$\gamma_{ij}$. The induced metric
is simply the spatial components of the spacetime metric
$\gamma_{i j} = g_{i j}$. On the other hand, the 4-dimensional metric
can be specified by providing the lapse function, shift vector, and
spatial metric. The 4-metric has the form
$$
g_{\mu \nu} = \left(\begin{array}{llll} g_{00} & \beta_1
  & \beta_2 & \beta_3 \\
  \beta_1 & \gamma_{11} & \gamma_{12} & \gamma_{13} \\
  \beta_2 & \gamma_{21} & \gamma_{22} & \gamma_{23} \\
  \beta_3 & \gamma_{31} & \gamma_{32} & \gamma_{33} \\
  \end{array}\right),
$$
where $g_{00} = -\alpha^2 + \gamma_{i
j}\beta^i \beta^j$ and $\beta_i = \gamma_{ij} \beta^j$.
While the surface normal has components $n^\mu = (1, -\beta^1, -\beta^2,
-\beta^3)/\alpha$ and $n_\mu = -\alpha \nabla_\mu t$.

The 3-metric $\gamma_{ij}$ and its matrix inverse $\gamma^{ij}$ can be used
to define a covariant derivative, Christoffel symbols, Riemann tensor, and
Ricci tensor. The formulas for these are nearly identical to those
presented in Sec.~\ref{sec:field_eq}, with the exception that indices only take on
the values 1, 2, and 3. We will distinguish these 3-dimensional tensor
(and tensor-like objects) from their 4-dimensional counterparts by
either using different symbols (e.g., using the symbol $D_i$ to
indicate the 3-dimensional covariant derivative), or by prepending a
superscript 3 surrounded by parentheses (e.g., ${}^{(3)}R_{ij}$).

Using the surface normal $n^\mu$, we can define a spatial tensor
$K_{ij}$ known as the extrinsic curvature, by projecting the tensor
$
(\nabla_\mu n_\nu + \nabla_\nu n_\mu)/2
$
onto the slice.
The resulting tensor is related to the time derivative of
$\gamma_{ij}$ by
$\partial_t \gamma_{ij} = - 2 \alpha K_{ij} + D_i \beta_j + D_j
  \beta_i$,
 where  $D_i$ is the covariant derivative associated with
$\gamma_{ij}$.

Finally,
with these choices, the ten Einstein equations become six evolution
equations for  $\gamma_{ij}$ and four constraint equations. This is
analogous to the way the Maxwell equations split into evolution
equations for $\vec E$ and $\vec B$ and two constraint equations for
${\rm div}\vec E$ and ${\rm div}\vec B$.

The resulting field equations, usually known as the
Arnowitt-Deser-Misner~\cite{Arnowitt62} (ADM)
equations, but are actually a reformulation of the standard ADM
equations by York~\cite{1979sgrr.work...83Y},  are given by (see,
e.g.,~\cite{BaumgarteBook2010, AlcubierreBook2008, 2016nure.book.....S})
\begin{eqnarray}
  \partial_t \gamma_{ij} &=& - 2 \alpha K_{ij} + D_i \beta_j + D_j
  \beta_i, \label{eq:ADM_EV1}\\
  \partial_t K_{ij} &=& -D_i D_j \alpha + \alpha({}^{(3)}R_{ij} - 2
  K_{ik}K^{k}_{\, j}) \nonumber\\
  && - 8 \pi \alpha (S_{ij} - \frac{1}{2} \gamma_{ij}
  (S-\rho)) \nonumber\\
  &&+ \beta^k D_k K_{ij} + K_{ik} D_j \beta^k + K_{kj}D_i
  \beta^k, \label{eq:ADM_EV2}\\
  16\pi\rho &=&{}^{(3)}R + K^2 - K_{ij}K^{ij}, \label{eq:ADM_CONST1}\\
  8\pi S^i &=& D_j(K^{ij} - \gamma^{ij}K), \label{eq:ADM_CONST2}
\end{eqnarray}
where
${}^{(3)}R_{ij}$ and ${}^{(3)}R$ are the Ricci curvature tensor and Ricci
scalar
associated with $\gamma_{ij}$ and the source terms are given by
\begin{eqnarray}
  \rho = n_\mu n_\nu T^{\mu \nu},\\
  S^i = - \gamma^{ij} n^\mu T_{\mu j},\\
  S_{ij} = T_{ij},\\
  S  = \gamma^{ij} S_{ij}. \label{eq:ADM_source}
\end{eqnarray}
Equations~(\ref{eq:ADM_EV1}) and (\ref{eq:ADM_EV2}) form the evolution equations, while
Eqs.~(\ref{eq:ADM_CONST1}) and (\ref{eq:ADM_CONST2}) are the Hamiltonian and momentum constraint
equations, respectively.
In Eqs.~(\ref{eq:ADM_EV1})-(\ref{eq:ADM_source}) Latin indices are
raised and lowered with the spatial metric $\gamma_{ij}$, i.e.,
$\beta_j = \gamma_{i j} \beta^i$ and $\beta^i = \gamma^{ij} \beta_j$.
The tensor $\gamma^{ij}$ is the matrix inverse of $\gamma_{ij}$.
Greek indices are raised and lowered with the full metric $g_{\mu
\nu}$ and its inverse $g^{\mu \nu}$.

\section{Numerical relativity formalisms and techniques}
\label{sec:NR}

There are two major classes of techniques used for numerical
simulations of the Einstein equations. These are finite-difference
methods, typically coupled to adaptive mesh refinement techniques, and
pseudospectral methods. To understand how these methods work, we will
consider some toy problems. Most of the finite difference codes are
based on modifications to the ADM system (see
Sec.~\ref{sec:well_posed}). These equations
are in a form with mixed second and first derivatives.
Basically, the system is such that only first time derivatives occur, but first and second
spatial derivatives occur.  (Of course, auxiliary evolution variables can be introduced
so that the system only has first spatial derivatives, but at the cost of
introducing additional constraints.)  A good
toy problem to illustrate how such equations are evolved is thus
\begin{eqnarray}
  \partial_t \Pi - \beta^i \partial_i \Pi = -\nabla^2 \Phi,\nonumber\\
  \partial_t \Phi - \beta^i \partial_i \Phi = \Pi,\label{eq:toy}
\end{eqnarray}
where the spatial coordinates will be denoted by $x,y,z$. We will
explore numerical techniques for solving Eq.~(\ref{eq:toy}) in the
next section.

\subsection{Finite differencing}

To solve Eq.~(\ref{eq:toy}), we consider a discrete grid labeled with
3 integer indices
$(i,j,k)$, where the values of $x$, $y$, and $z$ at a point $(i,j,k)$
are given by
$(x_0 + i\ dx, y_0 + j\ dy, z_0 + k\ dz)$. Furthermore, we denote the
values of a function $f(x,y,z)$ on this grid by $f_{i, j, k}$. We then
approximate spatial derivatives using these points. For example,
\begin{equation}
\partial_x^2 f(x,y,z) =  \left(f_{i+1,j,k} + f_{i-1,j,k} - 2 f_{i, j,
k}\right)/dx^2 + {\cal O}(dx^2), 
\end{equation}
is an approximation to $\partial_x^2 f$ using a three point stencil.
By using more points in the stencil, this derivative can be made more
accurate in the sense that the error will scale with higher powers of
$dx$. Modern numerical relativity codes tend to use sixth-order to eighth-order finite
differencing~\cite{Husa:2007hp, Lousto:2007rj, Loffler:2011ay}. With
these approximations, Eq.~(\ref{eq:toy}) becomes
\begin{eqnarray}
  \partial_t \Pi_{i,j,k} = \sum_{l,m,n} (C^{i,j,k}_{l, m, n}
  \Pi_{l,m,n} + D^{i,j,k}_{l, m, n} \Phi_{l,m,n}),\\
  \partial_t \Phi_{i,j,k} = \sum_{l,m,n} (E^{i,j,k}_{l, m, n} \Pi_{l,m,n}
  + F^{i, j, k}_{l,m,n} \Phi_{l,m,n}),
\end{eqnarray}
where the coefficients $C, D, E, F$  and the values of $(l,m ,n)$ in
the sums are determined by the finite
difference stencil used. Thus, the partial differential equation
(\ref{eq:toy}) becomes a set of coupled ordinary differential
equations for $\Pi_{i,j,k}$ and $\Phi_{i,j,k}$. These equations are
then typically solved using standard Runge-Kutta
techniques~\cite{1995tpdm.book.....G}.

\begin{figure}
  \includegraphics[width=.9\columnwidth]{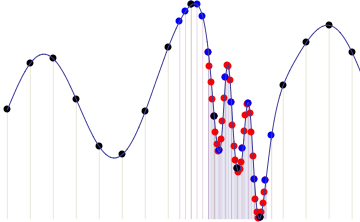}
  \caption{Schematic of how  mesh refinement works in one dimension.
    Shown are the values of a function at discrete points along the
    $x$ axis. A coarse grid
    (black) covers the entire domain and progressively finer grids  (blue and red)
    cover the parts of the domain where the function varies rapidly.  
}
\label{fig:amr}
\end{figure}

A major drawback of the above technique is that it is extremely
computationally wasteful. For convenience here, we will assume that
$dx = dy = dz = h$. The smaller the grid spacing $h$
the smaller the error, but the maximum value of $h$ one could use and
still have an acceptably accurate solution generally varies quite
strongly over space (e.g., many points are needed to resolve black
holes, but few needed far away). Furthermore, the regions where high
resolution (i.e., small values of $h$) are needed tend to be quite
small. Thus, if one used a uniform grid capable of resolving the
entire space, essentially all the calculation time would be spent
evolving the overresolved regions. This problem can be ameliorated to
some extend by choosing special coordinates~\cite{Baker:2001sf,
Zlochower:2005bj, Zilhao:2013dta} that concentrate gridpoints in
certain regions. The state-of-the art technique for overcoming this
inefficiency is the use of adaptive meshes~\cite{Berger:1984zza,
  Schnetter:2003rb, Imbiriba:2004tp, Bruegmann:2006at, Anderson:2007kz}. In an adaptive mesh code,
a coarse grid covers the entire computational domain, with
increasingly finer grids placed in location where high resolution is
required (see Fig.~\ref{fig:amr}).

Evolutions also involve a discretization in time, and the smaller
the timestep $dt$, the more steps are needed to cover a given time interval,
and the more expensive the simulation.  Explicit time integration methods
are subject to the Courant-Friedrichs–Lewy stability
condition~\cite{Courant1928}, which limits
$dt$ on a mesh to be less than around $h/v_s$, where $v_s$ is the maximum
signal speed, which for spacetime evolution is the speed of light.  The
effect on $dt$ is a price to be paid for smaller $h$.

\subsection{Pseudospectral methods}

The other major techniques used in black-hole simulations fall under
the category of pseudospectral methods~\cite{Kidder:2000yq,
Boyle:2006ne,
Scheel:2006gg, Ansorg:2004ds, Tichy:2006qn}. In spectral methods the
evolved fields are expressed in terms of a finite sum of basis
functions. An example of this would be to describe a field on a sphere
in terms of an expansion in spherical harmonics. These methods have
the advantage that if the fields are smooth\footnote{more precisely the fields
are smooth on the real axis and can be analytically continued into the
complex plane} then the error in truncating the expansion converges to
zero exponentially with the number of basis functions used in the
expansion.  In pseudospectral methods, values of functions are stored
at special gridpoints, the colocation points, corresponding to
Gaussian quadrature points of the basis functions~\cite{Boyd1999}. 
Codes can then
transform between spectral and colocation-point representations via
Gaussian quadrature.  This is especially useful for computing products
and other pointwise operations which are much simpler using gridpoints. 
In pseudospectral form, spectral methods can be thought of as a
particular limit of finite differencing, the limit that uses the
entire domain as its stencil so as to make the highest-order derivative
operator~\cite{Fornberg1996}.  The order of this operator will then
increase with the
number of colocation points, giving the method faster convergence than
a fixed polynomial order.  The payoff is that differentiations and
interpolations become more expensive much more quickly than fixed-stencil
finite difference methods as resolution is increased.  Also, exponential
convergence is lost for functions that are only smooth to finite order.
Perhaps the best known numerical relativity code that uses pseudospectral techniques is
the {\it Spectral Einstein Code}~\cite{Kidder:2000yq, Boyle:2006ne,
Scheel:2006gg}, or SpEC, used by the SXS (simulating
extreme spacetimes)\footnote{{\tt{https://www.black-holes.org/}}}
collaboration.

The choice of finite difference versus spectral methods affects which method
of handling black holes is easier to implement.  A black hole interior
presents a major challenge to any numerical technique because of the
curvature singularity it harbors (see, e.g., Wald~\cite{Wald:1984rg} for a
discussion on the inevitability  of forming curvature
singularities).  Fortunately, the singularity is
concealed behind an event horizon.  The region inside the horizon cannot
affect the exterior solution, so numerical simulations need not evolve
it accurately.  They only need to keep it from causing the simulation to
crash.  One way to do this is to simply not evolve a region inside the
horizon, i.e., to {\it excise} this region.  Spectral methods can do this
naturally, because even near the inner edge of the grid, no points are
needed from the other side of the excision boundary to take derivatives. 
A boundary condition physically should not be needed (because no information
flows out of a region where all characteristic speeds go inward),
and none is required.  The other method, the {\it puncture} method, described
in detail below, involves allowing singularities in the computational
domain.  In the appropriate gauge, these singularities are sufficiently
benign that finite difference methods can handle them.  It would be more
difficult to evolve a puncture stably with a spectral
code~\cite{Tichy:2006qn}.

\subsection{Making the problem  well posed}
\label{sec:well_posed}
As will be discussed later, the ADM equations by themselves proved to
be unstable for many strong-field  problems, including black-hole
mergers.
With all the difficulties encountered trying to implement the ADM
equations in the 1990s, emphasis changed to developing new 3+1 systems
and analyzing their well-posedness~\cite{Brodbeck:1998az, Tiglio:2003xm, Calabrese:2002xy, Sarbach:2002gr,
Calabrese:2001kj, Calabrese:2002ei, Kelly:2001kj, Frittelli:2006ap,
Frittelli:2004ye, Frittelli:2000uj, Neilsen:2004aj, Lehner:2004cf,
Calabrese:2002xy, Sarbach:2002bt, Szilagyi:2002kv}. Informally speaking, a hyperbolic system of equations is well
posed if the solution depends continuously on the initial and boundary
data. Ill-posed systems can have solutions that grow without bound
even for very small evolution times. 

To understand how reformulation of the basic evolution equations can
make or destroy well-posedness, consider a simple vector wave equation 
\begin{eqnarray}
  \left(\frac{\partial \vec E}{\partial t}\right) = \vec \nabla \times
  \vec B ,\nonumber \\
  \left(\frac{\partial \vec B}{\partial t}\right) = -\vec \nabla \times
  \vec E ,\label{eq:EM_ev}\\
    \mbox{subject to}\nonumber\\
  {\cal C}_E = \vec \nabla \cdot \vec E = 0,\nonumber\\
  {\cal C}_B = \vec \nabla \cdot \vec B = 0.\label{eq:EM_con}
\end{eqnarray}
This system is well posed in the sense that the solution $(\vec E(t),
\vec B(t))$
depends continuously in the initial data. Any constraint violation
(failure of ${\cal C}_E $ or ${\cal C}_B$ to be zero) will be
preserved (not grow or decrease) by the evolution system.

This system can be transformed into two separate  identical second-order
equations for $\vec E$ and $\vec B$ of the form
\begin{equation}
  \left(\frac{\partial^2 \vec A}{\partial t^2}\right) + \vec
  \nabla\times\vec\nabla\times\vec A = 0.
\end{equation}
This latter system is not quite equivalent to the original system in
that constraint violations now grow linearly in time.
A better system is obtained by noting
that
$\vec \nabla \times \vec\nabla \times \vec A = - \nabla^2 \vec A +
\vec \nabla\left(\vec \nabla \cdot \vec A\right)$. Using this, and the
assumption that $\vec \nabla\cdot \vec A=0$ we get
\begin{equation}
  \Box \vec A
         = 0,
         \label{eq:simple_pde2}
\end{equation}
where 
\begin{equation}
  \Box  = \frac{\partial^2}{\partial t^2} - \nabla^2.
\end{equation}
If we solve Eq.~(\ref{eq:simple_pde2}) with a small divergence, the
norm of the divergence will remain bounded. 
More generally,
if the constraint equations~(\ref{eq:EM_con}) are satisfied, then
any solution to Eq.~(\ref{eq:simple_pde2}) is also a solution to
\begin{equation}
  \Box \vec A + \kappa \nabla\left(\vec \nabla \cdot \vec A\right).
  \label{eq:simple_pde_fixed}
\end{equation}
If $\kappa$  is chosen larger than one, small violations of the 
$\vec \nabla \cdot \vec A=0$ constraint can blow up arbitrarily quickly.
Thus, with seemingly inconsequential changes, we can turn a system
from one with a minor blowup in the constraints to either one with
a catastrophic blowup in the constraints, or no blowup at all.
However, with the addition of an auxiliary field, we can do even
better. Consider the system~\cite{2002JCoPh.175..645D}
\begin{eqnarray}
  \left(\frac{\partial \vec E}{\partial t}\right) = \vec \nabla \times
  \vec B + \vec \nabla \psi_E,\nonumber\\
  \left(\frac{\partial \vec B}{\partial t}\right) = -\vec \nabla \times
    \vec E + \vec \nabla \psi_B,\nonumber\\
    \left(\frac{\partial \psi_E}{\partial t}\right) = - \vec \nabla
    \cdot \vec E - \psi_E,\nonumber\\
    \left(\frac{\partial \psi_B}{\partial t}\right) = - \vec \nabla
    \cdot \vec B - \psi_B.
\end{eqnarray}
For this system, the constraints satisfy
\begin{eqnarray}
  \Box {\cal C}_E - \left(\frac{\partial {\cal C}_E}{\partial
  t}\right) = 0,\nonumber\\
  \Box {\cal C}_B - \left(\frac{\partial {\cal C}_B}{\partial
  t}\right) = 0. \label{eq:damp_const}
\end{eqnarray}
Solutions to Eqs.~(\ref{eq:damp_const}) decay exponentially in time.
Thus if numerical effects introduce constraint violations, these will
be damped away.
The main message  here is that even for a physically motivated
evolution system, the addition or removal of terms nominally equal to
zero can make the difference between a solvable and an insolvable
system. Furthermore, for a constrained system, like the equations of
electromagnetism and general relativity, enlarging the system of
equations by adding new fields  can suppress unphysical constraint
violations.

The standard ADM formulation  is
now known to be ill-posed in the nonlinear 
regime~\cite{Frittelli:2000uj, Frittelli:2004ye}. On the other hand,
many well-posed formulations have been proposed but turned out not to
be an immediate panacea for unstable black hole simulations. Some of
the most influential of these formations are the 
Bona-Masso family~\cite{Bona:1994dr, Arbona:1999ym, Bona:1998dp, Bona:1997hp}, the
NOKBSSN family~\cite{Nakamura:1987zz, Shibata:1995we, Baumgarte:1998te},
Z4 family~\cite{Bona:2002fq, Bona:2002ft, Bona:2003fj, Bona:2004yp,
Bona:2004ky, Alic:2011gg, Gundlach:2005eh, Bernuzzi:2009ex},
the Kidder-Scheel-Teukolsky family~\cite{Kidder:2001tz}~\cite{Alcubierre:2003ti}, and the generalized-harmonic
family~\cite{Garfinkle:2001ni, Pretorius:2004jg, Szilagyi:2002kv}.

 One of the key improvements in numerical simulations prior to the
breakthroughs of 2005 was the introduction of the so-called NOKBSSN
formulation of the Einstein equations in 3+1. This system, which is named  after its developers
Nakamura, Oohara, Kojima, Shibata, Baumgarte, and
Shapiro~\cite{Nakamura:1987zz, Shibata:1995we, Baumgarte:1998te},
modifies the standard ADM equations in several crucial ways. Firstly,
the spatial metric $\gamma_{ij}$ is split into an overall conformal
factor $e^{\phi}$ and a conformal metric $\tilde \gamma_{ij}$, where
$e^{4 \phi} \tilde \gamma_{ij} = \gamma_{ij}$ and the determinant of
$\tilde \gamma_{ij}$ is unity.  This conformal metric has its
corresponding Christoffel symbols ${}^{(3)}\tilde {\Gamma^{k}}_{ij}$ and
Ricci tensor ${}^{(3)}\tilde R_{ij}$.  Second, the three combinations $\tilde
\gamma^{ij}\ {}^{(3)}\tilde {\Gamma^{k}}_{ij} = {}^{(3)}\tilde \Gamma^k$ $(k=1,2,3)$,
as well as the $K = \gamma^{ij} K_{ij}$ are promoted to evolved
variables.  Third, the remaining evolved extrinsic curvature variables are trace-free conformal extrinsic curvature variables
$\tilde A_{ij} = e^{-4 \phi} \left[K_{ij} - (1/3) K \gamma_{ij}\right]$. 
Finally, the momentum constraint equations are used to
modify the evolution equations for ${}^{(3)}\tilde \Gamma^k$, which introduces
a constraint damping quality to the system. These changes, in
conjunction with a particular choice of gauge conditions, namely
the use of certain Bona-Masso~\cite{Bona:1994dr}
type lapse conditions  (known as 1+log slicing) and $\Gamma$-driver
shift conditions~\cite{Alcubierre:2002kk} led to the first genuinely
stable, fully nonlinear implementations of the Einstein equations for
systems without symmetries, at least for non-black-hole spacetimes.
The factoring out of the conformal factor $\phi$ proved to be
particularly advantageous for collisions of black
holes~\cite{Alcubierre:2000ke, Alcubierre:2004hr}. However, the state
of the art for black-hole evolutions in the early 2000s only allowed
for \headon 
collisions and grazing collisions~\cite{Alcubierre:2000ke,
Alcubierre:2004hr}, where the black holes merge well before completing
one orbit. The NOKBSSN system also proved to be stable using higher-order
finite-differencing methods~\cite{Zlochower:2005bj} for \headon
collisions, as well.

\begin{figure}
  \includegraphics[width=\columnwidth]{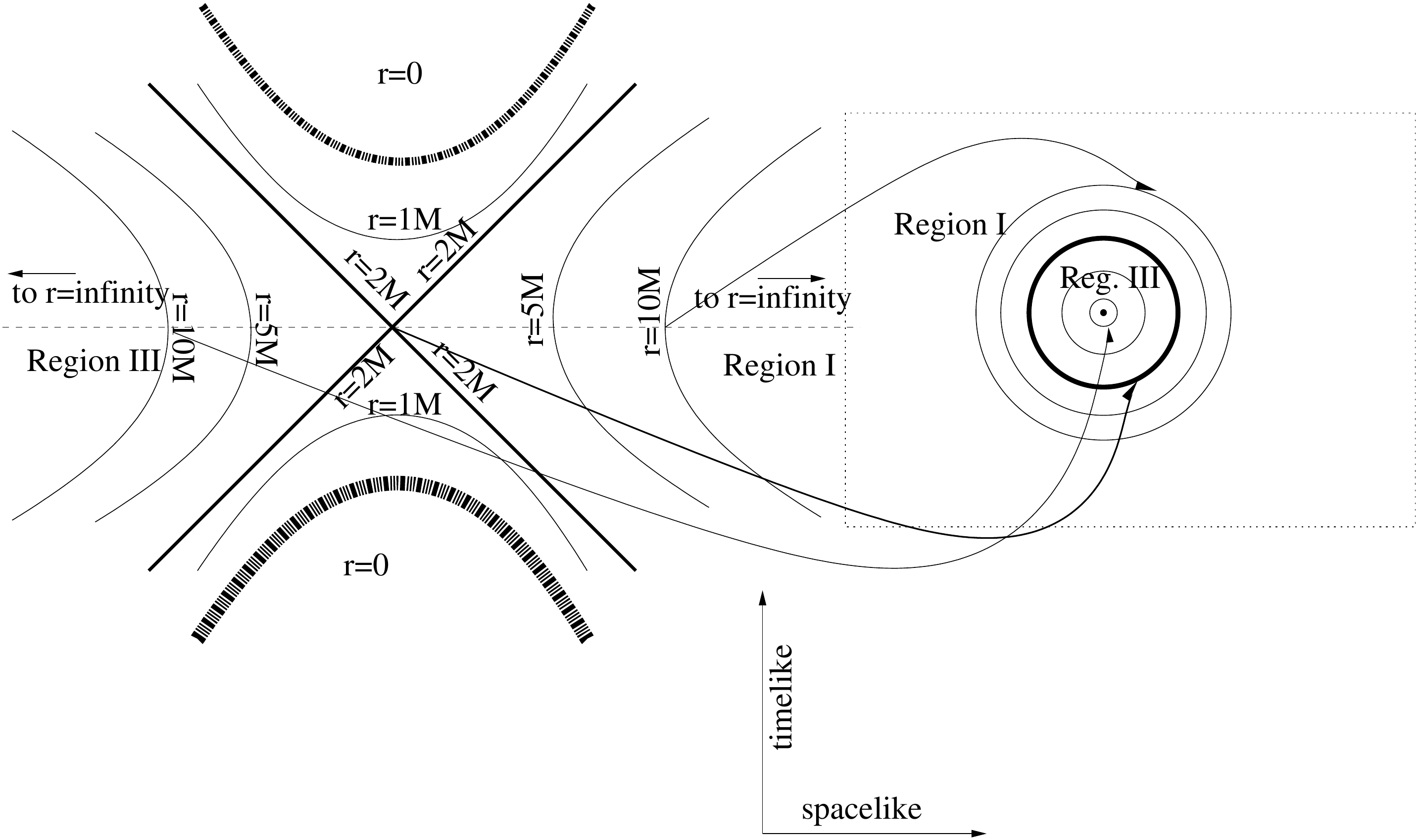}
  \caption{(Left) A spacetime diagram of a Schwarzschild black hole with the
    $(\theta, \phi)$ angular coordinates suppressed. Each point represents
a sphere of radius $4 \pi r^2$. Note that there are two curves
corresponding to each value of $r$. Radially ingoing and outgoing
light rays travel along $45^\circ$ lines in this diagram. The event
horizon(s) correspond to the $r=2M$ diagonal curves. The dotted line
represents a spatial slice with a two-sheeted topology. (Right)
The spatial surface corresponding to the dotted line shown with one
spatial dimension ($z$) suppressed. Points on the dotted line
correspond to circles here. The
horizons correspond to the thick circle and the $r=5M$ and $r=10M$
curves each map to one circle inside the horizon and one outside. The
central point maps to the $r=\infty$ of Region
III.}\label{fig:puncture}
\end{figure}

A key technique used in many of these early evolutions was the fixed-puncture
formalism~\cite{1988CMaPh.119...51F, Brandt:1997tf}. In this
formalism, a two-sheeted Einstein-Rosen bridge associated with a single black
hole is mapped into a single sheet with a singularity at the center.
As shown in Fig.~\ref{fig:puncture}, the standard Schwarzschild spacetime
can be recast as a puncture by taking a spatial slice that passes
through the bifurcation sphere\footnote{In the extended Schwarzschild
spacetime, also called the Kruskal extension, there is both a black
hole and a white hole. The bifurcation sphere is the
point where the two horizons meet(see Fig.~\ref{fig:puncture})}.
On either side of the sphere, the
slice extends infinitely far. Next we introduce a coordinate
${\cal R}$, which is related to the usual Schwarzschild coordinate $r$ by
\begin{equation}
  r = {\cal R} \left(1+\frac{M}{2{\cal R}}\right)^2.
\end{equation}
The spatial metric then takes in the form
\begin{equation}
ds^2 = \psi^4 \left(d{\cal R}^2 + {\cal R}^2 d\Omega^2\right),
\end{equation}
where
$\psi = 1+M/(2{\cal R})$. Here ${\cal R}=0$ and ${\cal R}=\infty$ both correspond to
$r=\infty$ and the metric is singular at ${\cal R}=0$.
This singularity is not the curvature blow-up singularity at the {\it
center} of the black hole, that singularity is in the future of this
slice, rather the singularity is entirely gauge and results from us
{\it stuffing} an entire asymptotically flat universe into the sphere
${\cal R}=M/2$.
This singularity is further only present
in the NOKBSSN function $\phi$, the components of NOKBSSN
conformal metric $\tilde \gamma_{ij}$ are
nonsingular. In the fixed puncture approach, there are singularities
of this type associated with each black hole and the gauge conditions
are chosen so that these singularities do not move.  Each of these
singularities is called a ``puncture''.

Keeping the puncture fixed has several advantages. First, the
singularity in the conformal factor can be handled analytically.
Second, by keeping the black holes fixed in coordinate space, one can
use the much simpler fixed excision techniques. Using these techniques,
Br{\"u}gmann, Tichy, and Jansen~\cite{Bruegmann:2003aw} and
later Diener {\it et al.}~\cite{Diener:2005mg} were able to evolve a
quasicircular binary for roughly one orbit. However, they were still
not able to get the merger waveform using these techniques.

One approach that was able to get the merger waveform from these early
simulations was the Lazarus method~\cite{Baker:2000zm, Baker:2001nu,
Baker:2001sf, Baker:2002qf, Baker:2003ds, Campanelli:2005ia}, which
used the numerical simulation to generate initial data for a
subsequent perturbative evolution of the radiative scalar $\psi_4$. 
(Measures of gravitational radiation, including $\psi_4$, are
described in Section~\ref{sec:waveforms}.)
The key to the success of the Lazarus approach was that when two black
holes are close enough, even though they have not merged yet, the
exterior spacetime can be well described by black hole perturbation
theory (i.e., the close-limit approximation~\cite{Price:1994pm}).
A subsequent evolution of $\psi_4$ on a carefully chosen black-hole
background is then used to evolve the gravitational radiation to
infinity.

\subsection{Breakthroughs in numerical relativity}
\label{sec:Breakthrough}
Perhaps one of the most significant breakthroughs in numerical
relativity occurred when Pretorius, who had previously developed an
adaptive-mesh-refinement (AMR) code based on the generalized harmonic
system~\cite{Pretorius:2004jg}, included constraint damping
techniques~\cite{Brodbeck:1998az} developed for the Z4 system by
Gundlach {\it et al.}~\cite{Gundlach:2005eh}. The key development
there was that the general relativistic action can be extended to
include terms that vanish when the constraints are satisfied, but act
to damp these constraint violations when they are nonzero.  When
applied to the generalized harmonic system used by Pretorius, the
constraint violations for a \bbh remained bounded and
Pretorius was thus able to perform the first successful evolution of an
orbiting binary~\cite{Pretorius:2005gq}.
Pretorius presented his initial results at the Banff International
Research Station Workshop on
Numerical Relativity in April
2005~\footnote{http://bh0.physics.ubc.ca/BIRS05/}.

\begin{figure}
  \includegraphics[width=\columnwidth]{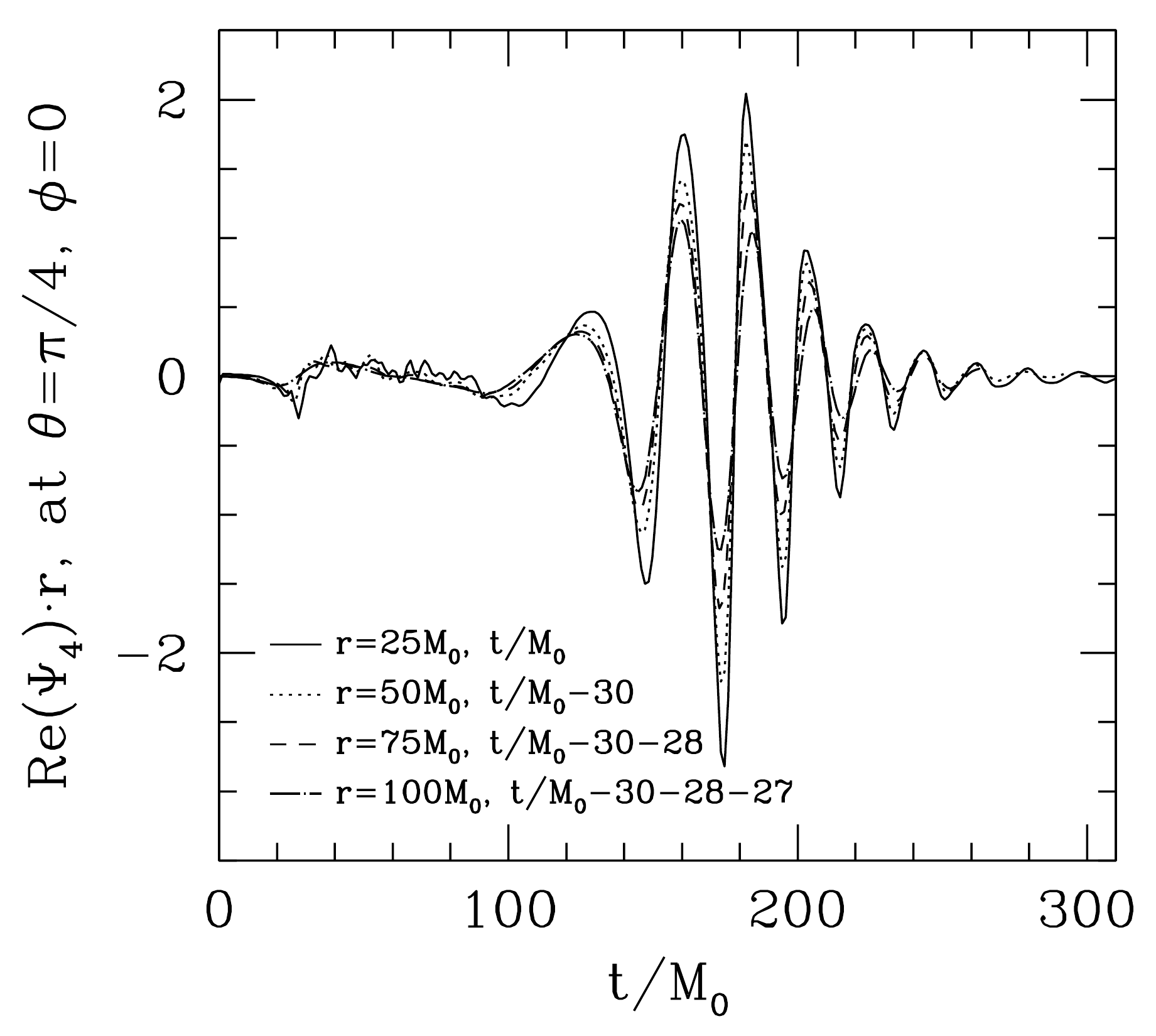}
  \caption{A reproduction of the waveform shown in Figure 3 of
    Ref.~\cite{Pretorius:2005gq} courtesy of the author. This was from the
      first fully nonlinear numerical simulation of the
        last orbit, merger, and ringdown of a
        \bbh. It was generated by
      Pretorius
    using the Generalized Harmonic Coordinate approach. The curves
  show the waveforms as calculated at various radii and translated in
time.}\label{fig:pretorius}
\end{figure}

The actual system Pretorius evolved was the Einstein equations coupled
to a scalar field. The initial data consisted of two scalar boosted
{\it stars} with supercritical densities. The stars collapsed into
two black holes that then orbited and merged.
Figure~\ref{fig:pretorius} is a reproduction of Figure 3 in of
Pretorius' paper~\cite{Pretorius:2005gq}. It shows the very first merger waveform from
an orbiting \bbh ever published. 

Just four months after Pretorius submitted his groundbreaking paper, a
new breakthrough was announced that became known as the {\it Moving
Punctures Approach}~\cite{Campanelli:2005dd, Baker:2005vv}. This new
method was developed independently by the groups then at the
University of Texas at Brownsville (UTB) and NASA Goddard Space Flight
Center (GSFC) and first demonstrated publicly in the {\it Numerical
Relativity 2005: Compact Binaries} workshop at NASA
GSFC~\footnote{https://astrogravs.gsfc.nasa.gov/conf/numrel2005/}.
Notably, the {\it moving punctures} approach allowed groups worldwide
to evolve \bbhs.  It is based on an extension of the
standard NOKBSSN system, with several important changes. (1) The
singular conformal factor is replaced by a non-singular function $\chi
= e^{-4 \phi}$ (although evolutions with $\phi$ itself are also
used~\cite{Baker:2005vv}) which is evolved fully numerically.  (2) The
gauge conditions explicitly allow the punctures to move.  Previously,
the shift condition was chosen so that the punctures could not move.
(3) The standard 1+log lapse condition 
\begin{equation}
  \partial_t \alpha = -2 \alpha K,
\end{equation} 
would require $K$ to be singular at the puncture in order for the
lapse to change from a zero value when the puncture is on a given
point to  a non-zero value when the puncture has passed. This {\it
singular} behavior is removed by changing the lapse condition to an
advection equation 
\begin{equation}
  \partial_t \alpha - \vec \beta \cdot \vec\nabla \alpha = -2 \alpha
  K.
  \end{equation}
Finally, within the evolution equations for $\tilde \Gamma^i$ there is
a singular term on the puncture location  proportional to the lapse.
By choosing an initial lapse that is identically zero on the puncture,
this singularity is also removed.
 With these changes, the gauge naturally evolves such that the
black holes orbit each other and inspiral in coordinate space. The
pathologies seen with the fixed-puncture approach vanished with these
new dynamic punctures. Furthermore, because the moving punctures
approach was similar to existing codes, groups around the world were
able to rapidly develop their own versions. These include the BAM
code~\cite{Bruegmann:2006at} developed at Jena, the Maya-Kranc code
developed at Penn State~\cite{Herrmann:2007cwl}, as well as the original
codes, LazEv developed at The University of Texas at Brownsville, and Hahndol, developed at GSFC. More
recently, the publicly available EinsteinToolkit~\cite{Loffler:2011ay,
einsteintoolkit}
code included an open-source implementation, known as McLachlan. 
LazEv, Maya-Kranc, and McLachlan all used the Cactus Computational
Toolkit~\cite{cactus_web}, originally developed at the Albert Einstein
Intsitute in Golm, Germany.

There were several
significant differences between Pretorius' techniques and the {\it Moving Punctures approach}.
Unlike in the {\it Moving Punctures approach}, the system Pretorius
developed used excision to handle the black hole
singularities, compactified the computational domain to include
spatial infinity, and of course, used the generalized harmonic
system with constraint damping. This system was
sufficiently unlike the other techniques used by numerical relativists
at the time that it was only slowly adopted. 

\begin{figure*}
  \includegraphics[width=0.48\textwidth]{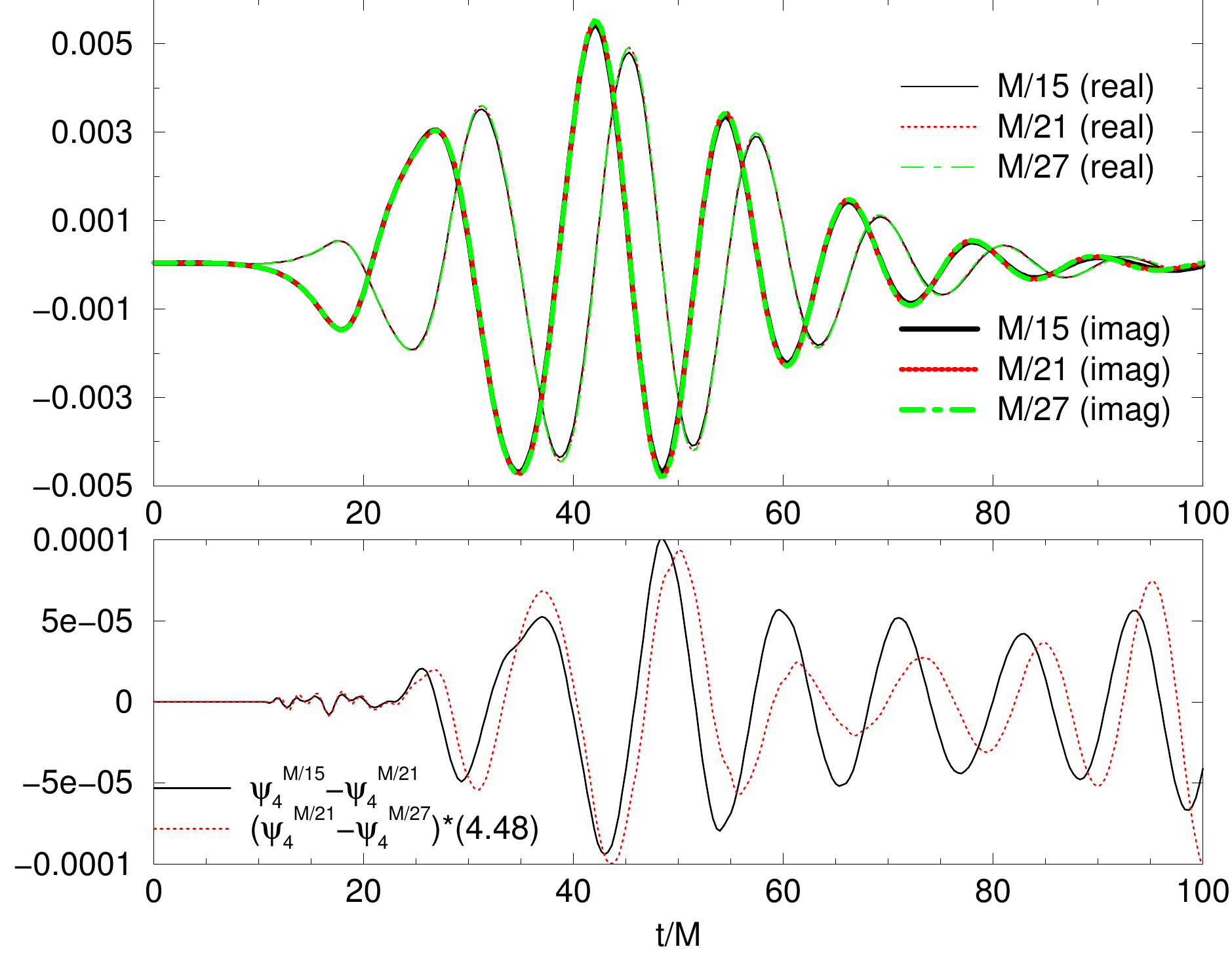}
  \includegraphics[width=0.48\textwidth]{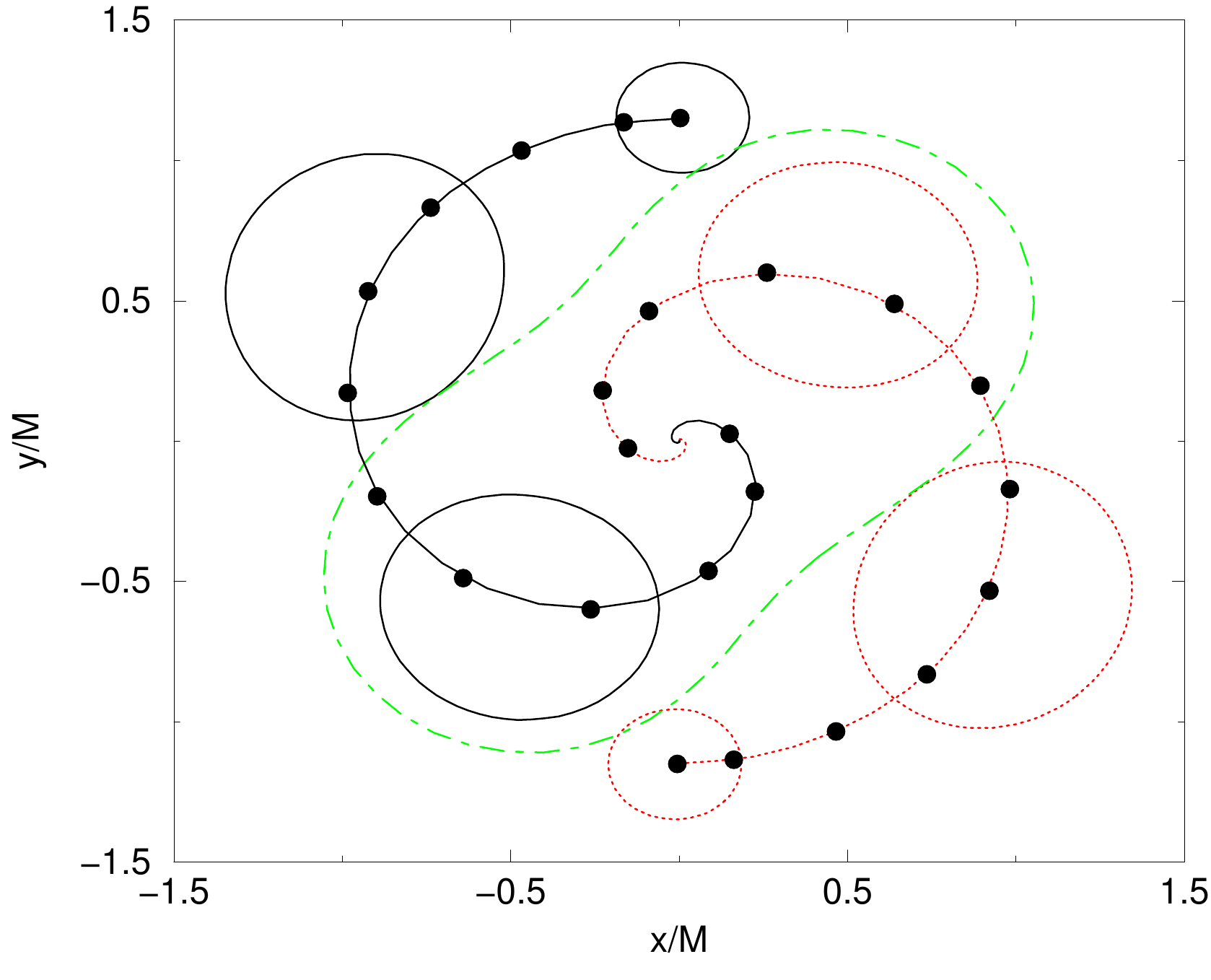}
\\
  \includegraphics[width=0.48\textwidth]{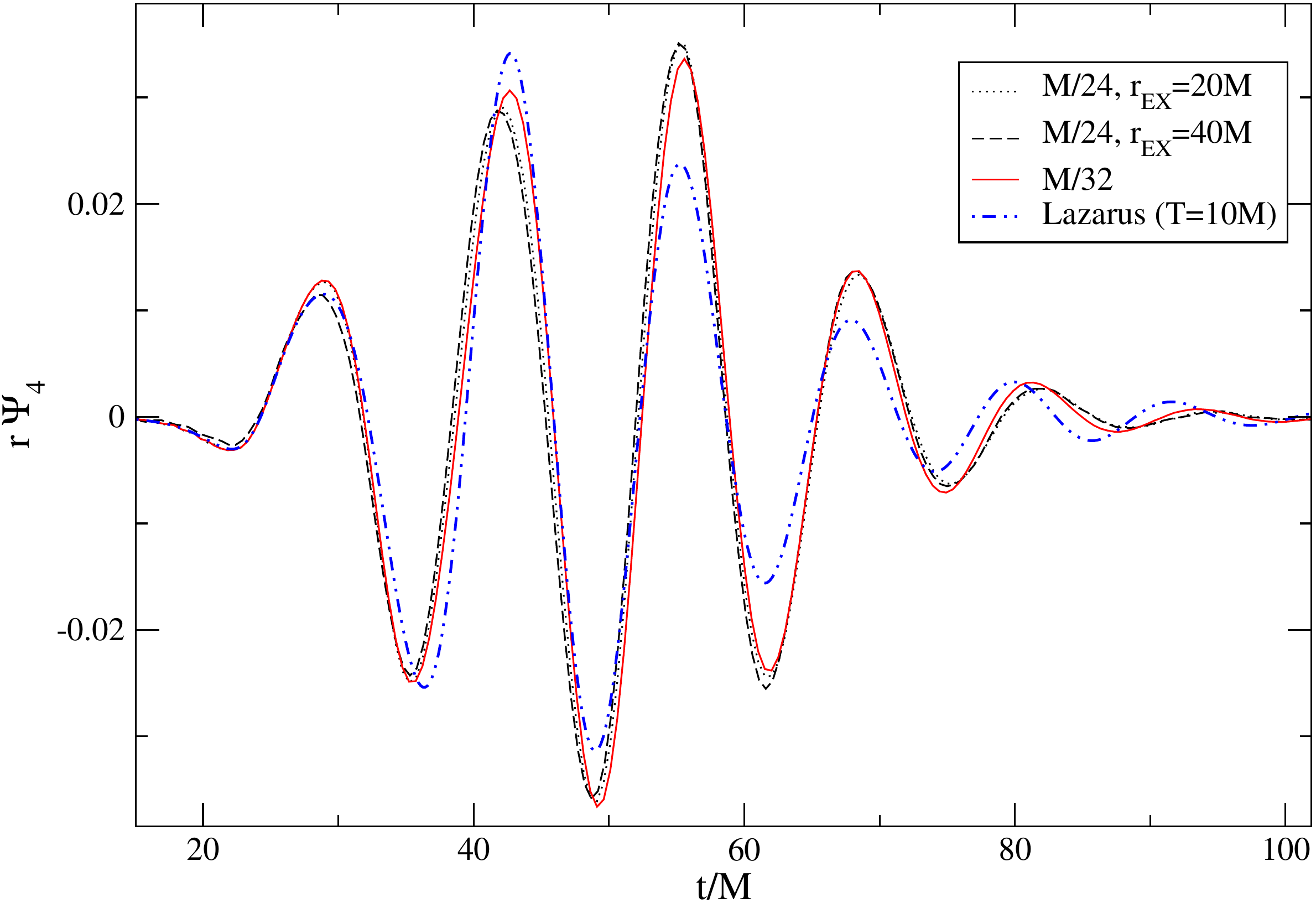}
  \includegraphics[width=0.48\textwidth]{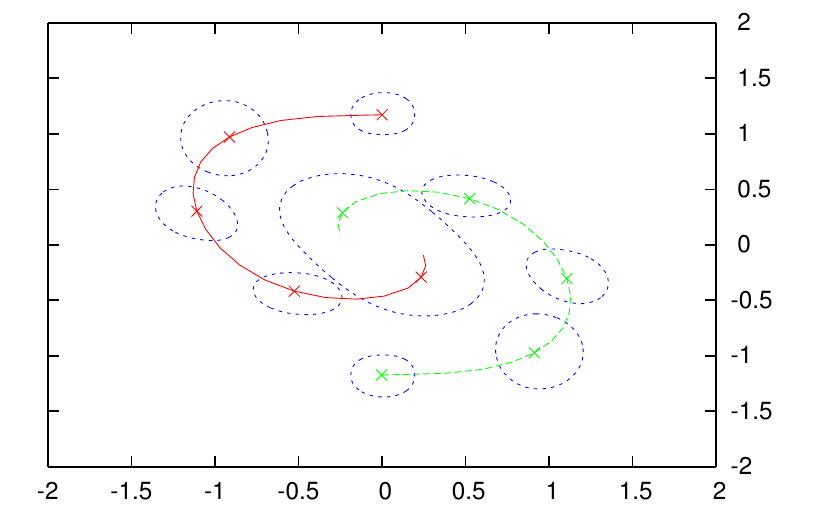}
  \caption{A reproduction of the waveform, black-hole trajectories, and horizons
    calculated in the Moving Punctures breakthrough
  papers~\cite{Campanelli:2005dd, Baker:2005vv} courtesy of the
  authors.
    The top panels are from~\cite{Campanelli:2005dd}, while the lower ones are
from~\cite{Baker:2005vv}. The two papers were published in the same
volume of Physical Review Letters. The top-left panels show the
real and imaginary parts of the $(\ell=2, m=2)$ mode of the waveform
at various resolutions, as well as a convergence study. The top-right
panel shows the individual horizons, first common horizon, and puncture
trajectory. The bottom-left panel shows a comparison of waveforms at
different extraction radii and resolution with the prediction of the
Lazarus approach. The bottom right panel also shows the individual horizons,
first common horizon, and puncture 
trajectory.}\label{fig:mp}
\end{figure*}

Figure~\ref{fig:mp} shows reproductions from the breakthrough papers
using the moving punctures approach. The waveforms, with various
analyses, and the horizons are shown.

Soon after the announcement of the {\it moving punctures} breakthrough,
simulations were reported where the binary completed more than one
full orbit~\cite{Campanelli:2006gf}
and then multiple orbits~\cite{Baker:2006yw}.
The latter, in particular, compared the merger waveforms
from simulations starting at various separations and found that the
merger waveform was insensitive to initial conditions.
This was followed shortly afterwards by the discovery of the orbital
hangup
effect~\cite{Campanelli:2006uy} for spinning binaries. This effect
either delays or accelerate the merger depending on whether the spins of
the two black holes are (partially) aligned or counteraligned with the
orbital angular momentum, and proved to be important for
parameters estimation of LIGO
sources~\cite{TheLIGOScientific:2016wfe}.  Some of the important
discoveries that proceeded from these simulations will be described below
in Sec.~\ref{sec:bbh}.

\subsection{Methods for evolving the fluid equations}

\subsubsection{Conservative formulation}

Many of the early numerical relativity hydrodynamic
simulations used a formulation introduced by Wilson in
1972~\cite{1972ApJ...173..431W}.  In Wilson's scheme, the
evolution variables are
\begin{equation}
  (\rho_{\star}=W\sqrt{\gamma}\rho_0,E=\rho_{\star}\epsilon,S_i
  =\rho_{\star}h u_i)\ ,
\end{equation}
a variable for rest-mass density, internal energy density, and
momentum density, respectively.  Often, there are numerical advantages
to evolving entropy rather than internal energy, so some codes
(e.g.~\cite{Shibata:1999hn,Duez:2002bn}) specializing to Gamma-law EoS
evolved a variable
$e_{\star}=W\sqrt{\gamma}(\rho_0\epsilon)^{\Gamma}$.  The equations
can be finite differenced in a conservative form:  fluxes are
calculated at cell interfaces; the same flux added to one grid cell is
removed from its neighbor, and no truncation error accrues to the
total rest mass.  However, the variables evolved (in particular $E$)
are not those that
are physically conserved, so an explicit artificial viscosity must be
added to correctly account for shocks.

Shock handling is accommodated more naturally if one evolves the
physically conserved variables, meaning that one should evolve the
total energy density rather than internal energy density.  The
resulting equations can be solved using established high-resolution
shock capturing methods.  This is the path followed in what has come
to be called the Valencia
formulation~\cite{1997ApJ...476..221B,Font:1998hf,Font:2001ew}, which
all current numerical relativity hydrodynamics codes essentially follow.  The evolution
equations take conservative form
\begin{equation}
  \partial_t {\bf U} + \nabla\cdot {\bf F} = {\bf S}
\end{equation}
where the conservative variables $U$ are
\begin{eqnarray}
  {\bf U} &=& (\rho_{\star}=W\sqrt{\gamma}\rho_0, X_i\rho_{\star}, \\
          & &  \tau=\sqrt{\gamma}\alpha^2 T^{00} - \rho_{\star},
                S_i=\sqrt{\gamma}\alpha T^0{}_i)
\end{eqnarray}
Conservative shock-capturing hydrodynamics codes in numerical
relativity have achieved at best 3rd-order convergence~\cite{Radice:2013hxh}.

After computing ${\bf U}$ at a new timestep, it remains to recover the
original (``primitive'') variables such as $\rho_0$ and $u_i$.  It
turns out to be sufficient to recover $W$ and $T$, but this will
involve some sort of root-finding process.

\subsubsection{Excision and punctures in the presence of matter}

If the black hole interior is removed via excision, matter must be
able to flow into the black hole and ``disappear'' in a stable way.
For problems involving collapse to a black hole, one must maintain
accuracy inside the collapsing object until an apparent horizon is
located, after which a region inside this horizon can be excised.
Scheel~{\it et al.}~\cite{Scheel:1994yr} did this in 1D for spherical
collisionless matter.  Next, Brandt~{\it et al.}~\cite{Brandt:1998cv}
introduced a code for evolving nonvacuum black hole spacetimes using
an isometry inner boundary condition at the apparent horizon.
Techniques for matter excision using horizon penetrating coordinates
were introduced roughly simultaneously by Duez~{\it et
  al.}~\cite{Duez:2004uh} and Baiotti~{\it et
  al.}~\cite{Baiotti:2004wn}.  Optimal methods for one-sided
differencing of the fluid equations near the excision boundary are
investigated by Hawke~{\it et al.}~\cite{Hawke:2005zw}, although these
initial simulations were less sensitive to this than to the gauge
choices needed to keep the coordinates horizon penetrating.  Excision
is still the method used for simulations in the generalized harmonic
formulation (e.g. SpEC).  A particularly elegant grid structure for
hydrodynamic excision is provided by the cubed-sphere
arrangement~\cite{Zink:2007xn}.

In NOKBSSN, it is much simpler to use moving puncture gauges and avoid
explicit excision.  One might worry that material inflow into the
puncture would cause numerical problems, but fortunately this turns
out not to be the case.  Numerical experiments showed that puncture
simulations can handle stellar collapse to a black hole~\cite{Baiotti:2006wm}
and spherical accretion into a black hole~\cite{Faber:2007dv}
with no code changes except a small extra dissipation in the metric
evolution (in~\cite{Baiotti:2006wm}) and a means of resetting fluid
variables near the puncture where conservative to primitive variable
recovery fails (in~\cite{Faber:2007dv}).  Shortly after this realization,
Shibata and Uryu carried out the first NOKBSSN black hole-neutron star merger
simulations~\cite{Shibata:2006bs}.

\subsubsection{More physics:  equations of state, neutrinos, magnetic fields}

Information about the properties of the matter enters through the equation
of state.  An extremely simple but nevertheless useful equation of state
is the polytropic law $P=\kappa\rho_0^{\Gamma} = (\Gamma-1)\rho_0\epsilon$,
where $\kappa$ and $\Gamma$
are constants.  Higher $\Gamma$ means stiffer EoS.  A notable feature
of this EoS is that it is barotropic; there is no explicit temperature
dependence, meaning the matter must be degenerate or the temperature
must itself be a function of density (as, for example, in an isentropic
gas).  In many cases, we may wish to allow an initially polytropic gas
to pick up added thermal pressure and internal energy via shock heating. 
In this case, one uses the more general Gamma-law EoS
$P=(\Gamma-1)\rho_0\epsilon$, where $\epsilon$ is now given not by the
polytropic law but by the energy density evolution equation.  For adiabatic
evolution, the polytropic law should be
maintained.  One gets a surprising amount of mileage out of this simple
EoS family.  Nonrelativistic ideal degenerate Fermi gases have $\Gamma=5/3$;
relativistic ideal degenerate Fermi gases have $\Gamma=4/3$; stars with
both radiation and gas pressure with a constant fraction of the total from
each have $\Gamma=4/3$.  Neutron stars are not polytropes, but much of
the early numerical relativity work involving neutron stars modeled them as $\Gamma=2$
polytropes.

Ultimately, an accurate treatment of dense matter is needed.  For a
general astrophysical gas, there may be many composition variables
$X_i$, each in need of its own evolution equation.   However, when
dealing with high densities and temperatures above $\sim $MeV, the
matter can be assumed to be in nuclear statistical equilibrium, in
which case there is only one composition variable, the proton fraction
or electron lepton number fraction $Y_e= n_p/(n_p + n_n)$.  This
variable evolves due to charged-current weak nuclear interactions,
which do not always have time to equilibrate.  Thus, for our equation of
state, we are left with functions of three variables,
e.g. $P(\rho,T,Y_e)$.  Unfortunately, they are unknown functions for
densities much above nuclear saturation, so numerical relativity simulations of neutron
stars must explore the range of equations of state consistent with
known nuclear and astrophysical constraints.  For the problem most
relevant for gravitational waves, compact binary inspirals, the
situation simplifies.  The nuclear matter is very degenerate and in
beta equilibrium, so the EoS is  effectively one-dimensional:
$P=P(\rho_0)$.

These 1D EoS are conveniently parameterized as piecewise-polytropes,
for which the density is divided into intervals, and each interval has
its own polytrope law.  For example, in the $i$-th interval, covering
the density range $\rho_{i-1}<\rho_0<\rho_i$, the pressure is $P =
\kappa_i\rho_0^{\Gamma_i}$.  The polytropic indices $\Gamma_i$ and
transition densities $\rho_i$ are free parameters.  $\Gamma_0$ and
$\kappa_0$ covers the low-density range where 
the pressure, dominated by relativistic electrons, is known.  The
other $\kappa_i$ are set by requiring $P$ to be continuous at
$\rho_i$.  Fortunately, only a few free parameters needed to
adequately cover the range of plausible EoS~\cite{Read:2008iy}.

At the end of inspiral, tidal disruption breaks beta equilibrium, as
the matter decompresses faster than charged weak interactions can
adjust $Y_e$.  Also, the occurrence of shocks heats the matter so that it
is no longer degenerate.  A number of studies add a Gamma-law
thermal piece to the pressure to allow shocks to heat the gas.  Any
cold EoS can by thus augmented as follows:
\begin{eqnarray}
  \epsilon &=& \epsilon_{\rm cold}(\rho_0) + \epsilon_{\rm th} \\
  P &=&  P_{\rm cold}(\rho_0) + (\Gamma_{\rm th}-1)\rho\epsilon_{\rm th}
\end{eqnarray}
where now $\epsilon_{\rm th}$ comes from the energy density evolution.
The thermal Gamma law may not capture important parts of the true 3D
EoS.  This has been tested in the context of \bns
mergers by Bauswein~{\it et al.}~\cite{Bauswein:2010dn}.  They find
that the thermal Gamma law approximation can alter the post-merger
gravitational wave frequency by 2--8\%, post-merger torus mass by
30\%, and delay time to collapse to a black hole by up to a factor of
2.  In addition, only 3D EoS provide the physical temperature
information needed for neutrino calculations.

The evolution of the lepton number and $Y_e$ are given by weak nuclear
processes such as electron and positron capture, which emit neutrinos
that travel some distance, as well as the reverse absorption processes. 
Also, neutrino cooling is the dominant source of cooling in most
simulations with neutron stars, and neutrino absorption above the
neutrinosphere can be an important driver of winds.  Newtonian
simulations, especially in the supernova context, have long concerned
themselves with these effects, and around 2010 they began to be
incorporated into numerical relativity simulations.  At first, neutrino emission effects
were approximated by local sink terms for the energy and lepton number
(``neutrino
leakage'')~\cite{2010CQGra..27k4107S,Sekiguchi:2011zd,Deaton:2013sla,Galeazzi:2013mia,Neilsen:2014hha}.
Effective emission rates differ in optically thick and optically thin
regions; the neutrino optical depth can be computed by an inexpensive
iterative procedure~\cite{Perego:2014qda,Neilsen:2014hha}.  The
current state-of-the-art for numerical relativity is neutrino transport in an
energy-integrated moment closure
approximation~\cite{Shibata:2011kx,Shibata:2012zz,Wanajo:2014wha,Foucart:2015vpa},
which is impressive progress in so short a time but still far from a
full solution to the 6D Boltzmann equation.

A final major piece of realistic matter numerical relativity simulations is the
electromagnetic field evolution.  Neutron star interiors have plenty
of free charges and very high electrical conductivity, so the
magnetohydrodynamic (MHD) approximation is valid in most regions.
Thus, we must add the Maxwell stress tensor for the electromagnetic
field $T^{\rm EM}{}_{\mu\nu}$ to the total stress tensor $T_{\mu\nu}$,
and so magnetic terms appear in $\tau$ and $S_i$.  The evolution of
the magnetic field $B^i$ is given by the induction equation--in words,
that magnetic field lines are attached to (``frozen into'') fluid
elements.  The electric field is set by the MHD condition that the
electric field in the conducting fluid rest frame vanish:  $\alpha E_i =
-\epsilon_{ijk}(v^j+\beta^j)B^k$.  At all times, the magnetic field
should satisfy the constraint $\nabla\cdot B=0$.  In practice, magnetic
monopoles are avoided by constrained transport (staggering magnetic
and electric variables so that the change of $\nabla\cdot B$ exactly
vanishes~\cite{1988ApJ...332..659E}), by evolving a vector
potential~\cite{Giacomazzo:2010bx} (which, with appropriate
staggering, is equivalent to constrained
transport~\cite{DelZanna:2002rv,Etienne:2010ui}), or by divergence
cleaning (extending Maxwell's equations so that monopoles damp and
propagate off the
grid~\cite{2002JCoPh.175..645D,Liebling:2010bn,Moesta:2013dna}).

An interesting limit of the MHD equations occurs in magnetospheres,
where the Maxwell piece of $T_{\mu\nu}$ dominates, so that $\tau$ and
$S_i$ become essentially the electromagnetic energy density and
Poynting flux, respectively.  Magnetospheres differ from vacuum
electromagnetism because enough free charges remain to prevent
electric potential differences along field lines ($E\cdot B=0$).
These conditions are expected to obtain in the region around neutron
stars and the polar jet region around accreting black holes.
Specialized codes have been developed to evolve the relativistic
force-free equations, evolving either the electric and magnetic
fields~\cite{2004MNRAS.350..427K,Spitkovsky:2006np,Palenzuela:2010xn}
or the magnetic field and Poynting flux~\cite{McKinney:2006sc}.
Lehner~{\it et al.}~\cite{Lehner:2011aa} introduce a scheme for
evolving the full MHD equations in high-density regions and the
force-free equations in low-density regions.  This scheme was
successfully used to study the collapse of magnetized neutron stars,
but for future applications a single set of equations able to handle
both fluid and field-dominated regimes was desirable.  This was done
by Palenzuela~\cite{Palenzuela:2012my} in the context of a resistive
MHD code.  Resistivity and the force-free limit might sound like
different issues, but in fact the inhibition of flow by charged
particles across field lines in a magnetosphere can be modeled as an
anisotropic resistivity~\cite{2004MNRAS.350..427K}, so by allowing
sufficiently general Ohm's laws, Palenzuela's code can both handle
resistivity inside stars and impose the force-free limit outside.
Also, Paschalidis~{\it et al.}~\cite{Paschalidis:2013gma} have adjusted
their MHD code to extend to the force-free limit, showing that MHD and
force-free $(B^i,S_k)$ evolution just differ in the primitive variable
recovery.  These codes have been used to study magnetosphere
interaction in the late inspiral of neutron star--neutron
star~\cite{Palenzuela:2013hu,Palenzuela:2013kra,Ponce:2014sza} and
\bhnss~\cite{Paschalidis:2013jsa}, which has
been suggested as a mechanism to create precurser signals to short
duration gamma ray bursts.

\section{Black hole--black hole binary simulations}
\label{sec:bbh}
There have been several recent reviews of the history of numerical
relativity~\cite{Sperhake:2014wpa, Centrella:2010mx,
AlcubierreBook2008, BaumgarteBook2010, Holst:2016gmn,
Eisenstein:2018zpw}. Here we will
briefly cover some of the major highlights.

\subsection{Early efforts}
Attempts at numerical simulations of \bbhs date back to
the 1960s with the pioneering work of Hahn and
Lindquist~\cite{1964AnPhy..29..304H},
who were able to simulate
two initially stationary black holes for a short time. Later, with
faster computers and improved algorithms, Smarr {\it et al.} were~\cite{Smarr:1976qy, 1975PhDT.......168S, 1975PhDT........51E}  able to
simulate \headon collisions through merger. It was not until 1993 that
computers were powerful enough to calculate accurate waveforms from
such mergers~\cite{Anninos:1993zj}.

In the 1990s the National Science Foundation of the United States 
supported a large
collaboration, the Binary Black Hole Grand Challenge Alliance, with
the goal of advancing numerical relativity to the point where
evolutions of orbiting black holes became feasible.
There were several important developments enabled by the grand
challenge. These include the full 3D evolutions of boosted single
black holes using excision~\cite{Cook:1997na}, perturbative techniques
to extract gravitational waveforms from numerical
simulations~\cite{Abrahams:1997ut}, stable evolutions of single
black-hole spacetimes using characteristic
techniques~\cite{Gomez:1998uj}, and the development of toolsets for
parallel simulations.
The
alliance~\cite{Brandt:2000yp}, and
independently Br\"ugmann~\cite{Bruegmann:1997uc}, were able 
to evolve grazing collisions of \bbhs in 3D. However, these
simulations crashed after a short evolution time ($\lesssim 50M$).

As mentioned above,
the Lazarus approach~\cite{Baker:2000zm, Baker:2001nu,
Baker:2001sf, Baker:2002qf, Baker:2003ds, Campanelli:2005ia}, could be
used to extend these simulations and generate waveforms from closely
separated binaries.

These grazing collisions were performed using the ADM system of
equations evolved as a standard Cauchy problem. Using characteristic
evolution techniques,  the PITT Null code~\cite{Gomez:1996ge,
Bishop:1997ik, Gomez:1998uj} (developed at the University of
Pittsburgh) could evolve highly
distorted spacetimes for arbitrary lengths of time. For these long
evolutions, the PITT code used a coordinate system based on outgoing
null (i.e., lightlike) geodesics. In a two black-hole spacetime, these
geodesics would form caustics in the vicinity of the two black holes, making the associated coordinate system
singular. Thus the PITT code could not evolve the interior of a \bbh. However, this code has since proven to be very useful for
evolving the exterior region in which gravitational waves propagate
away from the central system.  An interior Cauchy evolution an exterior
characteristic evolution can be combined to produce
highly-accurate waveforms in a technique known as Cauchy-Characteristic
extraction~\cite{Bishop:1996gt, Bishop:1997ik, Bishop:1998uk,
Winicour:2005ge, Winicour:2005eoq, Babiuc:2005pg, Babiuc:2010ze}.

\subsection{Post-breakthrough results}
With the breakthroughs of 2005, there was rapid progress in our
understanding of the physics of \bbhs.

\subsubsection{Recoils}

One of the most remarkable results that came from these simulations is that
the merger remnant can recoil at thousands of kilometers per second.
Determining just how fast the remnant can recoil took several years and required many hundreds of individual
simulations.

Perhaps the most straightforward way to conceptualize why the emitted
power due to an inspiral can have (instantaneously) a preferred
direction is to consider the case of unequal-mass black holes.
The asymmetry of the system leads to a small excess of radiation along
the direction of the linear momentum of the smaller black hole.
For perfectly circular orbits, this effect would average out to zero
over an orbit,
but since the
binary will also be inspiraling, the cancellation will not be exact
and a net recoil will be generated. The net recoil only becomes
significant during the fast plunge phase. 

\begin{figure}
  \includegraphics[width=\columnwidth]{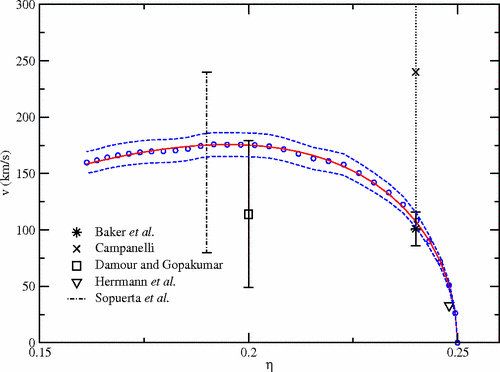}
  \caption{A reproduction of Fig 2. of Ref.~\cite{Gonzalez:2006md}
    courtesy of the authors.
    The results of the first large-scale numerical relativity study.
    Shown are the measured recoils for over 30 binary simulations,
    the estimated errors (the region between dotted curves),  a fit
    (red curve), and various older approximations for recoils. The
    horizontal axis is the symmetric mass ratio defined as $\eta = m_1
  m_2 /(m_1+m_2)^2$.}\label{fig:gonzalez}
\end{figure}

Initial measurements of the recoil concentrated on analyzing
individual
configurations~\cite{Baker:2006vn}, or several configurations, but at
very close separations~\cite{Herrmann:2006cd}.  The study of recoils
started in earnest with Gonzalez {\it et al.}~\cite{Gonzalez:2006md}.
Theirs was the first to do what was previously unheard of, a large
number of relatively long-term accurate simulations. In the case of~\cite{Gonzalez:2006md}, they performed over 30 individual simulations
and determined that the recoil very nearly obeys the simple formula
$V = 16\,A\eta^2 \sqrt{1-4 \eta}(1+B \eta)$, where $A = 750\KMS$ and
$B=-0.93$, $\eta = q/(1+q)^2$ is the symmetric mass ratio, and
$q=m_1/m_2$ is the usual mass ratio. 
Based on the formula provided by Gonzalez {\it et al.}, the maximum
recoil generated by unequal mass binaries is $175\pm 11 \KMS$.
As we will see, this contribution to the recoil can be vastly swamped
by contributions due to the spins of the black holes themselves.
Figure~\ref{fig:gonzalez} shows the results of their study.

Soon after, other groups showed that the maximum recoil for spinning
binaries, where the spins are aligned and antialigned with the angular
momentum, is much larger. In
Ref.~\cite{Herrmann:2007ac}~and~\cite{Koppitz:2007ev}, it was shown
that the maximum recoil for an equal mass, spinning binary with one
black hole
spin aligned with the orbital angular momentum and other antialigned
is $\sim475\ \KMS$.  A still larger recoil of $V_{max}\sim525\ \KMS$
for a mass ratio of $q\approx0.62$  was found in~\cite{Healy:2014yta}
when they extended the analysis of aligned/counteraligned spin binaries to unequal
masses.

The recoils induced by unequal masses and aligned/counteraligned spins
is always in the orbital plane of the binary (which, by symmetry,
does not precess). Ref.~\cite{Campanelli:2007ew} performed a
set of simulations that showed that the out-of-plane recoil, which is
induced by spins lying in the orbital plane, can be much larger.
These {\it superkicks}~\cite{Campanelli:2007ew, Gonzalez:2007hi,
Campanelli:2007cga, Dain:2008ck, Lousto:2010xk} were found to be up to
$4000\ \KMS$ when the spins were exactly in the orbital plane.

\begin{figure}
  \includegraphics[width=.75\columnwidth]{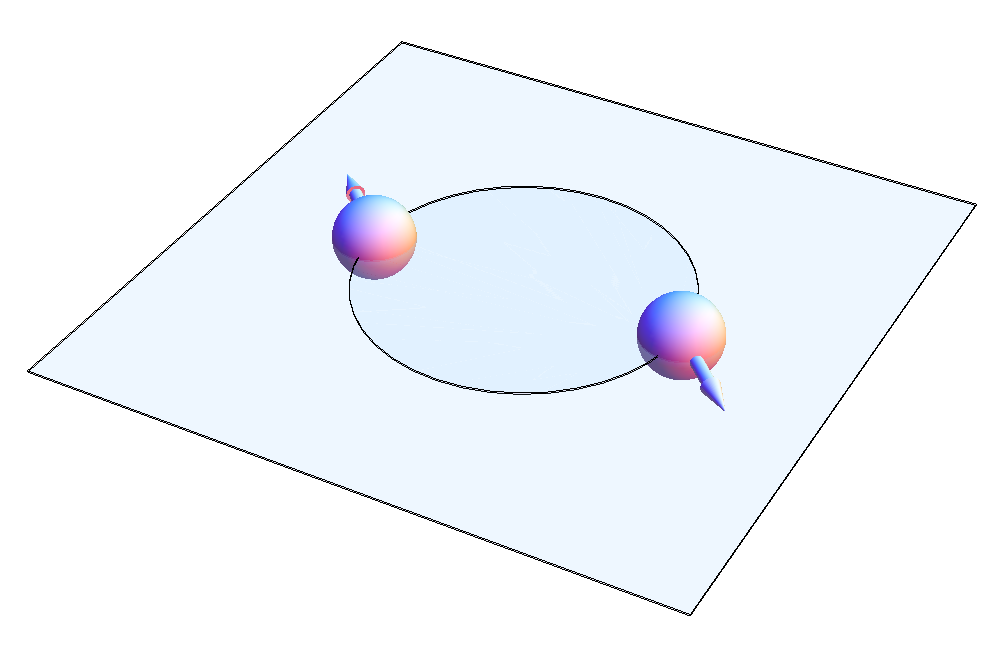}
  \caption{A sketch of the superkick configuration. Spins are entirely
  in the plane and anti-aligned.}\label{fig:superkick}
\end{figure}

\begin{figure}
  \includegraphics[width=\columnwidth]{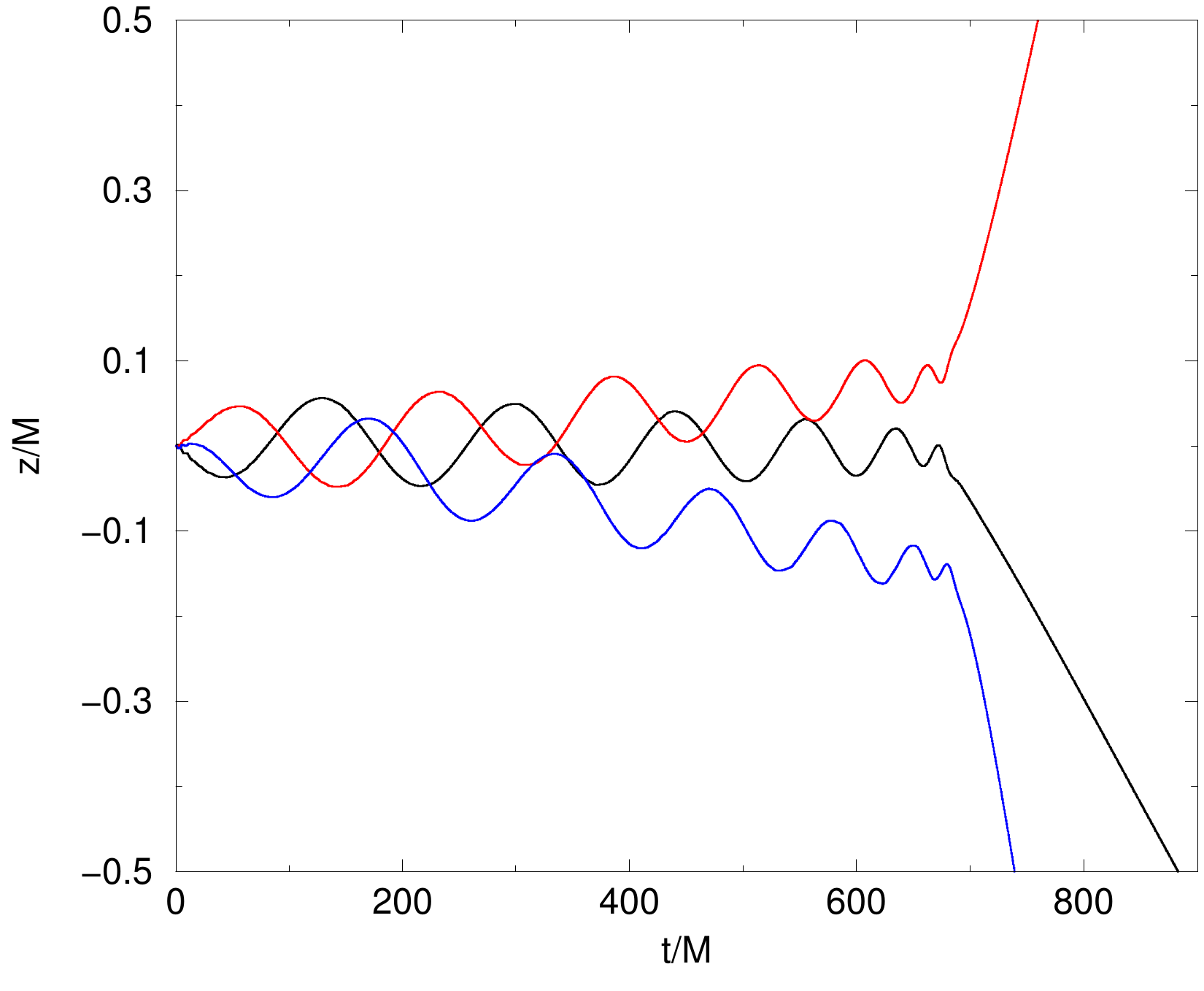}
  \caption{The bobbing of a binary studied in~\cite{Lousto:2012gt}.
    The coordinate distance of the binary from the original orbital
    plane is shown for three azimuthal variations of the hangup-kick
  configuration.}\label{fig:bob}
\end{figure}

\begin{figure}
\includegraphics[width=.66\columnwidth]{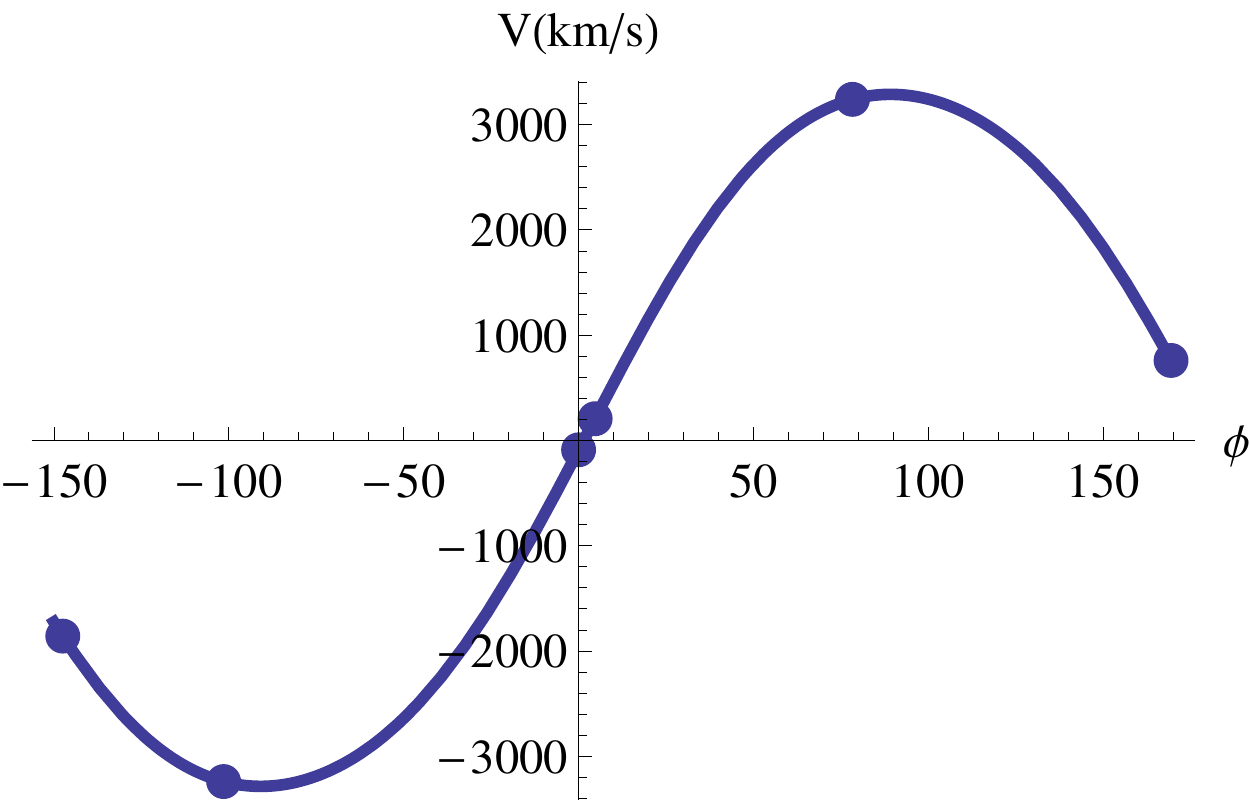}
  \includegraphics[width=.30\columnwidth]{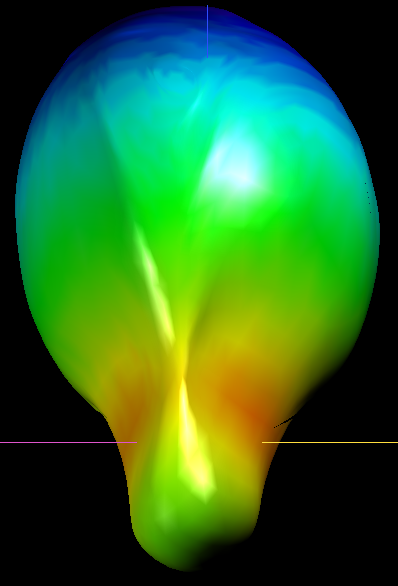}
  \caption{(Left) The measured recoil for an equal-mass, superkick
    configuration~\cite{Lousto:2010xk} with
  spins $\chi = 0.9$ and various azimuthal orientations. Note that
very large and very small recoils are both
possible. (Right) The radiated power $\frac {dP}{d\Omega}$ per unit
solid
    angle for a configuration studied in~\cite{Campanelli:2010ac}.
  Note the large
  excess of power directed upwards, which leads to a downward kick.
}\label{fig:superkick2}
\end{figure}

\begin{figure}
  \includegraphics[width=.9\columnwidth]{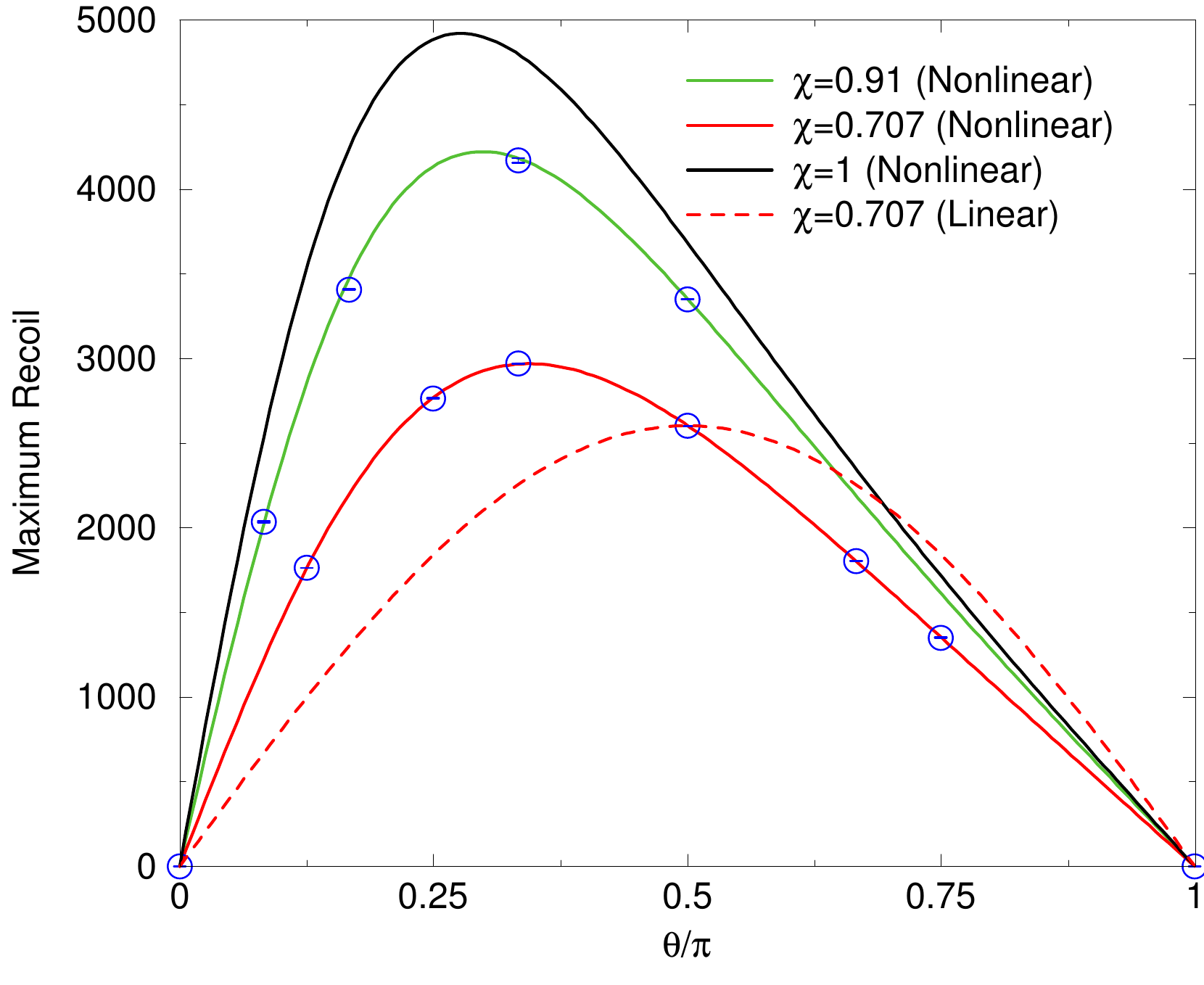}
  \caption{The measured maximum recoil for equal-mass binaries in a
    hangup-kick configuration as a function of the polar orientation of the
  spins~\cite{Lousto:2011kp, Lousto:2012su}.
}\label{fig:hangupkick}
\end{figure}

The superkick configuration is quite interesting, not only because of
the large recoil, but also because of the direction of the recoil.
Figure~\ref{fig:superkick} shows the basic setup. 
The spins are
anti-aligned with each other and in the in the orbital plane. Such a
system will not precess and the orbital angular momentum will always
point in the $z$ direction. Furthermore, the system has $\pi$-rotation
symmetry about the $z$-axis. This means the recoil cannot lie in the
orbital plane. What actually happens in this case is the binary bobs up
and down long the orbital axis at
ever increasing speeds until it merges. This bobbing is controlled by
the orientation of the spins, as shown in Fig.~\ref{fig:bob}. The net effect is quite unexpected. The
magnitude and direction of the recoil depends sinusoidally on the azimuthal
orientation of the spins (see Fig.~\ref{fig:superkick2}).
Originally, it was thought that these in-plane spins maximized the
recoil, however, it was later found out
in~\cite{Lousto:2011kp, Lousto:2012su, Lousto:2012gt} that,
due to the hangup and other nonlinear-in-spin effects~\cite{Campanelli:2006uy},
having partially miss-aligned spins actually leads to a substantially
larger
recoil (up to $5000\ \KMS$). The basic setup of this hangup-kick
configuration is very similar to the superkick, with the exception
that the out-of-plane components of the spins are aligned. For small
spins, the recoil depends sinusoidally on the polar orientation (i.e.,
$V\propto \sin\theta$). However, for larger spins, the recoil is
substantially larger for smaller angles, as shown in
Fig.~\ref{fig:hangupkick}.
  
Of critical importance for modeling the superkick is the dependence on
mass ratio.  It is perhaps surprising that even though the central
supermassive black holes in galaxies with a bulge can range in mass
from under a million to tens of billions of solar masses, roughly $\sim 93\%$~\cite{Yu:2011vp,
Stewart:2008ep, Hopkins:2009yy}  of galactic mergers are expected to
produce supermassive black hole merges with mass ratios
in the range $1/10 < q < 1$. If the falloff of the recoil with mass
ratio is steep, then even a 10:1 binary may have a negligible recoil.
Based on post-Newtonian theory, the group formerly at UTB, now at the
Rochester Institute of Technology (RIT), argued that this
dependence should vary as $q^2$~\cite{Campanelli:2007ew,
Lousto:2007db}. This was first put to the test in~\cite{Baker:2007gi},
where the authors found the recoils fall faster than $q^2$, at least
in certain symmetric configurations. A follow-up series of papers
by the RIT group~\cite{Lousto:2008dn, Zlochower:2015wga} modeled the
spin and mass ratio dependence for more generic configurations and
found a leading $q^2$ dependence on the recoil for these more generic
configurations.

Modeling the recoil for generic configurations is complicated by the
need to perform hundreds of simulations. Even for a fixed mass ratio
and zero eccentricity, the recoil depends on six spin degrees of
freedom. In order to tackle this problem, the dimensionality of the
free parameters needs to be reduced.  The basic procedure developed by
the RIT group in a series of papers~\cite{Lousto:2008dn, Lousto:2012gt,
Zlochower:2015wga} starts with a family of simulations related by a
rotation of the azimuthal direction of the spins at the initial
separation (the spins of both black holes are rotated the same
amount). This leads to a 1-parameter family of configurations with
(typically) a nearly sinusoidal dependence of the recoil. The
amplitude of this sinusoidal dependence is then measured as a function
of the polar spin orientations. This procedure works well when the
azimuthal spins are antialigned, but is of unknown utility for
completely generic configurations.  Another method used there was to
consider the subspace where only one of the black holes is spinning
(either the smaller or larger one). This again reduces the
dimensionality of the problem. Even
so, the number of individual simulations required to generate the
latest models for the recoil was $~200$.

The discovery that merging black holes can recoil at thousands of
$\KMS$ sparked many searches for recoiling supermassive black holes.
These searches are ongoing.  For example, QSO 3C
186 was recently proposed as a candidate recoiling black hole with
recoil of order $2000\ \KMS$~\cite{Chiaberge:2016eqf}.
An active galactic nucleus (AGN) is thought to be a candidate for a
recoiling supermassive black hole if there is a red/blue shift between
broad line and narrow line emissions. Gas tightly bound to the
central black hole would have much higher velocity dispersion (which
is a function of the kinetic energy of the gas) than gas further out.
If a binary merges and the remnant recoils, gas close to the remnant
will remain bound and recoil with the remnant, while gas further out
is left behind. This then would lead to two different redshifts for
gas that remains bound to the central black hole and the rest.
To date, no source has been definitively shown to be a recoiling
remnant black hole. For a recent history of these searches,
see~\cite{Komossa:2012cy}.

If large recoils are common, then  why are there not more
candidates? Since only gas-rich mergers lead to luminous signals that
can be detected electromagnetically, it may well be that a recoiling
AGN cannot be luminous.
Newtonian and post-Newtonian simulations appear to
indicate that accretion will tend to align or counteralign the black
hole spins with the orbital angular momentum~\cite{Bogdanovic:2007hp,
Dotti:2009vz, Miller:2013gya}. Depending on the degree of alignment,
this may essentially suppress the superkick style recoils. There has
therefore been a resurgence of interest in modeling recoils for
spin-aligned systems.

The modeling of recoils from binaries with spins aligned and
counteraligned with the orbital angular momentum began soon after the
breakthroughs in numerical relativity. The first such simulations were
performed in~\cite{Herrmann:2007ac}~and~\cite{Koppitz:2007ev}, with
the first systematic studies of the recoil from such binaries
in~\cite{Pollney:2007ss} and~\cite{Schnittman:2007ij}. And the process
of generating
empirical models for the remnant masses, spins, and recoils from such
mergers were first performed in Refs.~\cite{Boyle:2007sz, Baker:2008md,
Boyle:2007ru, Rezzolla:2008sd, Lousto:2007db}.
In addition, these types of binaries have been studied for their use
in waveform modeling with examples in the SXS~\cite{SXS:catalog, Mroue:2013xna},
Georgia Tech~\cite{GT:catalog, Jani:2016wkt} and
RIT~\cite{RIT:catalog, Healy:2017psd} catalogs.
More recently, because of the apparent lack of observed
highly-recoiling AGN, the RIT group began a systematic study of the
spin and mass ratio dependence of the recoil for aligned/counter
aligned binaries in~\cite{Healy:2014yta, Healy:2016lce}.

\subsubsection{Modeling the remnant properties}
One of the important tasks required in order to make the wealth of
information from numerical simulations useful for astrophysics was to
model the radiated energy-momentum and the corresponding final mass,
spin, and recoil of the remnant black hole from \bbh mergers in terms
of the initial parameters of the binary. Developing these models
required thousands of computationally expensive simulations.

In developing these models, two different techniques were initially
used, but current models now combine aspects of both. The first
technique used post-Newtonian theory~\cite{Gonzalez:2006md, Baker:2007gi,
Campanelli:2007ew, Campanelli:2007cga, Lousto:2007db, Dain:2008ck,
Lousto:2008dn, Baker:2008md, Koppitz:2007ev, Pollney:2007ss}, or other physical
arguments~\cite{Rezzolla:2007rz, Rezzolla:2008sd, Barausse:2009uz} to
determine the functional form and free parameters of an approximate
model for the relevant quantity, and the other used ad-hoc
expansions~\cite{Rezzolla:2007rd, Campanelli:2006uy, Boyle:2007ru,
Boyle:2007sz}.
The work of~\cite{Boyle:2007sz, Boyle:2007ru} pioneered the technique
of using symmetry arguments to limit the degrees of freedom in the
models. Their construction only assumed that the remnant can be
described by the spin vectors of each black hole and the mass ratio.
The model then must obey the following two symmetries. If 
$F(\vec \chi_1, \vec\chi_2, q)$  is a formula for the remnant, mass, spin vector, or
recoil, then $F$ must obey $F(\vec \chi_1, \vec\chi_2, q) =
F(\vec\chi_2, \vec\chi_1, 1/q)$, i.e., the
physical outcome of a merger cannot depend on the labels ($1,2$) of
the two black holes. Second, if $F$ must transform appropriately under
parity. One, however, need not use the variables $(\vec \chi_1,
\vec\chi_2, q)$. Inspired by post-Newtonian expressions, the RIT
group has made extensive use of the variables $(\vec \Delta,
\vec S, \delta M)$ as well the variables $\eta$ and $\vec
S_0$~\cite{Healy:2016lce, Zlochower:2015wga, Healy:2014yta,
Lousto:2013wta, Lousto:2012gt, Lousto:2012su, Lousto:2011kp,
Zlochower:2010sn, Lousto:2010xk, Lousto:2009mf, Lousto:2008dn}.
These are defined as
\begin{eqnarray}
   \vec S_1 = m_1^2 \vec \chi_1,\\
   \vec S_2 = m_2^2 \vec \chi_2,\\
   \vec S = (\vec S_1 + \vec S_2)/m^2,\\
   \Delta  = (\vec S_2/m_2 - \vec S_1/m_1)/m,\\
   \delta m = (m_1 - m_2)/m,\\
   \vec S_0 = \vec S  + (1/2) \delta m \vec \Delta.
\end{eqnarray}
Any expansion in terms of one set of variables can be reexpressed in
terms of an other. However, since the goal is to model the remnant
with accuracy, one wants to use variables that minimize the number of free
parameters required to fit the known data.

State of the art models for remnant properties now combine results
from simulations of many different groups (see
Refs.~\cite{Barausse:2012qz, Hemberger:2013hsa,
  Healy:2014yta, Zlochower:2015wga, Hofmann:2016yih, Healy:2016lce, Jimenez-Forteza:2016oae})
  with the goal of reducing systematic biases (which may arise from
  different groups concentrating on different regions of parameters
  space).

\subsubsection{Numerical relativity at the extremes}
\label{sec:extreme_nr}

State of the art numerical relativity codes now routinely evolve
binaries with mass ratios as small as $q\lesssim
1/10$~\cite{Gonzalez:2008bi, Lousto:2010qx, Lousto:2010ut,
Sperhake:2011ik, Chu:2015kft, Jani:2016wkt}, moderately-to-highly precessing
systems~\cite{Campanelli:2006fy, Campanelli:2007ew, Campanelli:2008nk,
Lousto:2008dn, Schmidt:2010it, Lousto:2012gt, Hannam:2013pra,
Schmidt:2014iyl, Lousto:2014ida, Ossokine:2015vda, Zlochower:2015wga,
Lousto:2015uwa, Blackman:2017pcm}, and binaries with moderate spins.
However, much smaller mass ratios, and spins close to 1 are still
quite challenging. Prior to the work of~\cite{Lovelace:2008tw} it
was not even possible to construct initial data for binaries with
spins larger than $\sim 0.93$~\cite{Cook:1989fb}. This limitation was
due to the use of conformally flat initial data.\footnote{Initial data
are said to be conformally flat if the spatial metric associated
with the data is proportional to the flat space metric.} Conformal flatness of
the spatial metric is
a convenient assumption because the Einstein constraint system take on
particularly simple forms. Indeed, using the puncture approach, the
momentum constraints can be solved exactly using the Bowen-York
ansatz~\cite{Bowen:1980yu}. There were several attempts 
to generate data for highly-spinning \bbhs, while still preserving conformal
flatness~\cite{Dain:2002ee, Lousto:2012es}, but these introduced
negligible improvements. Lovelace {\it et al.}~\cite{Lovelace:2008tw}
were able to overcome these limitations by choosing the initial data
to be a superposition of conformally Kerr black holes in the
Kerr-Schild gauge. Using these
new data, they were soon able to evolve binaries with spins as large
0.97~\cite{Lovelace:2011nu}, and later spins as high as
0.994~\cite{Scheel:2014ina}.

While spins of $0.92$ may seem reasonably close to 1, the scale is
misleading. The amount of rotational energy in a black hole with spin
0.9 is only 52\% of the maximum. Furthermore, particle limit and
perturbative calculations show even more extreme differences between
spins of 1 and spins only slightly smaller. For example, Yang {\it et
al.}~\cite{Yang:2014tla} studied an analog to turbulence in black-hole perturbation
theory. 
For spins close to 1, there is an inverse energy cascade from
 higher azimuthal ($m$) modes to lower ones for $\ell$ modes that obey
 $\epsilon = |1 - \chi|  \lesssim \ell^{-2}$. 
This give
hints that a more informative measure of the spin is actually
$1/\epsilon$. Similarly, analysis of Kerr
geodesic~\cite{Schnittman:2014zsa,
Berti:2014lva} and  particle-limit calculations of
recoils~\cite{Hirata:2010xn, vandeMeent:2014raa} indicate that the
dynamics of nearly-extremal-spin black
holes cannot be elucidated with any degree of certainty using
lower spin simulations.

Recently, the group at RIT also introduced their version of
highly-spinning initial data, also based on the superposition of two
Kerr black holes~\cite{Ruchlin:2014zva, Healy:2015mla}, but this time
in a puncture gauge.~\footnote{Recall that initial data in a puncture gauge is
  constructed by mapping the two infinitely large spacelike hypersurfaces of an Einstein-Rosen
bridge into a single spacelike hypersurface with a singular point.
Initial data for a binary would then have two such singular points.}
The main differences between the two approaches
is how easily the latter can be incorporated into moving-punctures
code. They compared their results to the SXS results for both spins
of $\chi=0.95$~\cite{Ruchlin:2014zva} and $\chi =
0.99$~\cite{Zlochower:2017bbg}, and
found very good agreement.

The other type of extreme simulation concerns small mass ratios.
Because current numerical relativity codes use explicit algorithms to
evolve the spacetime, the
Courant-Friedrichs-Lewy condition, which determines how large a
timestep can be relative to the spatial discretization, severely limits
the run speed when one of the black holes is much smaller than the
other.  Basically, the number of timesteps required near the smaller
black hole is set by the size of that black hole, not by its dynamics.
This, coupled to the fact that the inspiral for a small-mass-ratio
binary is much slower than for a similar mass one, means that such
simulations are extraordinarily expensive. To date, the smallest
quasi-circular inspiral evolved so far had
$q=1/100$~\cite{Lousto:2010ut, Sperhake:2011ik}.

\begin{figure}
  \includegraphics[width=.75\columnwidth]{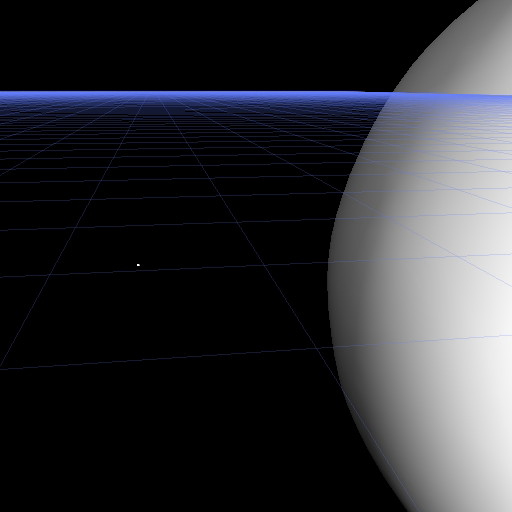}
  \caption{The merger of a 100:1 mass ratio
    binary~\cite{Lousto:2010ut}. The smaller black hole is a factor of
  nearly 4 times smaller than might be expected by a mass ratio of
1:100.}
\end{figure}

One promising method to overcome these limitations is to use
semi-implicit techniques~\cite{Lau:2011we, Lau:2008fb}, but these have
not yet been shown to work for small-mass-ratio binaries.

Finally, in Ref.~\cite{Lousto:2013oza} another {\it extreme} was
explored:  that of binaries at far separations. There the authors
used fully nonlinear numerical relativity to model several orbits of
binaries separated at $D=20M$, $D=50M$, and $D=100M$ and compared the
orbital dynamics to post-Newtonian and Newtonian predictions. Very
good agreement between post-Newtonian and numerical relativity
predictions for the orbital frequency was found for the $D\geq50M$
cases. The longest simulations to merger published to date was in
Ref.~\cite{Szilagyi:2015rwa}, where a binary was evolved for 175
orbits. For reference, an equal-mass binary at a separation of
$D=100M$ will complete over 2000 orbits before merging and requires
about $8.2\times10^{6}M$ of evolution time.  (If the
mass of each black hole in the binary is $30 M_\odot$, then the merger from
$D=100M$ would take about 40 minutes.)

\subsubsection{Waveform modeling}
\label{sec:waveforms}

One of the major goals of numerical relativity is to
produce accurate waveforms for gravitational wave data analysis. The actual process
of obtaining the waveform from a numerical simulation can be involved. 
See~\cite{Bishop:2016lgv} for a review of modern techniques for
extracting the gravitational waveform from a numerical simulation.

\begin{figure}[t!]
\includegraphics[width=\columnwidth]{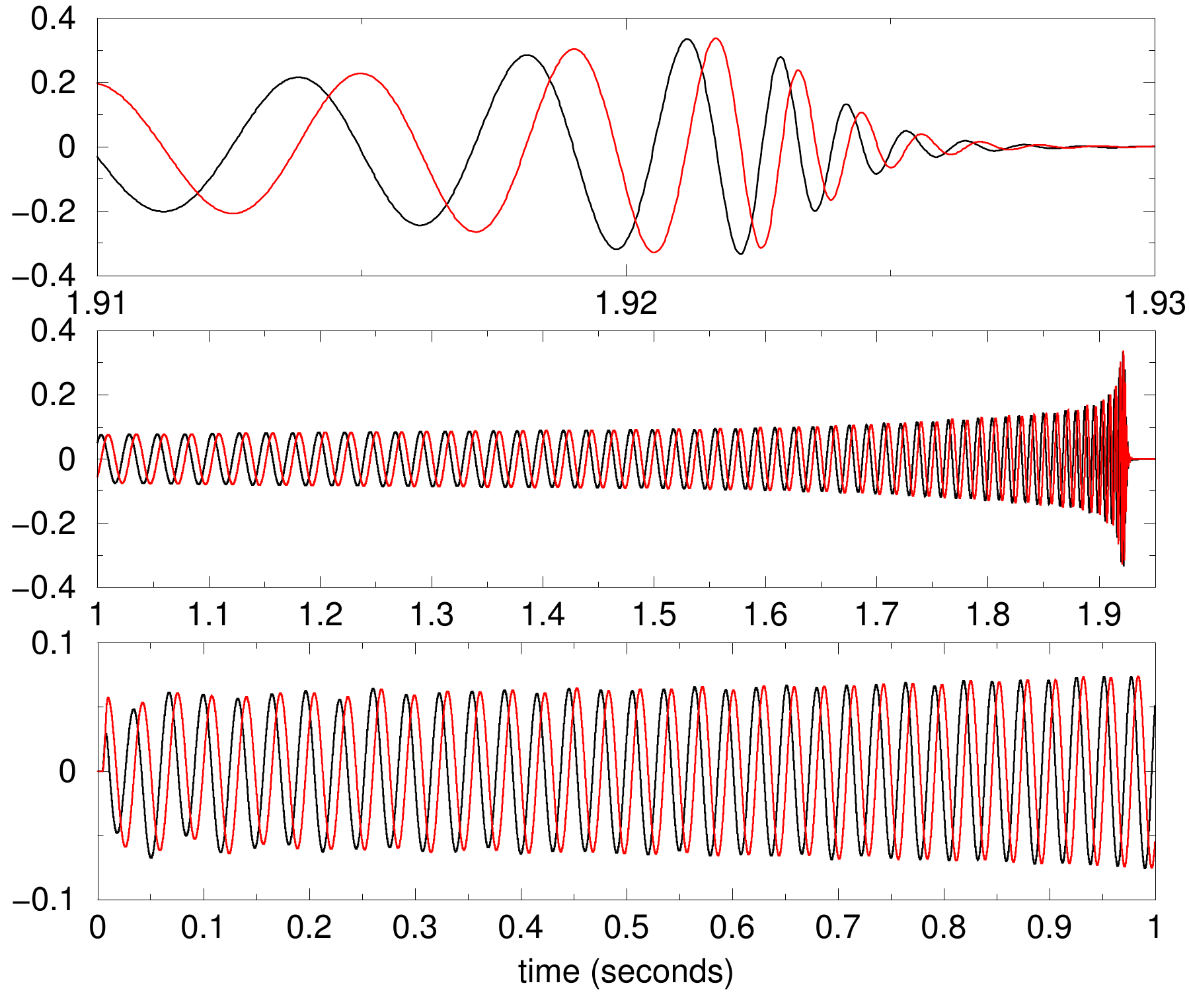}
\caption{The waveform (real and imaginary components of the (2,2) mode
  of the rescaled strain) from a precessing binary with total mass
  $20M_\odot$~\cite{Lousto:2014ida, Lousto:2016nlp}.
 To obtain the physical strain, the values plotted need to be rescaled
 by a factor of $9.7\times 10^{-16}/D$, where $D$ is the distance to the
 binary in units of kiloparsecs. 
 }\label{fig:wave}
\end{figure}

\begin{figure}[t!]
\includegraphics[width=\columnwidth]{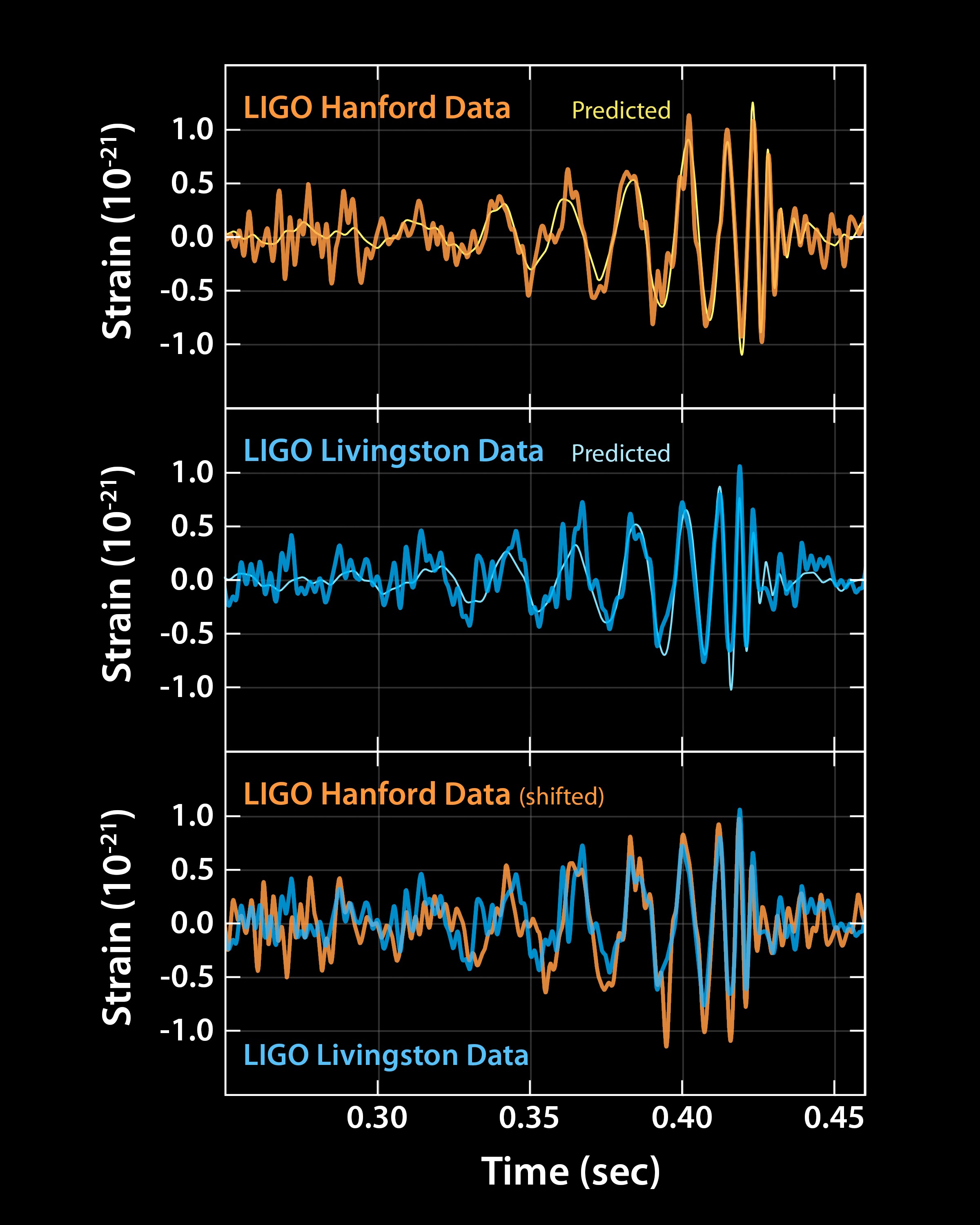}
\caption{  These plots show the signals of gravitational waves detected by the
 twin LIGO observatories at Livingston, Louisiana, and Hanford,
 Washington. The signals came from two merging black holes with masses
 30 and 35 
 times the mass of our sun, respectively, lying 1.3 billion light-years away.
 The top two plots show data received at Livingston and Hanford, along
 with the predicted shapes for the waveform. These predicted waveforms
 show what the waveform from two merging black holes with these
 masses should look like according to the
 equations of Albert Einstein's general theory of relativity, along
 with the instrument's ever-present noise. Time is plotted on the
 X-axis and strain on the Y-axis. Strain represents the fractional
 amount by which distances are distorted.
 As the plots reveal, the LIGO data very closely match Einstein's
 predictions.
 The final plot compares data from both detectors. The Hanford data
 have been inverted for comparison, due to the differences in
 orientation of the detectors at the two sites. The data were also
 shifted to correct for the travel time of the gravitational-wave
 signals between Livingston and Hanford (the signal first reached
 Livingston, and then, traveling at the speed of light, reached
 Hanford seven thousandths of a second later). As the plot
 demonstrates, both detectors witnessed the same event, confirming the
 detection. (Courtesy Caltech/MIT/LIGO Laboratory)
 }\label{fig:detection}
\end{figure}

In Fig.~\ref{fig:wave}, we show an example waveform from a recent \bbh
simulation~\cite{Lousto:2014ida, Lousto:2016nlp}.  In order to
use physical units, we consider a binary  that has a total
mass of $20M_\odot$. The plot
shows the waveform from the last 48 orbits for an equal-mass binary in
a precessing configuration. There are two distinct phases of the
waveform. The longest phase is due to the slow inspiral and eventual
plunge. In the
figure, this corresponds to the start of the waveform until about
1.92s. Small oscillations in the amplitude are apparent. These are due
to precession of the orbital plane. The overall ramp up of both the
amplitude and frequency is due to the inspiral and its associated
increase in the orbital frequency. Following a brief transition
between 1.915s and 1.925s, the waveform changes to a damped sinusoid.
This phase is due to the rapid equilibration of the now single black
hole.

Remarkably, on September 14, 2015 the twin LIGO observatories detected
the gravitational waveform from the inspiral and merger of two black
holes~\cite{Abbott:2016blz, TheLIGOScientific:2016wfe}. The resulting waveform is shown in Fig.~\ref{fig:detection}.

In this section we will review the history of fully
non-linear numerical simulations of black hole mergers to generate and
verify the waveforms. The current generation of numerical relativity
codes calculate the gravitational waveform of a merger simulations
by calculating the Regge-Wheler-Zerilli perturbations (which can be
related to the strain $h$), the Bondi News function $N$,  
or Weyl scalar $\psi_4$ (as well as combinations of the above). With the exception of the calculation of $N$
using Cauchy-Characteristic extraction~\cite{Babiuc:2010ze,
Bishop:1998uk, Reisswig:2009rx, Reisswig:2006nt, Winicour:2005ge}, the
waveform is calculated at a series of finite radii and extrapolated to
$r=\infty$ along an outgoing null (lightlike) paths using some form of either
polynomial extrapolation, or
perturbative expansion~\cite{Nakano:2015pta}. In a suitable gauge,
$\psi_4  = \dot {\bar N} = \ddot h$, where an overbar denotes complex
conjugation and  a dot represents a time derivative.
Note that gravitational wave detectors measure $h$ directly, while the
emitted power is directly related to $N$. The {\it points} at
$r=\infty$ along  outgoing null rays are collectively known as {\it future
null infinity}. While these points are formally {\it outside} the
spacetime, they are quite useful for defining gravitational radiation.

For an isolated source\footnote{more precisely, for an asymptotically
flat spacetime}, the Bondi News function $N$ is directly related to
the radiated power-per unit solid angle in the  gravitational radiation by
\begin{equation}
  \left(\frac{dP}{d\Omega}\right)  = \frac{1}{4 \pi} \left(N\bar
    N\right).
\end{equation}
$N$ itself is defined on future null infinity, and the only gauge
freedom in $N$ is associated with the supertranslation freedom
$N(\tau, x^A) \to N(\tau + \sigma(x^A), x^A + \tau \omega^A(X^A))$, where
$\tau$ is an affine parameter (a generalization of proper time along
lightlike curves), $x^A$ denotes angular coordinates, and $\sigma(x^A)$ and $\omega^A$ are constant functions of angle.

However, as standard Cauchy codes cannot include future null infinity
(but see~\cite{Vano-Vinuales:2017qij, Hilditch:2016xzh,
Vano-Vinuales:2015lhj} for an approach which may allow evolutions that
include future null infinity),
calculations of $N$ involve a matching procedure, where data from a
Cauchy code is used as boundary data for a characteristic evolution.
This matching requires the specification of unknown data from the edge
of the Cauchy domain to null infinity. This induces 
spurious radiation~\cite{Babiuc:2010ze}, which can be
controlled by moving the matching procedure to farther radii.

 In Chu
{\it et al.}~\cite{Chu:2015kft} the authors compared their
extrapolations of the Regge-Wheeler-Zerilli perturbations to the News
calculation. They found that  errors due to both gauge effects and extraction at
finite radii lead to mismatches of a $\sim 5\times 10^{-4}$. It is
interesting to note that this mismatch may be geometrical in nature and
arising from the difficulty in defining gravitational radiation at a
finite distance from the source.

The other main technique for extracting radiation involves the
calculation of the Weyl scalar $\psi_4$. Calculations of $\psi_4$ have
the advantage that there is a simple, well defined
procedure for calculating the Newman-Penrose scalars $\psi_4$ in a
class of tetrads where $\psi_4$ represents the outgoing radiation (and $\psi_0$
the ingoing radiation). This class of tetrads, known as
quasi-Kinnersley tetrads~\cite{Nerozzi:2004wv, Nerozzi:2005hz,
Campanelli:2005ia, Zhang:2012ky} is unique up to an overall phase
factor and normalization.

\begin{figure}
  \includegraphics[width=.42\textwidth]{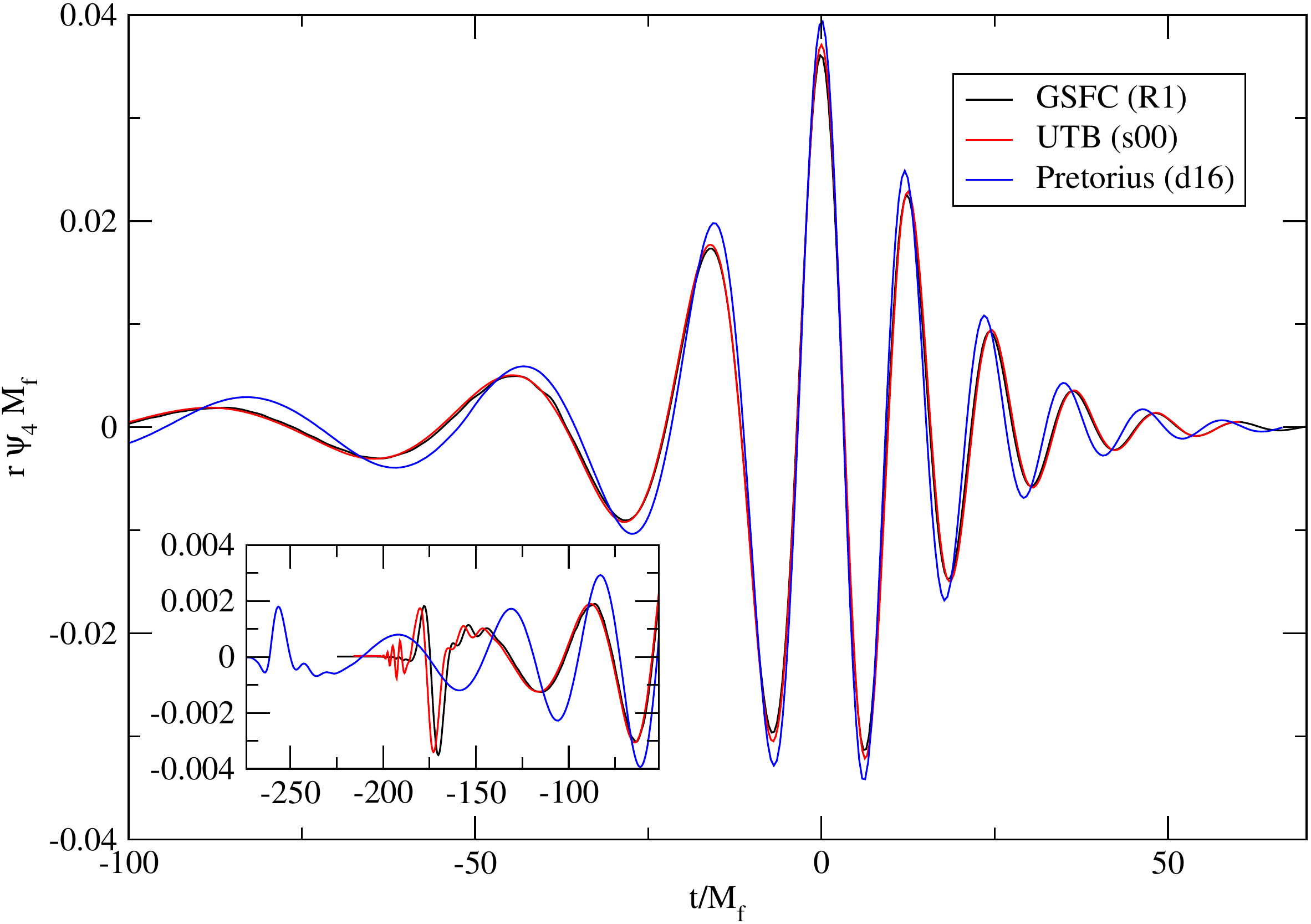}
  \caption{The first comparison of numerical relativity waveforms
    produced by different codes and different groups. The figure is a
    reproduction of Fig.~1 of Ref.~\cite{Baker:2007fb} and shows a
    comparison of waveforms generated by the groups at NASA GSFC, UT
    Brownsville and RIT,
    and Frans Pretorius. Note that the initial
    binary configuration used by Pretorius was a corotating
    configuration, which has a moderate spin, while
    the other two configurations were nonspinning. Hence the waveform
    amplitude and frequency are not expected to be identical to the
  other two waveforms.}\label{fig:first_comparison}
\end{figure}

Since there are no analytically known waveforms from the mergers of
black holes, in order to test the correctness of numerically derived
waveforms one needs to both carefully audit the codes and compare
results generated from different code bases. The first such
comparison~\cite{Baker:2007fb}
was performed early on with waveforms generated by the inspiral of
an equal-mass, low-spin binary\footnote{The binary evolved by
  Pretorius had a very small dimensionless spin of 0.02, while the
  binaries
evolved by the LazEv and Hanhdol codes were nonspinning.}
obtained using the LazEv code~\cite{Zlochower:2005bj,
Campanelli:2005dd} developed at Brownsville and Rochester Institute of
Technology, the 
Hahndol code~\cite{Baker:2006kr, Imbiriba:2004tp, vanMeter:2006vi}
developed at NASA-GSFC,
and Pretorius' original code~\cite{Pretorius:2004jg,
Pretorius:2005gq}.
For that test, each
group evolved similar binaries, but at slightly different initial
configurations. The results from this first comparison are shown in
Fig.~\ref{fig:first_comparison}.
Later on, as more groups
developed their own codes, more large-scale comparisons were
performed. For the Samurai~\cite{Hannam:2009hh} project, comparisons
were made between the SpEC~\cite{Boyle:2007ft, Scheel:2008rj} code developed by the SXS collaboration,
the Hahndol code,
the MayaKranc code~\cite{Vaishnav:2007nm} developed at Penn State
/ Georgia Tech, the CCATI code~\cite{Pollney:2007ss} developed at the
Albert Einstein Institute, and the BAM code~\cite{Bruegmann:2006at,
Husa:2007hp} developed at the University of Jena.  One of the major
differences between the Samurai project and~\cite{Baker:2007fb}
was the use of simulated LIGO noise data to determine if the
differences between the waveforms generated by the various codes is,
in practice, detectable.

Such comparisons took on more urgency with the detection of
gravitational waves in 2015~\cite{Abbott:2016blz,
TheLIGOScientific:2016wfe}. In the process of verifying that
detection two groups, the SXS collaboration and the group at RIT,
generated waveforms as part of their work for the LIGO scientific
collaboration. Despite the two codes sharing no common routines, and
using different initial data generation techniques, different
evolutions techniques, and different waveform extraction techniques,
the dominant $(\ell=2,m=2)$ modes produced by the two codes agreed to
better than 99.9\%~\cite{Lovelace:2016uwp}.

One of the main goals of waveform modeling is to infer the properties
of the source based on the observed waveforms. One of the first
systematic studies of how sensitive the waveform is to the parameters
of the binaries in the no-precessing case was performed
in~\cite{Reisswig:2009vc}, where they studies the effects of  spins on waveform detectability.

\subsubsection{Semi-analytic waveform models}

Because fully nonlinear numerical simulations of black hole mergers are computationally
expensive,
many semi-analytic and empirical approaches have been
developed in order to model the waveform of black hole mergers using
information from previously modeled simulations~\cite{Ajith:2007qp, Ajith:2007kx, Gopakumar:2007vh, Hannam:2007wf,
  Boyle:2007ft, Boyle:2008ge, Scheel:2008rj, Hannam:2009hh,
  Ajith:2009bn, Damour:2008te, Campanelli:2008nk, Buonanno:2009qa, Chu:2009md, Pan:2009wj,  Santamaria:2010yb, Ohme:2011zm,
  Lovelace:2011nu, Rezzolla:2010df, Barausse:2011kb, Lousto:2010tb,
  Lousto:2010qx, Nakano:2011pb, Kelly:2011bp, MacDonald:2012mp,
  Ajith:2012az, Schmidt:2012rh, Baird:2012cu, Hinder:2013oqa,
  Mroue:2013xna,Hemberger:2013hsa,Taylor:2013zia, Kumar:2013gwa,
  Hannam:2013oca, Hannam:2013pra, Aasi:2014tra, Schmidt:2014iyl,
  Varma:2014jxa, Husa:2015iqa, Szilagyi:2015rwa, Blackman:2015pia,
Chu:2015kft, Khan:2015jqa, Jimenez-Forteza:2016oae, Kumar:2016dhh,
Abbott:2016apu, Lovelace:2016uwp, Huerta:2016rwp, Bohe:2016gbl,
Blackman:2017dfb}.
As demonstrated in~\cite{Abbott:2016wiq}, the systematic errors
associated with waveforms models used by the LIGO Scientific
Collaboration led to negligible errors in the inferred parameters of
the source of GW150914.

One conceptually straightforward method for
constructing a waveform model would be to interpolate known numerical
waveforms as a function of the source parameters. The actual details
of how this interpolation is performed is quite sophisticated and
requires first representing waveforms in a reduced basis. The
resulting models~\cite{Field:2013cfa, Blackman:2015pia,
Blackman:2017dfb, Blackman:2017pcm} have been shown to be accurate, at
least under the conditions tested~\cite{Blackman:2017pcm}.

Another class of models, usually referred to as phenomenological
models~\cite{Pan:2007nw, Ajith:2007qp, Ajith:2009bn, Santamaria:2010yb, Husa:2015iqa, Khan:2015jqa}, are based on an expansion of the waveform in Fourier space.
The amplitude and phase of the waveform are expanded as  algebraic
functions of frequency. The coefficients in these expressions are then
modeled as functions of the binary's parameters by fitting to existing
waveforms. 

Yet another method for modeling the waveform from a binary, when the
two black holes are still far apart, is based on a series expansion in
the black-hole separation and velocity, known as the post-Newtonian
expansion.  For a review of post-Newtonian theory,
see~\cite{Blanchet:2013haa}.

Since post-Newtonian theory is expected to be accurate when the binary
separation is large, and become increasingly more inaccurate during
the inspiral, a natural question is to determine where the
post-Newtonian waveforms differ substantially from the numerical ones. The first direct comparisons of the waveform predictions from
numerical relativity and post-Newtonian theory were performed by
Buonanno, Cook, and Pretorius~\cite{Buonanno:2006ui}, shortly
thereafter by 
the NASA-GSFC group~\cite{Baker:2006ha, Pan:2007nw}, with the group
at Jena 
following soon after that~\cite{Hannam:2007ik, Gopakumar:2007vh,
Hannam:2007wf}. These early comparisons were between post-Newtonian
and numerical relativity predictions of the waveforms for
non-precessing systems. The first comparison of precessing waveforms
was done in Ref.~\cite{Campanelli:2008nk}. All of these comparisons
were for relatively short waveforms and in a regime where
post-Newtonian theory is not particularly accurate. The longest
comparison to date between post-Newtonian and numerical relativity
waveforms was a 175 orbit evolution performed in
Ref.~\cite{Szilagyi:2015rwa}.  
Other studies of much more separated binaries (where the evolutions
were not taken to merger) showed good agreement in the dynamics
between post-Newtonian and numerical relativity for several orbits
with separations of $100M$ and $50M$ for an  equal-mass,
nonspinning binary~\cite{Lousto:2013oza}.

While current post-Newtonian waveforms are not particularly accurate
during the late-inspiral phase, there are models {\it inspired} by
post-Newtonian theory that reproduce numerical waveforms with greater
accuracy. The Effective One Body formalism~\cite{Buonanno:1998gg,
Buonanno:2000ef, Damour:2000we, Damour:2001tu, Buonanno:2005xu}
recasts the problem of the evolution of the binary as an effective
field  theory for a single particle. This formalism contains free
parameters which can be modeled using numerical relativity
simulations. The resulting formalism EOB-NR can reproduce
gravitational waveforms, at least for non-precessing binaries, with
great accuracy. Furthermore, these waveforms are produced at a
fraction of the cost of the original numerical
simulations~\cite{Buonanno:2007pf, Damour:2007vq, Buonanno:2009qa, Pan:2009wj,
Pan:2011gk, Hinder:2013oqa, Pan:2013rra, Taracchini:2012ig,
Taracchini:2013rva, Babak:2016tgq}.

\section{Relativistic stars and disks}
\label{sec:stars-disk}

\subsection{Black hole--black hole binaries with accretion}
\label{sec:bbh-disk}

Astronomers expect that the environment of supermassive \bbh
mergers will often be gas-rich; therefore, there is hope
for an electromagnetic counterpart to the (low-frequency) gravitational
wave signal.  In a thin accretion disk around a single black hole,
angular momentum flows outward (an effect of MHD turbulence), causing
gas to slowly spiral inward, releasing energy radiatively as it falls
deeper into the gravitational potential.  (Thus, the more compact the
object, the more efficiently accretion onto it can release energy.) 
A \bbh near merger might be accompanied by gas orbiting the binary
itself, forming a ``circumbinary disk''. 
Early 1D studies of circumbinary disks predicted that
gravitational torques from the binary would clear out a region of radius about twice
the orbital separation (for binaries with mass ratio around unity),
suggesting that accretion onto the black holes
would be mostly frustrated.  As the binary inspirals, eventually the
inspiral timescale becomes smaller than the disk's viscous
timescale~\footnote{The viscous timescale is the timescale on which angular
momentum transport moves gas inward.  The name comes from the common practice
of modeling this transport process with a viscosity.},
presumably causing the disk to decouple from the binary, its inner
edge unable to keep up with the shrinking binary.  Newtonian
2D (vertically summed)~\cite{1994ApJ...421..651A,Macfadyen:2006jx} and
3D~\cite{Hayasaki:2006fq,Cuadra:2008xn,Roedig:2012nc}
simulations confirm the evacuation of the region around the binary,
but find that gas is efficiently carried in narrow accretion
streams from the inner disk to the black holes.
Meanwhile, perturbations to the disk caused by the merger itself,
with the associated mass loss and kick of the central system due
to gravitational waves, have been investigated by artificially
reducing the mass and adding linear momentum to the central
object around an equilibrium disk~\cite{ONeill:2008sat,2010MNRAS.404..947C,
2010MNRAS.401.2021R,Zanotti:2010xs,Ponce:2011kv}.  All of this
suggests that high luminosity can be maintained after decoupling
and through merger.

Numerical relativity studies of fluids near \bbhs began shortly after the
moving puncture revolution.  It is not clear that this had to be
the case.  Some of the most advanced recent works consider disks
around inspiraling binaries without dynamically evolved spacetimes. 
For example, Noble~{\it et al.}~\cite{Noble:2012xz} used a 2.5
post-Newtonian-order
approximation to the binary spacetime further than 10$M$ from the binary
combined with a 3.5 post-Newtonian approximation
to the binary orbital evolution, while Gold~{\it et al.}~\cite{Gold:2013zma}
simply rotated their conformal thin sandwich initial data.
Nevertheless, the first numerical relativity treatments did include spacetime evolutions
through merger.  These early studies by Bode~{\it et
  al.}~\cite{Bode:2009mt,Bogdanovic:2010he,Bode:2011tq} and Farris~{\it
  et al.}~\cite{Farris:2009mt} considered binaries immersed in
low-angular momentum gas (advection-dominated / Bondi-like inflow) and
calculated electromagnetic luminosity from bremsstrahlung and
synchrotron emission.

\begin{figure}
\centering
\subfloat[Part 1][$-11.0 \,M_8$
hrs]{\includegraphics[scale=0.25]{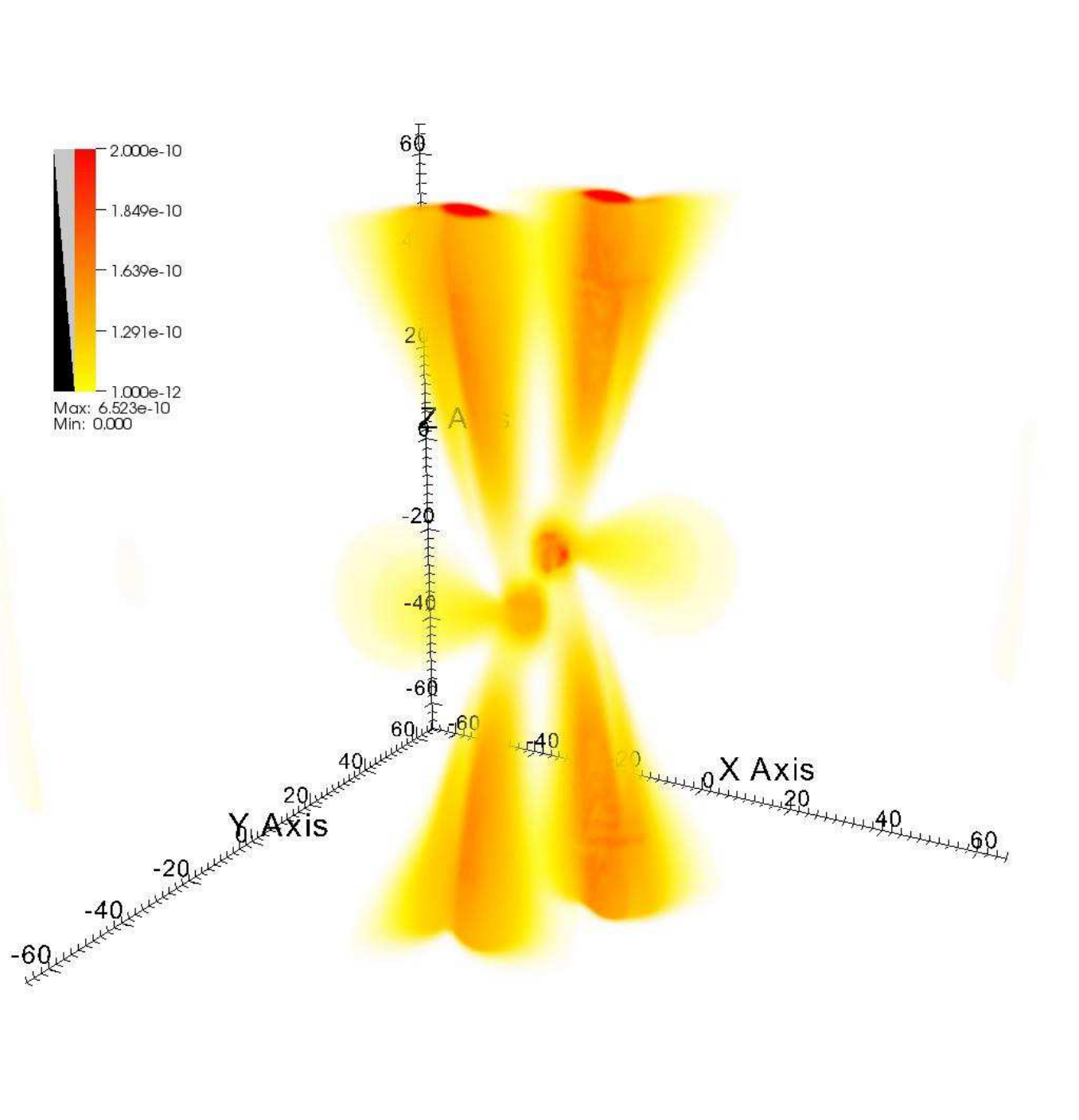} \label{fig:fourfigures-a}}
\subfloat[Part 2][$-3.0\, M_8$
hrs]{\includegraphics[scale=0.25]{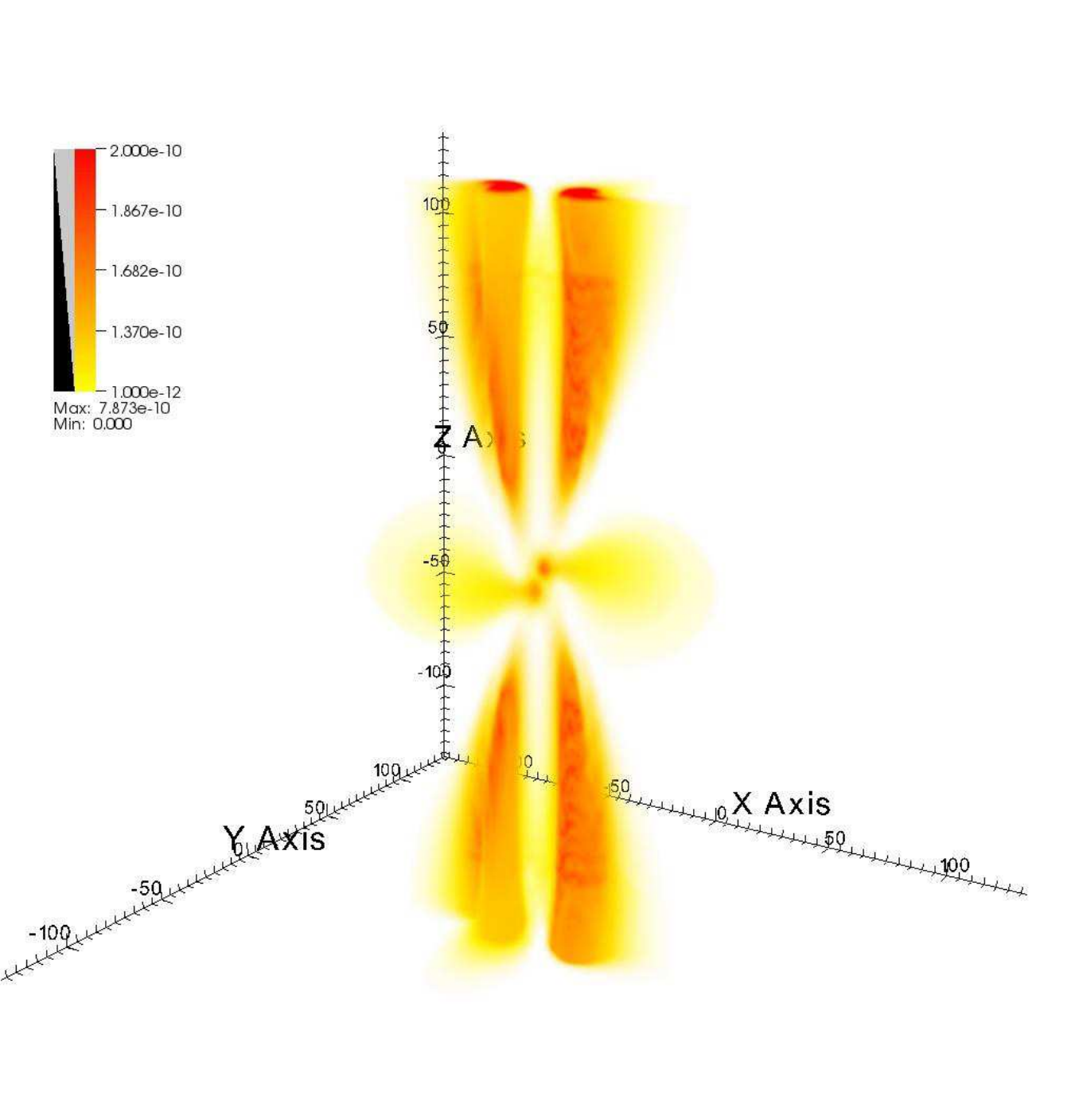} \label{fig:fourfigures-b}}\\
\subfloat[Part 3][$4.6 \, M_8$
hrs]{\includegraphics[scale=0.25]{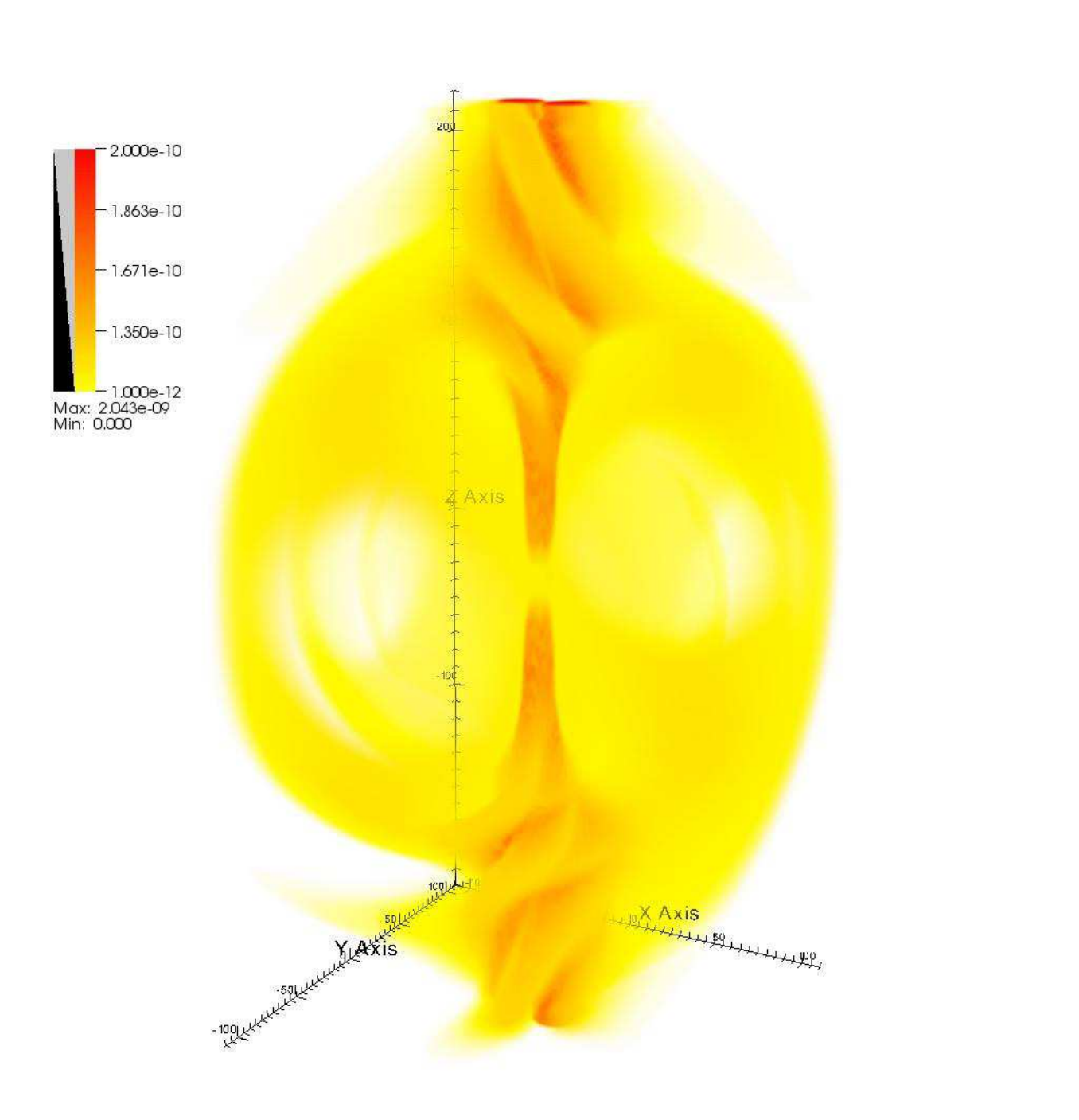} \label{fig:fourfigures-c}}
\subfloat[Part 4][$6.8\, M_8$
hrs]{\includegraphics[scale=0.25]{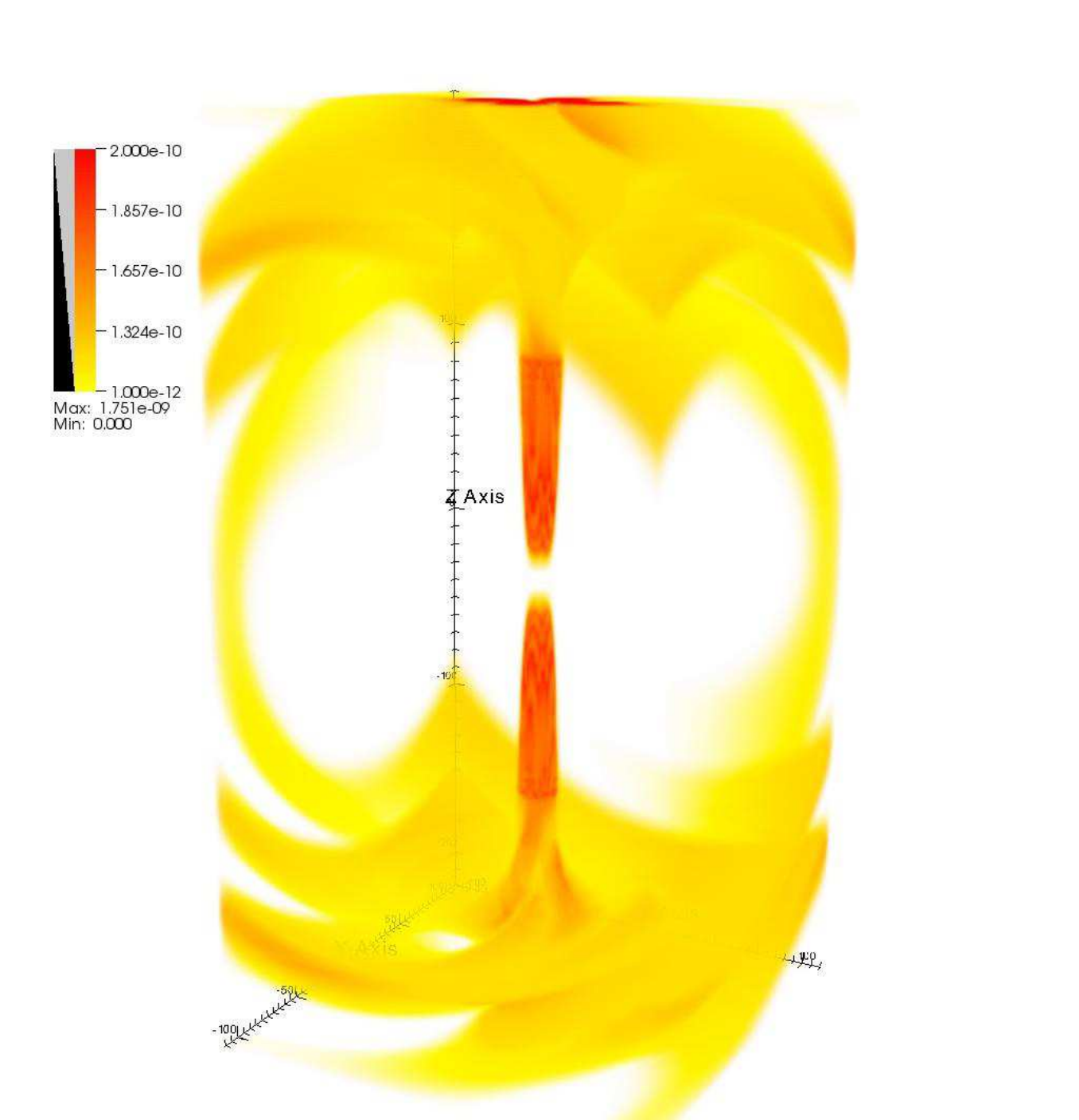} \label{fig:fourfigures-d}}
\caption{Electromagnetic energy flux at different times in the force-free magnetosphere
  surrounding a \bbh merger.  The collimated
  part is formed by two tubes orbiting around each other following the motion of the black holes.
  A strong isotropic emission occurs at the time of merger, followed by a single
  collimated tube as described by the Blandford-Znajek scenario.  Reproduced with permission
  from~\cite{Palenzuela:2010nf}.}
\label{fig:dualjets}
\end{figure}

Clearly, magnetic field effects might have important effects on these inflows.
Prior to MHD simulations, Palenzuela~{\it et al.}~\cite{Palenzuela:2010nf}
and Moesta~{\it et al.}~\cite{Moesta:2011bn} performed force-free simulations
of the effect of a \bbh merger on nearby magnetic field lines (presumed to
be anchored to a circumbinary disk outside the computational domain).
An interesting finding of these simulations, shown in Figure~\ref{fig:dualjets},
is the appearance of dual
jets by a sort of binary system generalization of the Blandford-Znajek
process;
in this case energy is extracted from the orbital motion of the binary,
rather than the spin energy of a black hole (the latter being the classic
Blandford-Znajek effect).  (It should be noted, though, that while multiple
studies confirm the presence
of dual jets, they also show that the emission is predominantly
quadrupolar~\cite{Moesta:2011bn,Alic:2012df}.)

Newtonian MHD simulations by Shi~{\it et
  al.}~\cite{2012ApJ...749..118S} found that magnetohydrodynamic disks
(as opposed to previous disks with alpha viscosity) accrete more
rapidly and experience stronger tidal torques.
Soon afterward, numerical relativity MHD simulations were carried out for the
low-angular momentum plasma case by Giacomazzo~{\it et
  al.}~\cite{Giacomazzo:2012iv} and for the circumbinary disk case by
Farris~{\it et al.}~\cite{Farris:2012ux}.  Using numerical relativity MHD and a
post-Newtonian \bbh metric, Noble~{\it et al.}~\cite{Noble:2012xz}
showed that neither binary torques nor decoupling reduce the overall
accretion rate by a large factor.  Gold~{\it et
  al.}~\cite{Gold:2013zma,Gold:2014dta} have carried out numerical
  relativity MHD
simulations varying the binary mass ratio between 1:1 and 1:10,
confirming that magnetic fields boost accretion rate, increase shock
heating, and produce dual jets merging into a single jet at large
distances.  After decoupling but prior to merger, the jets coalesce
into a single jet.  The merger leads to a one-time boost in the jet's
magnetic field strength and outflow velocity, which the authors hope
can provide a signature of \bbh merger (as opposed to a single black
hole disk
flare).

Computational cost has forced nearly all simulations to date to either
excise the inner region containing the binary or, for those cases that
do track flow into the separate black holes, evolve for less than a
viscous timescale.  To observe viscously settled accretion flows,
Farris~{\it et al.}~\cite{Farris:2013uqa} performed long-term
pre-decoupling 2D Newtonian evolutions resolving the inner region.
They find that individual ``mini-disks'' form around each black hole.
Bowen~{\it et al.}~\cite{Bowen:2016mci} have studied these mini-disks
in their post-Newtonian spacetime, inserting disks around each hole and evolving
them to an overall steady state.  The discovery of these
``mini-disks'' illustrates the possibility of further surprises as
future simulations incorporate long evolutions as well as radiation
transport with associated thermal effects.

\subsection{Relativistic stars}
\label{sec:stars-stability}

Numerical relativity is the main tool for studying rapidly rotating
relativistic stars, where it is used to test the stability of
equilibrium configurations and the nonlinear evolution driven by
instabilities.  Because the outcome of these instabilities often involve
black holes, relativistic stars will be a notable part of our story. 
For a full treatment of this topic, see the Living Review article by
Paschalidis and Stergioulas~\cite{Paschalidis:2016vmz,Stergioulas2003},
the book on this subject by Friedman and
Stergioulas~\cite{2013rrs..book.....F}, and
also Chapter 14 of Baumgarte and Shapiro~\cite{BaumgarteBook2010}.

Collapsing star simulations did not have to wait for the black hole
problem to be solved.  Simulations up to the turn of the century
could use singularity avoiding slicings such as maximal slicing
and its approximates to follow collapse a short while past
apparent horizon formation before grid stretching effects (increasing
distortion of slices needed to keep them from intersecting the
singularity) destroyed
the run's accuracy.  Evolutions to late times after collapse,
and evolutions of systems like \bhnss,
which have a black hole throughout, had to wait until general black
hole spacetimes could be stably evolved.

\subsubsection{Radial stability and collapse outcome}

Supramassive and hypermassive stars can be created using two-dimensional stellar
equilibrium codes (cf.~\cite{Stergioulas2003}).
Knowing that these equilibria exist, we next consider whether they are
dynamically stable, and if not whether the instabilities are of a kind to destroy
the equilibrium on a dynamical timescale or if they just introduce
some small-scale ''churning'' with effects on a secular timescale.
Stability concerns the behavior of initially small perturbations.
Numerical error provides perturbations on its own, but it is
resolution-dependent, so stability studies often seed perturbations.
A popular method for studying radial stability in stars is pressure
depletion, a slight reduction of pressure below the equilibrium
requirement.  This can be done in a way that preserves the constraints
and respects the equation of state by simply holding $\tau$ and $S_i$
fixed and slightly increasing $\rho_{\star}$~\cite{Haas:2016cop}.  To
study the stability of nonaxisymmetric modes, these modes can be
seeded by nonaxisymmetric perturbations of the density.

Because no black hole is involved (at least until after instability has
clearly manifested itself), numerical relativity could begin
addressing these questions even before the breakthroughs numerical
relativity in 2005 that allowed for the evolutions of orbiting
\bbhs.
Already in 2000, simulations by Shibata {\it et
  al.}~\cite{Shibata:1999yx} showed that the dynamical instability
point for uniformly rotating supramassive $n=1$ polytropes nearly
coincides with the secular instability point.  Despite the rapid
rotation, when pressure-depleted unstable stars collapse to black
holes, they leave almost no disk.  A follow-up study of supramassive
polytropes with various polytropic index $n<2$ near their
mass-shedding limit also found almost no disk mass around the
post-collapse Kerr black holes~\cite{Shibata:2003iw}.  These
simulations used singularity avoiding slicings and could not evolve
long after horizon formation.  Subsequent studies using excision after
apparent horizon location confirmed the no-disk result for uniformly
rotating stars with stiff polytropic
EoS~\cite{Duez:2004uh,Baiotti:2004wn}.  Based on angular momentum
distribution at the maximum mass configuration, it is expected that
this result carries over to realistic neutron-star EoS~\cite{Margalit:2015qza}.
(In section~\ref{sec:collapse}, we will see that the story is quite different for
non-compact stars with soft EoS.)  After moving puncture gauge
conditions were discovered, it was possible to evolve to a final black
hole state and confirm that the metric matches the spinning puncture
form~\cite{Dietrich:2014wja}.

\begin{figure*}[thb]
\begin{center}
\leavevmode

\leavevmode
\hspace{-0.7cm}\includegraphics[width=1.8in]{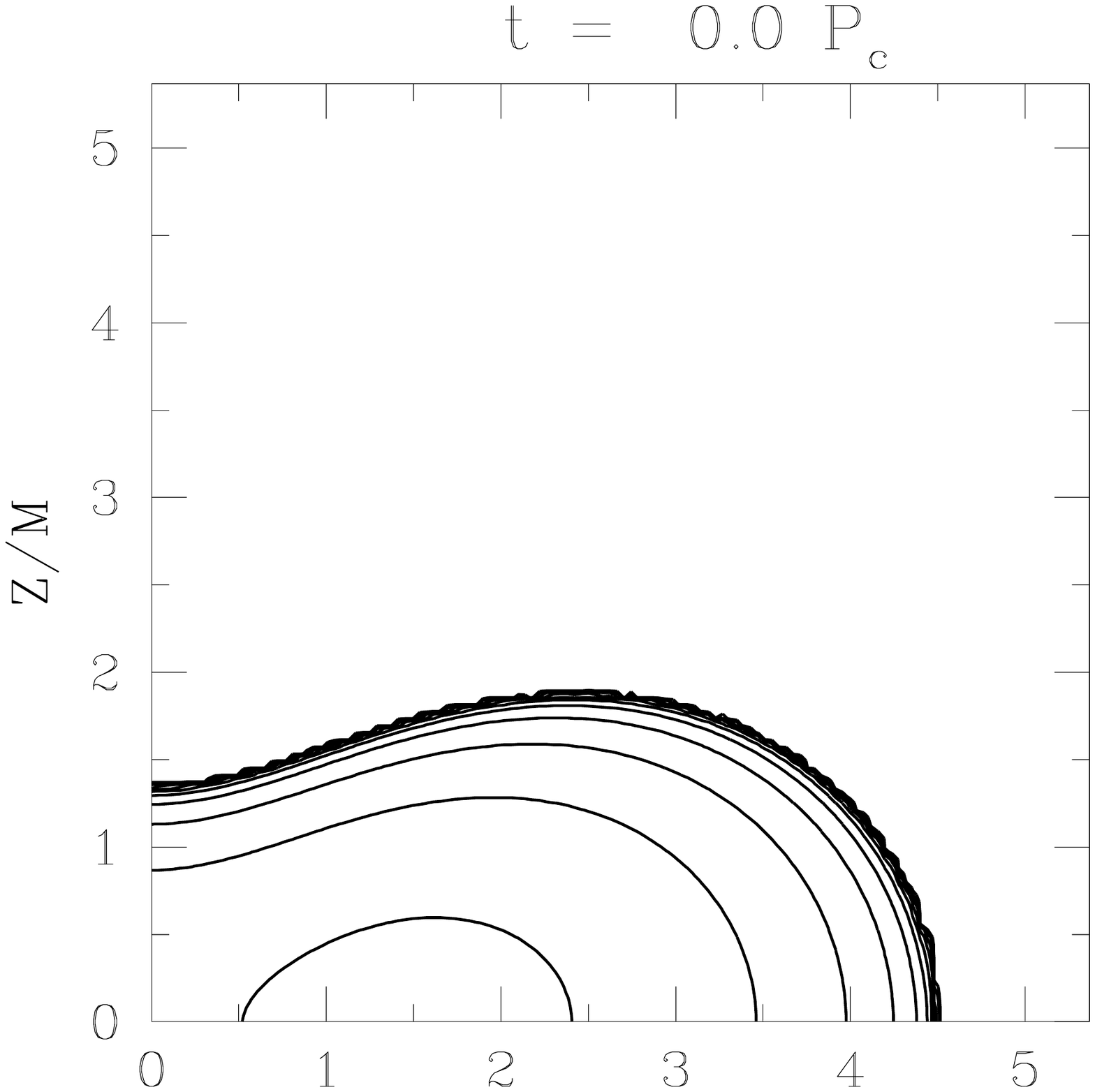}
\leavevmode
\hspace{-0.7cm}\includegraphics[width=1.8in]{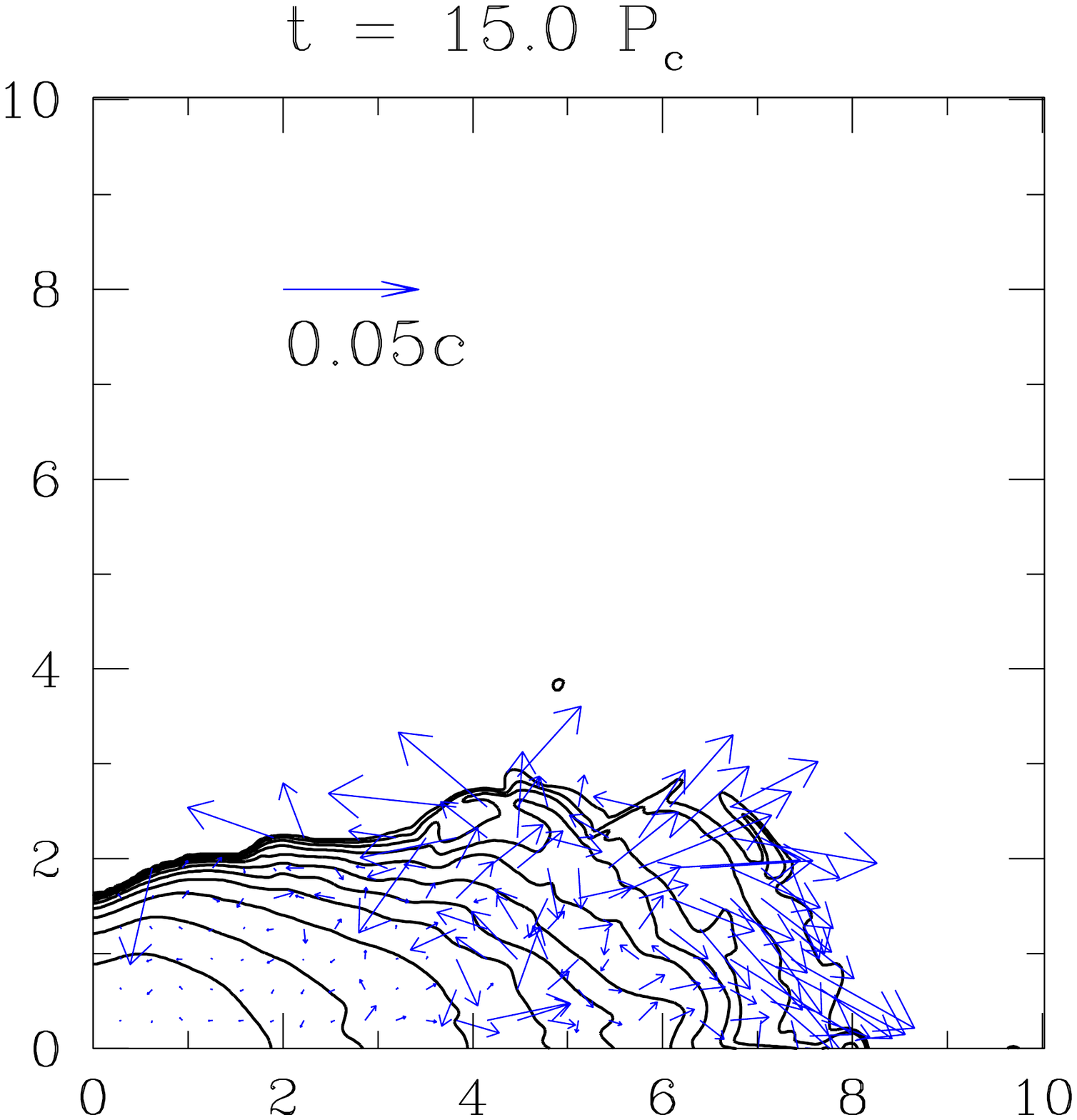}
\leavevmode
\hspace{-0.7cm}\includegraphics[width=1.8in]{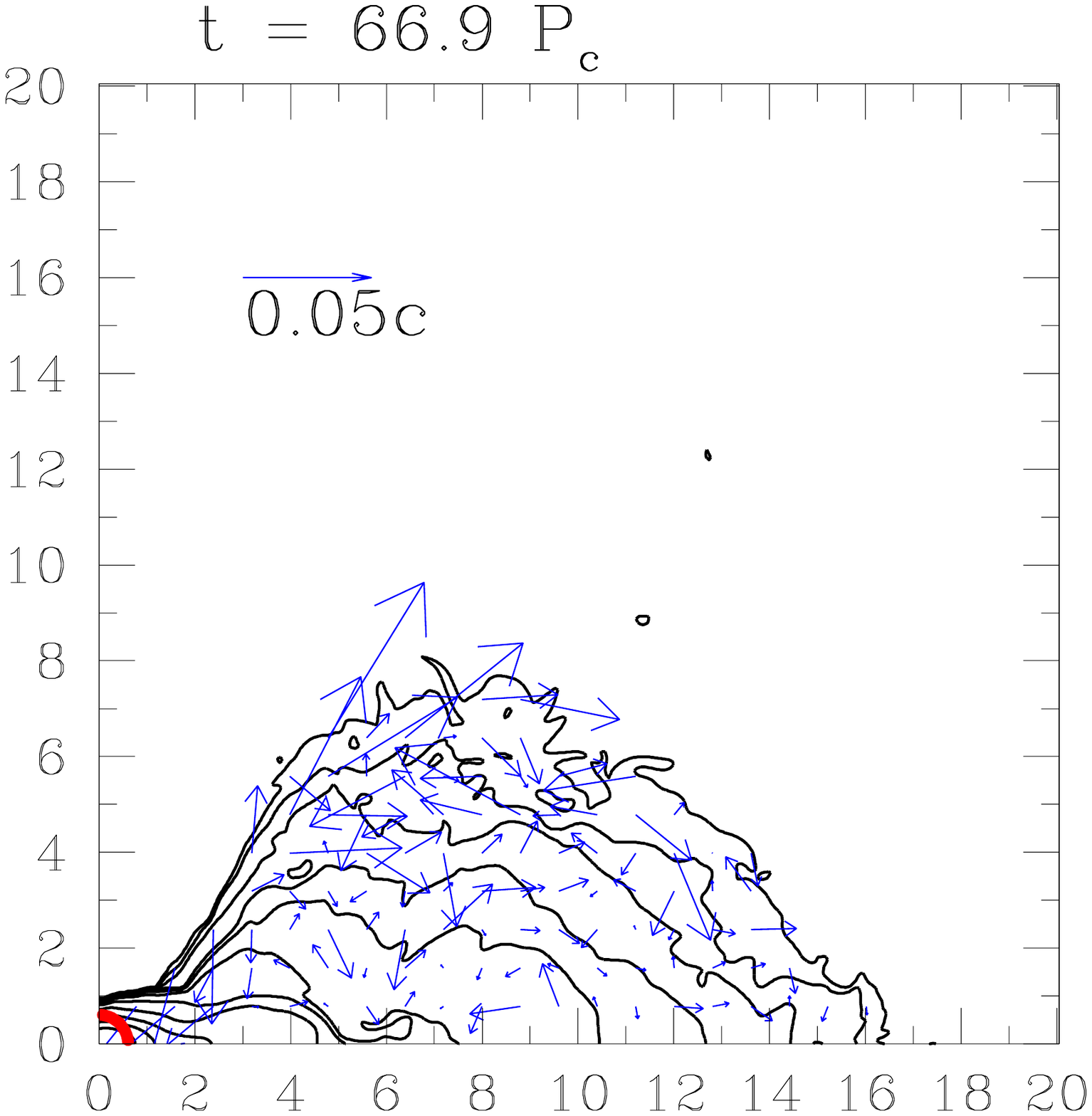}
\leavevmode
\hspace{-0.7cm}\includegraphics[width=1.8in]{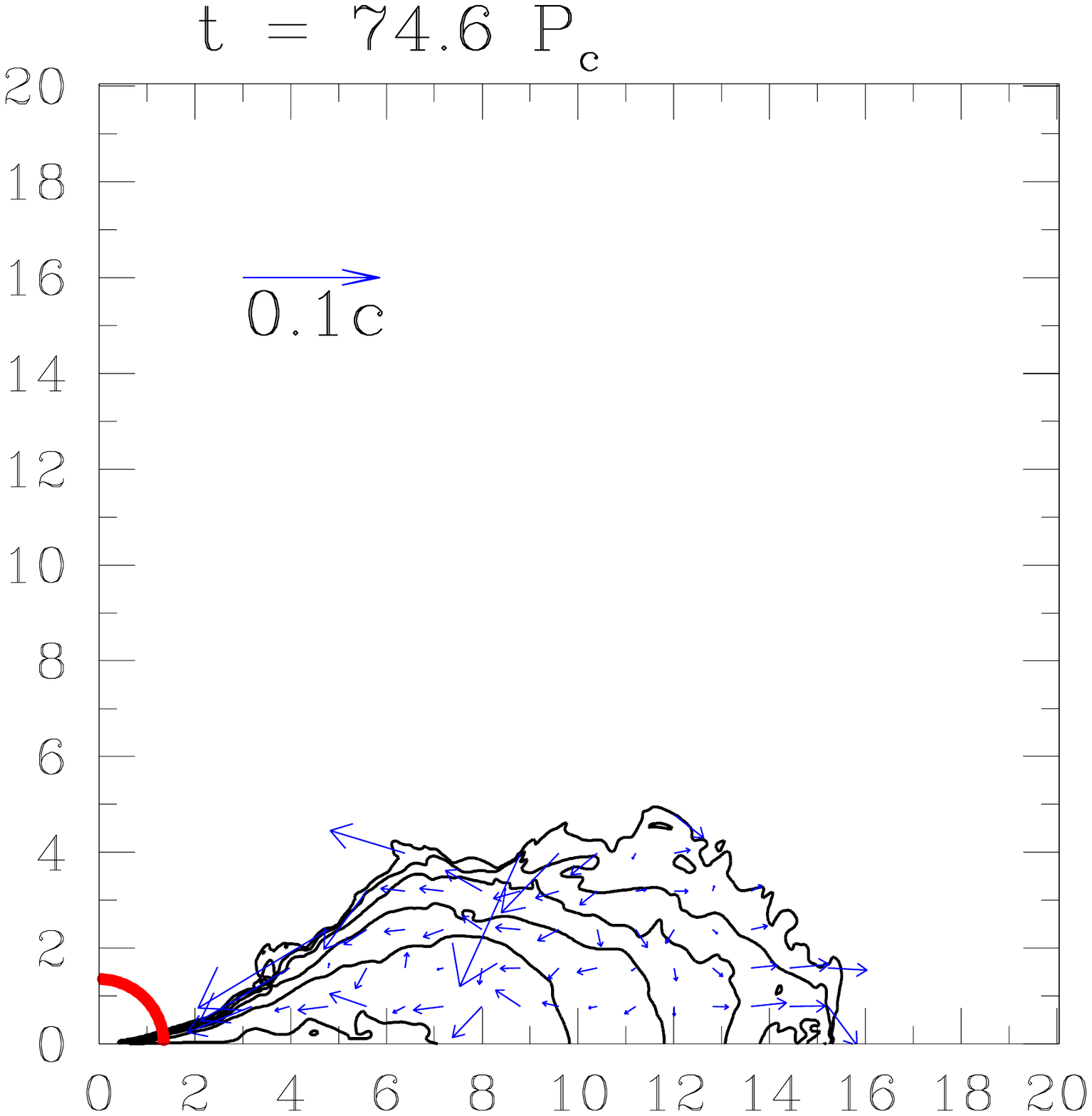} \\
\vspace{-0.5cm}
\leavevmode
\hspace{-0.7cm}\includegraphics[width=1.8in]{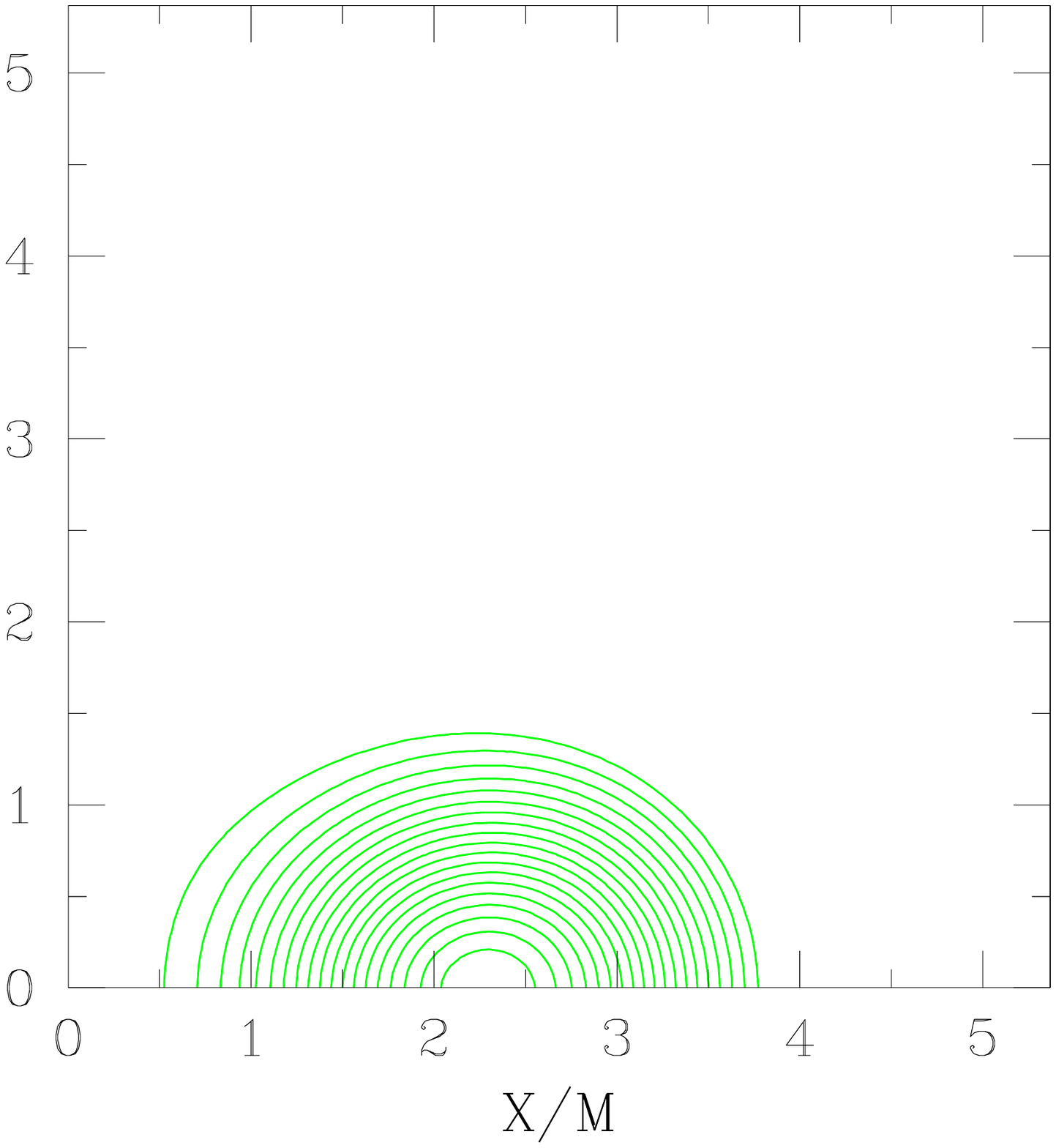}
\leavevmode
\hspace{-0.7cm}\includegraphics[width=1.8in]{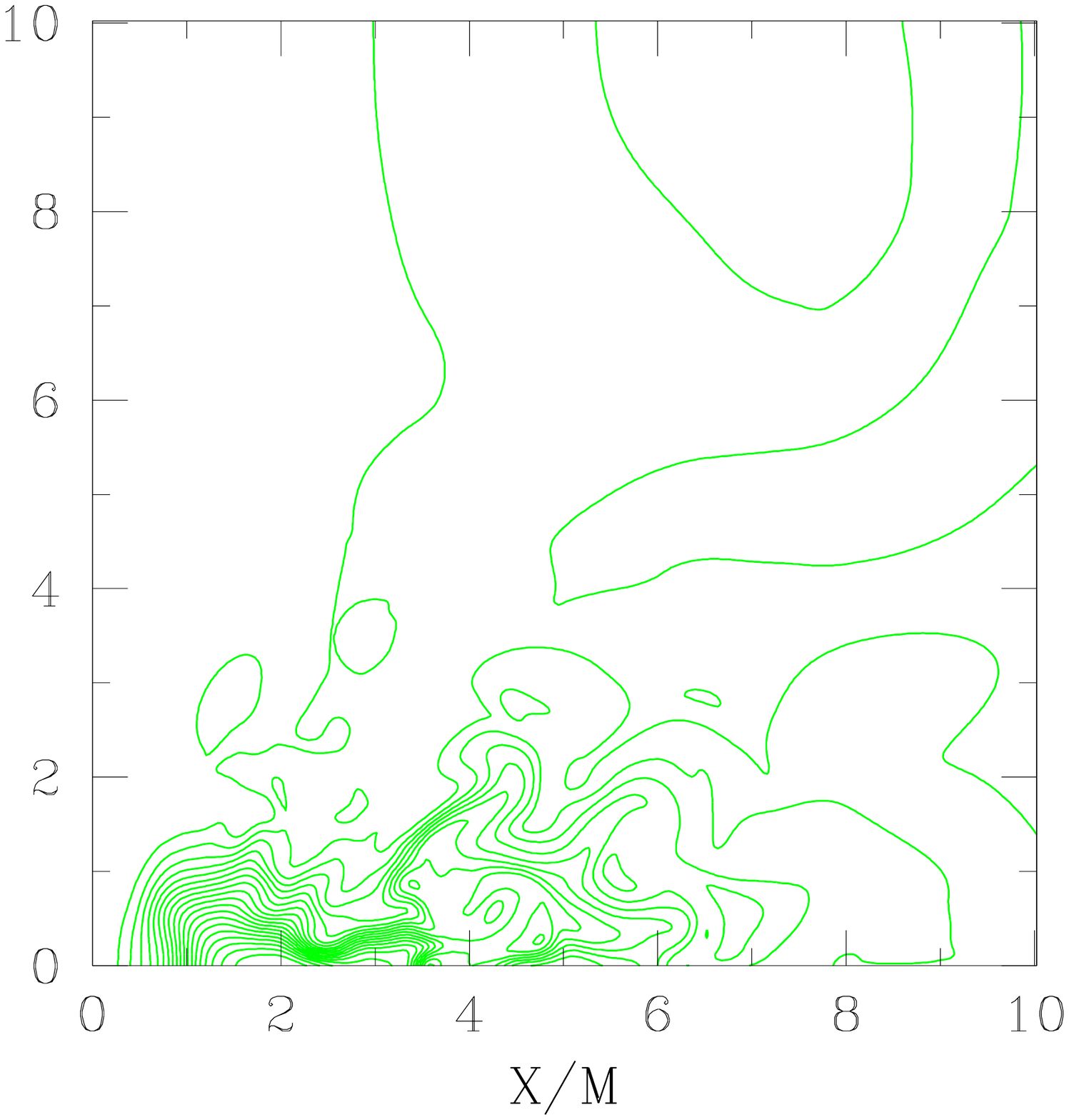}
\leavevmode
\hspace{-0.7cm}\includegraphics[width=1.8in]{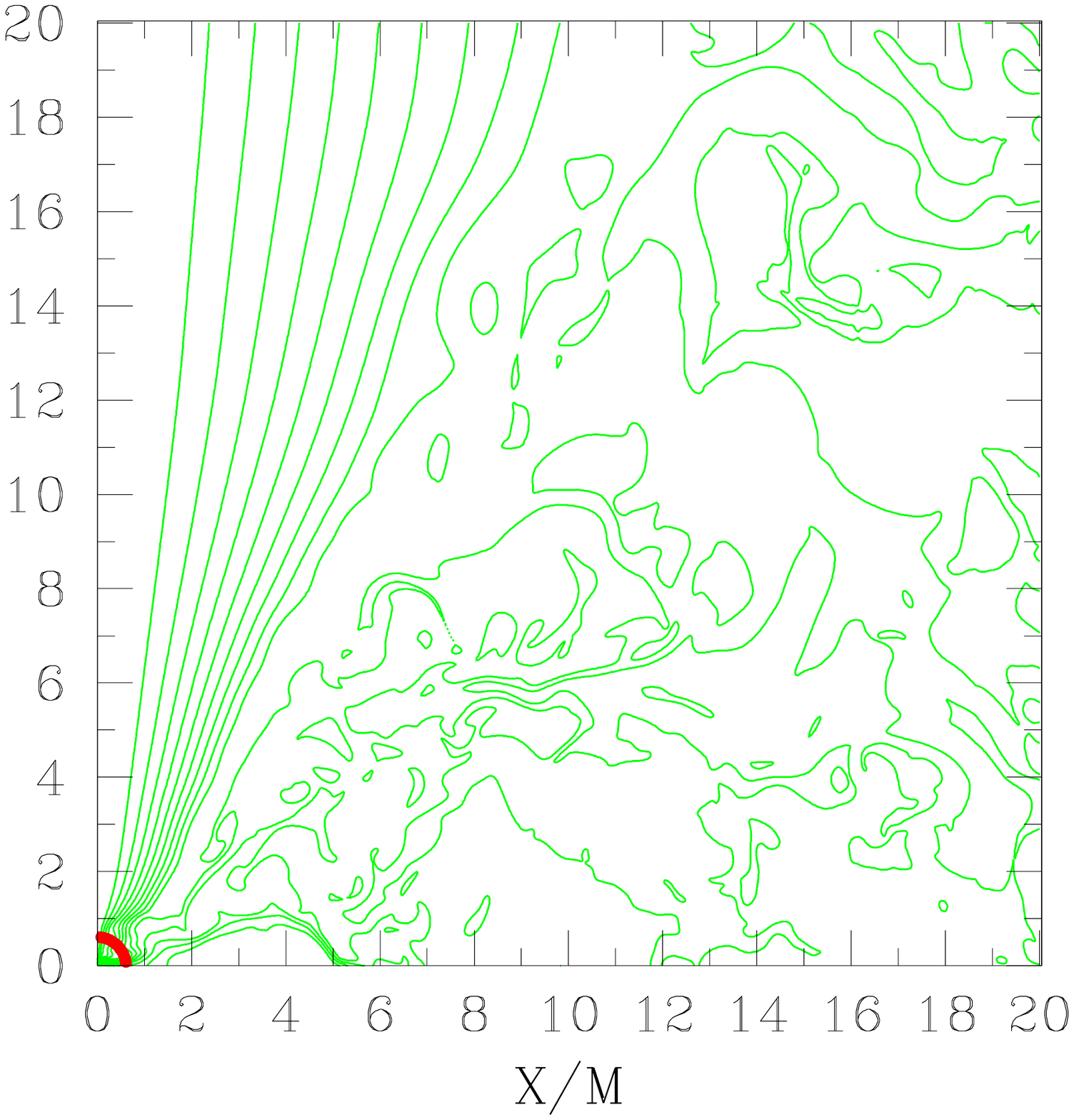}
\leavevmode
\hspace{-0.7cm}\includegraphics[width=1.8in]{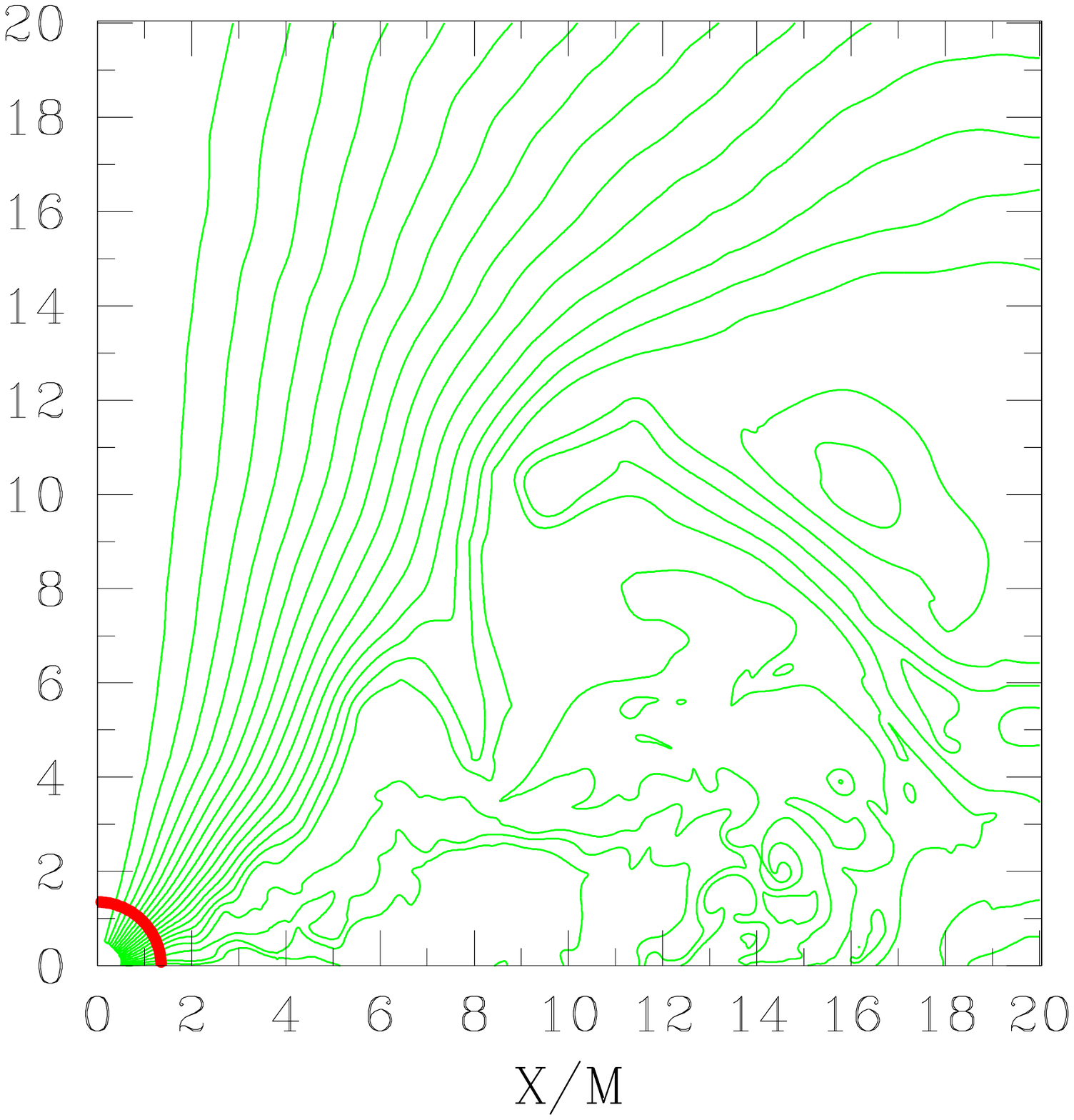}

\caption{The evolution of a hypermassive star under the
influence of a seeded magnetic field.  
The upper 4 panels show snapshots of the rest-mass density
contours and velocity vectors on the meridional plane. The lower
panels show the field lines 
for the poloidal magnetic field at the same times
as the upper panels. 
The thick solid (red) curves denote the apparent horizon
which appears when the central region collapses.
 Reproduced with permission from~\cite{Duez:2005cj}.
\label{mhd_collapse}}
\end{center}
\end{figure*}

Numerical relativity provides a straightforward way to test the
dynamical stability of hypermassive neutron stars:  just evolve for
several dynamical times (perhaps with some initial perturbation).  By
this test, Baumgarte~{\it et al.}~\cite{Baumgarte:1999cq} demonstrated
the stability of a model with mass around $1.6 M^{\rm max}{}_{\rm TOV}$.
It is easy to construct differentially rotating compact stars with
spin angular momentum on
either side of the Kerr limit, so inducing their collapse by
sufficient pressure depletion provides a test of cosmic censorship;
unsurprisingly, it passes~\cite{Duez:2004uh,Giacomazzo:2011cv}.  The
super-Kerr systems undergo a centrifugal bounce, and the fluid forms a
torus which then fragments due to nonaxisymmetric instabilities.

\subsubsection{Magnetohydrodynamic evolution}
\label{sec:stars_mhd}

These hypermassive neutron stars, although dynamically stable, are
presumably driven to collapse on a secular timescale by processes that
transport angular momentum outward, robbing the core
of its rotational support.  The ultimate source of angular momentum
transport is most likely turbulence driven by the magnetorotational instability.
To study secular evolution from first principles requires magnetohydrodynamic
(MHD) simulations.

Such simulations must specify an initial state for the magnetic field.
One injects a small magnetic field into the equilibrium state, usually
not chosen to be an MHD equilibrium but thought of as a ``seed'' of
the more physical magnetic field that will grow through shear and
turbulence.  MHD simulations suggest that the magnetic field in a star
with random initial state tends to settle to a helical mixture of
poloidal and toroidal field~\cite{2006A&A...450.1077B}.  Equilibrium
fields can be constructed in relativity (e.g.~\cite{Bocquet:1995je,
Kiuchi:2008ch,Frieben:2012dz,Ciolfi:2013dta}),
but simpler seed fields are more often used, e.g. poloidal fields
moving along isodensity contours constructed from an azimuthal vector
potential
\begin{equation}
  A_{\phi} = \left\{\begin{array}{cc}
  A_b\varpi^2 (\rho_0 - \rho_{\rm cutoff})^n &
  {\rm for} \rho_0 > \rho_{\rm cutoff} \\
  0                  & {\rm otherwise}
  \end{array}\right. \ ,
  \label{seed-field}
\end{equation}
where $\varpi$ is the cylindrical radius; $A_b$, $n$, and $\rho_{\rm
  cut}$ are freely specifiable constants.  The expectation was that,
for magnetorotationally unstable systems, no memory of the seed field
would long survive.  Black hole-torus simulations give some support to
this assumption for the interior of the torus but find that the
appearance and strength of polar jets is very sensitive to seed field
geometry~\cite{Beckwith:2007sr}.

Differential rotation with $d\Omega/dr<0$ (the usual case for
differential rotation) is unstable to the magnetorotational
instability (MRI)~\cite{1998RvMP...70....1B}.  The fastest growing MRI
mode has size $\lambda_{\rm MRI}\sim v_A/\Omega$, where $v_A\sim
B/\sqrt{\rho_0}$ is the Alfven speed, and growth timescale
$\sim\Omega^{-1}$.  For realistic $B$ fields, resolving $\lambda_{\rm
  MRI}$ can be a serious computational challenge.  Numerical
  relativity MHD studies
sometimes avoid this problem by using magnetar-strength seed fields.
The MRI has been identified in relativistic
disk (e.g.~\cite{DeVilliers:2003gy,Gammie:2003rj,
  Anton:2005gi,Anninos:2005kc}) and star~\cite{Siegel:2013nrw}
simulations.  The effect of the MRI is to initiate turbulence and thus
on a secular timescale to dissipate energy as heat and transport
angular momentum outward.  The MRI's role in driving disk accretion is
the subject of a vast amount of work; the distinctive role of
numerical relativity has
been to study its effect on differentially rotating neutron stars.

Axisymmetric simulations of magnetized hypermassive neutron stars were
undertaken by Duez~{\it et al.}~\cite{Duez:2005cj} using a $\Gamma=2$
EoS and initial angular velocity about three times higher at the center
than at the equator.  These simulations were
notable for the narrative of this article as being among the first
astrophysically interesting numerical relativity simulations to prolong evolution past
collapse using excision.  [Shortly afterward, both groups in the
collaboration--University of Illinois at Urbana-Chanmpaign (UIUC) and
Kyoto University--switched to moving punctures.] 
A combination of magnetic
winding/braking and MRI turbulence transports angular momentum outward,
causing the envelope to expand and the core to contract.  After about
$10^2$\,ms, the core undergoes collapse on a dynamical timescale. 
The collapse leaves a massive torus surrounding
the black hole, a promising setup for a short-duration gamma ray
burst~\cite{Shibata:2005mz}.  Snapshots from this evolution are shown
in Figure~\ref{mhd_collapse}.  The same collaboration then performed
similar studies for other differentially rotating neutron star initial
configurations~\cite{Duez:2006qe}.  Use of a more realistic EoS produces a
qualitatively similar outcome, but collapse is averted (unsurprisingly)
if the star is not hypermassive.  A non-hypermassive star with angular
momentum too high for a uniformly rotating star settles to an
equilibrium uniformly rotating star plus torus configuration.

\subsubsection{Nonaxisymmetric mode instability}

Nonaxisymmetric modes, which have the form $\delta\propto
e^{i(M\phi-\omega t)}$, are interesting as gravitational wave sources.
Since the background equilibrium is often differentially rotating, there is
a clear conceptual difference between the fluid's rotation and the
perturbation mode, which rotates with constant pattern speed
$\Omega_p=\omega/M$ everywhere.  In fact, rotating perfect fluid
stars are generically unstable because of gravitational waves
via the Chandrasekhar-Friedman-Schutz instability, although in
most realistic cases this is suppressed by viscosity or other
effects.  (See~\cite{Stergioulas2003} and references therein.)  

It is well-known that for rotating stars, the fundamental $L=M=2$
(bar) mode becomes unstable for sufficiently high $T/|W|$:  around
0.14 for a secular instability and around 0.27 for a dynamical
instability.  The unstable bars grow to nonlinear amplitude and lead
to the shedding of high angular-momentum material.  It is thus hard to
imagine $T/|W|>0.27$ stars persisting in nature.  The dynamical bar
mode instability (often called the ``high $T/|W|$ instability'') has
been confirmed in numerical relativity simulations of differentially rotating
stars~\cite{Shibata:2000jt,Baiotti:2006wn,Manca:2007ca}.

Numerical relativity investigations also found unexpected unstable growth of low (but
nonzero) $M$ modes in strongly differentially rotating or toroidal
stars at $T/|W|$ well below the dynamical bar mode
threshold~\cite{Centrella:2001xp,Shibata:2002mr,Shibata:2003yj,Saijo:2003hj}.
The instability has been seen in numerical relativity rotating stellar core collapse
simulations~\cite{Ott:2006eh} and so may be an important gravitational
wave source from a galactic supernova.  As the mechanism was not at
first understood, the instability was called the ''low $T/|W|$
instability'' or the ''one-armed spiral instability'', names that
sometimes persist.  We now know that it is caused by a corotation
resonance~\cite{Watts:2003nn}.  The corotation radius $r_c$ of a mode
is the radius where $\Omega(r_c)=\Omega_p$.  The mode has positive
energy for $r>r_c$ and negative energy for $r<r_c$, and thus energy
can be transferred outward at $r_c$ to strengthen the mode on both
sides.  The corotation instability takes many crossings to grow, so
the mode must be trapped in a region containing $r_c$.  In stars, a
minimum of the vortensity can act like a trapping
potential~\cite{Ou:2006yd}, and this can be produced by a toroidal
density structure or extreme differential rotation.  Subsequent
numerical simulations are consistent with this
model~\cite{Corvino:2010yj,Muhlberger:2014pja}.  Bar mode growth from
both the low-$T/|W|$~\cite{Muhlberger:2014pja} and the
high-$T/|W|$~\cite{Franci:2013mma} instabilities has also been simulated
in the presence of magnetic fields, where magnetic tension fails to
suppress the instabilities for any realistic field strength.

\subsubsection{Stability of self-gravitating black hole accretion disks}

Given the ability to evolve matter in dynamical black hole spacetimes,
we can carry out a similar analysis to that of relativistic stars,
this time for self-gravitating tori around black holes.  Once again
we require axisymmetric constraint-satisfying equilibrium initial
data, with a central black hole introduced either by a horizon inner
boundary condition~\cite{1994ApJ...427..429N} or a
puncture~\cite{Shibata:2007zzb}.  Then we just watch perturbations
evolve.

Self-gravity can lead to instabilities
in black hole-torus systems.  Even in Newtonian physics, a disk will
break apart if it violates the Toomre stability criterion.  A
potential axisymmetric dynamical instability that has received much
attention from numerical relativists is the runaway
instability~\cite{1983Natur.302..597A}.  In this scenario, an
equilibrium torus filling its Roche lobe will be unstable because a
small amount of accretion into the black hole increases the black
hole's mass, pushing the Roche lobe into the torus.  The mass transfer
into the black hole is then unstable and destroys the torus in a
dynamical timescale.  Analysis of stationary configurations suggest
that the disk's self-gravity enhances the
instability~\cite{1996MNRAS.278L..41N}, but disks are stabilized by a
positive radial angular momentum gradient in the disk and by black
hole spin~\cite{1998A&A...331.1143A}.  The first relativistic
simulations, by Daigne and Font~\cite{Font:2002bi,Daigne:2003tf},
neglected torus self-gravity, treating evolution of the metric by
allowing the mass and spin of the Kerr black hole to increase by
accretion.  These confirmed that even rather small angular momentum
gradients prevent the instability.  However, only live-metric
simulations could properly include torus self-gravity.  These were
first undertaken by Montero, Font, and Shibata~\cite{Montero:2010gu},
and they did not find a runaway stability in any of their models.
However, simulations by Korobkin~{\it et al.}~\cite{Korobkin:2012gj}
did demonstrate the existence of the instability in disks particularly
prone to it.

There are also nonaxisymmetric instabilities present in black
hole-torus systems analogous to those found in rotating stars.  Before
they were identified in stars, corotation instabilities were already
known to exist in nearly-constant angular momentum disks (the
Papaloizou-Pringle
instability~\cite{1984MNRAS.208..721P,1985MNRAS.213..799P}) and in
disks with vortensity maxima (the Rossby wave
instability~\cite{Lovelace:1998kv}).  Self-gravity is not an essential
feature of the instability, and in fact self-gravity tends to suppress
the Papaloizou-Pringle instability~\cite{1988MNRAS.231...97G}.

Torus self-gravity can trigger nonaxisymmetric instabilities not
present in nonself-gravitating disks.  These instabilities have been
studied systematically by 3D simulations in Newtonian
physics~\cite{1994ApJ...420..247W} and numerical relativity~\cite{Korobkin:2010qh}.  The
latter study, by Korobkin~{\it et al.}, constitutes one of the first
notable applications of cubed-sphere multipatch technology to
numerical relativity. 
Simulations show that moderate self-gravity triggers
``intermediate mode'' instability~\cite{1988MNRAS.231...97G},
spontaneous elliptic deformations of the disk that, in fact, can be
considered the disk analogue of the high-$T/|W|$
instability~\cite{1992ApJ...388..451C}.  An interesting effect of
self-gravity on the $m=1$ Papaloizou-Pringle mode is momentum transfer
between the disk and black hole, leading to an outspiraling motion of
the black hole~\cite{Korobkin:2010qh}.  The nonlinear development of
this instability was explored in numerical relativity by Kiuchi~{\it et
  al.}~\cite{Kiuchi:2011re}, who suggest it may be a significant source
of gravitational waves (e.g. from a GRB central engine or the aftermath
of a supermassive star collapse).

\section{Black hole formation}
\label{sec:collapse}

In Section~\ref{sec:stars-stability}, we considered the collapse of uniformly rotating
compact, stiff stars, finding that they tend to collapse to Kerr black
holes with no significant leftover material to form an accretion disk.
Much more common astrophysically is the formation of black holes from
non-compact stars with soft EoS around $n=3$ (the marginal stability
limit for Newtonian polytropes).  Scenarios can be divided by the mass
of the progenitor star.  Population I and II stars with masses $\sim
10^1$--$10^2M_{\odot}$ form iron cores of mass $\sim M_{\odot}$, where
$n=3$ comes from the dominance of relativistic degenerate electrons to
the pressure.  The first generation of stars, the metal-free
Population III stars, may have had masses $\sim
10^2$--$10^3M_{\odot}$.  If metal-free gas is unable to cool and does
not fragment into Pop III stars, $\gtrsim 10^5 M_{\odot}$ supermassive
stars may form.  These very massive stars are radiation-pressure
dominated and (because of convection) isentropic, leading them to also
take the form of $n=3$ polytropes.  Like black hole spacetimes,
polytrope systems can be scaled to any mass.  However, this scale
invariance is broken when one takes into account EoS stiffening and
nuclear reactions, which depend on the actual density and not just the
dimensionless compaction.  Numerical relativity is needed to determine the collapse
outcome: the mass and spin of the black hole and the properties of any
accompanying disk.  Even before these simulations could be carried to
post-collapse equilibrium, it was possible to guess from a trick
introduced by Shapiro and Shibata~\cite{Shapiro:2002kk} that these
collapses would be more likely to form massive disks.  The
post-collapse disk is roughly the matter with high enough initial
angular momentum to orbit outside the innermost stable circular orbit
(ISCO)\footnote{In general relativity, the effective potential
  associated with orbits (i.e., timelike geodesics) around a black
  hole is similar to the effective potential in Newtonian gravity, but
  with additional attractive terms proportional to $1/r^3$. Because of
  this term, there is a region from the black-hole horizon to about
  three times the Schwarzschild radius of the black hole were there
  are no stable circular orbit. At the boundary of this region is the
  innermost stable circular orbit, or ISCO. A particle following an inspiraling
  quasicircular orbit will plunge into the black hole once it crosses
this ISCO.}
of the black hole to be formed, which can be of order 10\% of a
supermassive star progenitor.

\subsection{Population I/II core collapse and collapsars}

Stellar mass black holes are thought to originate in the core collapse
of massive stars for cases where, for some reason, the process of
permanently expelling the gas around a protoneutron star fails. 
(The case of successful supernova explosion and neutron star formation
has been the subject of much numerical work, the discussion of which
would take us too far afield.)  If
the progenitor has sufficient angular momentum, the newly formed black
hole may be surrounded by an accretion torus.  This is the explanation
of long-duration gamma ray bursts in the collapsar model of Woosley
and
MacFadyen~\cite{1993ApJ...405..273W,1993AAS...182.5505W,MacFadyen:1998vz}.
Formation of a black hole-torus system may occur in several ways.  The
inner iron core may originally collapse to a protoneutron star, but
a supernova may fail to occur (the shock stalls and does not
sufficiently re-energize), so that the star eventually collapses under
its accumulating mass; this is a Type I
collapsar~\cite{MacFadyen:1998vz}.  A mild explosion may occur, but
enough material falls back onto the star to trigger collapse--a Type
II collapsar~\cite{MacFadyen:1999mk}.  Finally, the inner core might
collapse directly to a black hole--a Type III
collapsar~\cite{Fryer:2000my}.

The first 1D numerical relativity stellar collapse simulations were carried out in 1966
by May and White~\cite{1966PhRv..141.1232M} using the formulation of
Misner and Sharp~\cite{Misner:1964je}: a Lagrangian method with,
however, a slicing that does not avoid singularities.  Thus,
simulations could not be continued long after black hole formation,
but this was enough to determine that a black hole in fact forms,
rather than a neutron star.  This problem can be overcome using a
retarded time coordinate, which avoids the black hole
interior~\cite{1966ApJ...143..452H,1995ApJ...443..717B}.  The same basic
methods have continued to
be used for subsequent spherical collapse simulations with
increasingly sophisticated microphysics and neutrino
transport~\cite{1971ApJ...163..209W,1978ApJ...225L.129V,
1996ApJ...468..823B,OConnor:2010moj},
including detailed studies of the neutrino signals from failed supernovae
solving the Boltzmann transport
equation~\cite{Liebendoerfer:2002xn,Sumiyoshi:2007pp,Fischer:2008rh}.
One-dimensional general relativity simulations find that black hole formation tends to happen
for high progenitor mass (which, due to stellar winds, may be much
lower than the zero age main sequence mass, so black hole formation
is more likely for low-metallicity stars, which suffer less mass loss). 
Prompt
collapse may only occur for very high mass, low metallicity (perhaps
only Population III stars)~\cite{Fryer:1999mi,Heger:2002by,OConnor:2010moj}. 
2D simulations are needed to study rotating collapse.  Some such
simulations were carried out beginning with
Nakamura~\cite{1981PThPh..65.1876N} but lacked realistic initial
conditions and EoS.  In 2005, on the eve of the numerical relativity revolution in black
hole treatment, Sekiguchi and Shibata~\cite{Sekiguchi:2005bi}
attempted greater realism using a set of two-component piecewise
polytrope EoS.  The paper limited itself to the criterion for prompt
black hole formation because subsequent evolution could not be
followed.

While waiting for numerical relativity, 2D post-collapse simulations were being used to
study the evolution of the post-black hole formation torus and the
possible initiation of a gamma ray burst.  Lacking the true
post-collapse configuration, these first simulations had to insert a
black hole into a collapsing flow by hand.  For Newtonian simulations,
the black hole is a Newtonian or pseudo-Newtonian point-mass addition
to the gravitational potential and an inner absorbing
boundary~\cite{MacFadyen:1998vz,MacFadyen:1999mk,Proga:2003ic,Fujimoto:2006eh,LopezCamara:2008ba}.
For relativistic MHD simulations, a fixed Kerr metric was
used~\cite{Mizuno:2003it,Mizuno:2004kw,Barkov:2007us,Nagataki:2009ni}.
A key input parameter is the initial angular momentum.  To produce a
promising torus, this is usually chosen to be large enough for
circular orbit well outside the nascent black hole's ISCO but small
enough that the disk is compact and can lose energy efficiently by
neutrinos.  However, Lee and Ramirez-Ruiz~\cite{Lee:2005et} find
promising behavior even for somewhat lower angular momenta; shocked
gas on the equator forms a dwarf disk which accretes rapidly due to
general relativistic
effects even without magnetic fields or viscosity.  Follow-up
simulations by Lopez-Camara~{\it et al.}~\cite{LopezCamara:2008ba}
suggest that low-$j$ collapsars might differ from high-$j$ collapsars
by the former not producing an accompanying supernova.  Hydrodynamic
simulations, such as the original study by MacFadyen and
Woosley~\cite{MacFadyen:1998vz} add an alpha viscosity and tend to
find that the polar regions free-fall into the black hole while inside
an accretion shock a thick torus forms and viscous heating-driven
outflows are launched.  Subsequent simulations with MHD for both
high-$j$~\cite{Proga:2003ic} and
low-$j$~\cite{Mizuno:2004kw,Harikae:2009dz} cases differ primarily in
the quick appearance of magnetically-driven polar jets.

\begin{figure}[t]
\includegraphics[width=8.5cm]{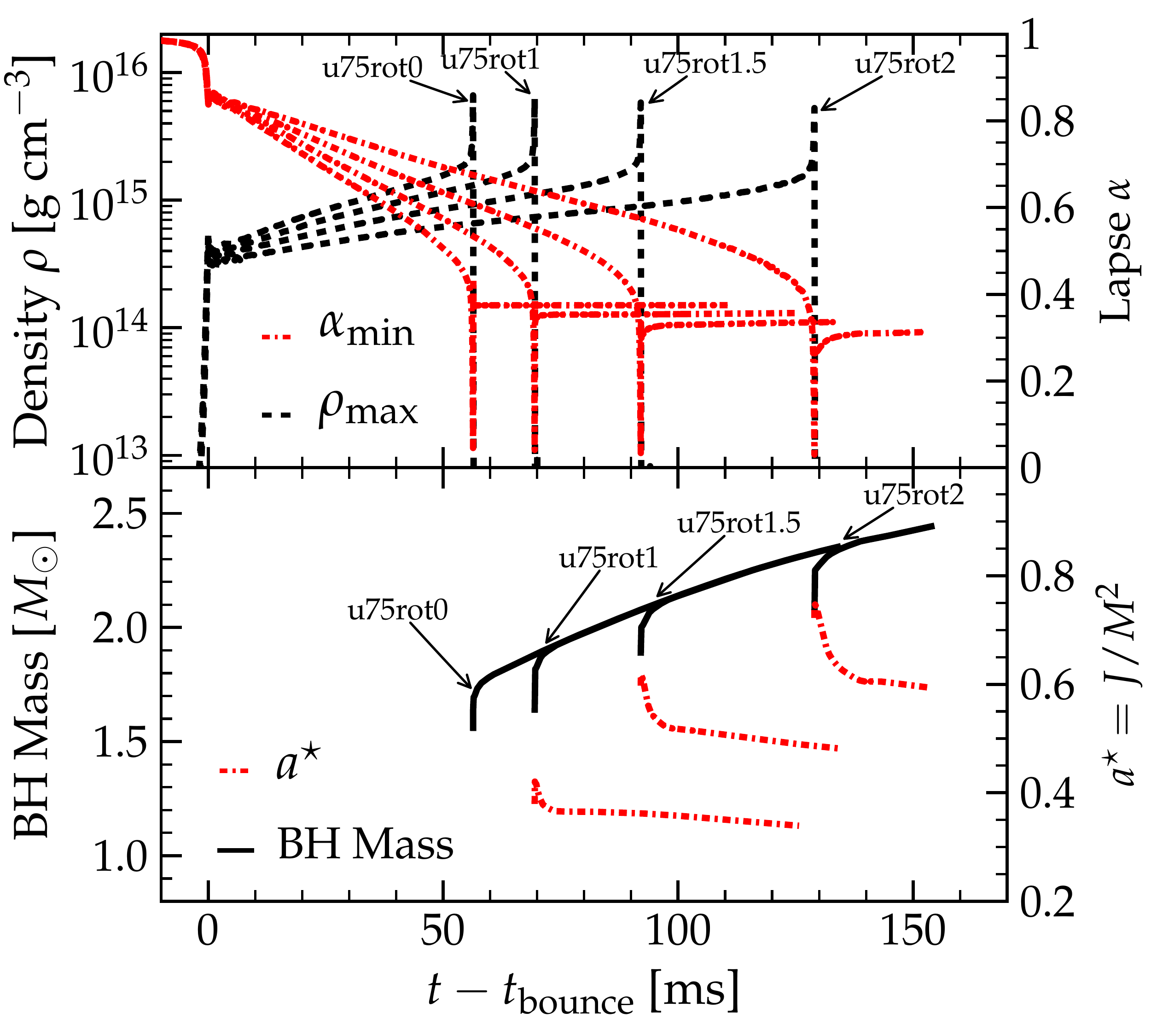}
\caption{
The postbounce evolution of the center of a collapsar in
3D numerical relativity.  Different models correspond to
different choices for the progenitor spin. 
{Top}: Maximum density $\rho_\mathrm{max}$ and central
ADM lapse function $\alpha_\mathrm{min}$ as a function of postbounce
time in all models. After horizon formation, the region interior to it
is excluded from min/max finding.  {Bottom}: black-hole mass and
dimensionless spin $a^\star$ as a function of postbounce time. All
models follow the same accretion history once a black hole forms and settles
down.  Reproduced with permission from~\cite{Ott:2010gv}.
\label{fig:collapsar}}
\end{figure}

Clearly, numerical relativity simulations were needed which include the collapse of a
realistic rotating stellar core, self-consistent black hole formation,
and evolution long past black hole formation to study the dynamics of
the torus.  With the ability to stably form and evolve black holes in
numerical relativity, this became possible.  The first such simulation was carried out
by Sekiguchi and Shibata~\cite{Sekiguchi:2010ja}.  They evolved a
high-entropy core from collapse through a second past black hole
formation using a finite temperature equation of state and neutrino
leakage.  A range of initial $j$ were used; low-$j$ cores produced
geometrically thin shocked disks, while high-$j$ cores produce thick
tori.  The first 3D numerical relativity collapse simulations were performed by Ott~{\it
  et al.}~\cite{Ott:2010gv}.  Octant symmetry and eleven levels of
adaptive mesh refinement made it possible to follow the collapse in
3D, but the simulation was only followed for $\sim 0.1$\,s after black
hole formation.  Core collapse is followed by a bounce, but the
accretion shock stalls, and the protoneutron star collapses to a black
hole.  Several key quantities in these simulations are plotted in
Figure~\ref{fig:collapsar}.  The dimensionless spin peaks at 0.75 for the most rapidly
rotating case and then rapidly decreases as lower-$j$ material is
accreted.  The collapse-bounce-collapse sequence of events leads to a
distinct gravitational wave signal.  In addition to this collapse
waveform, gravitational waves may be produced by inhomogeneities in
the collapsing matter, self-gravitational instabilities in the torus,
and anisotropic neutrino radiation.
(See~\cite{2011LRR....14....1F,Kotake:2012it} and references therein.)

A difficulty for 3D collapsar simulations is the long (multi-second)
timescale on which rapid accretion occurs.  The inclusion of MHD and
neutrino transport will make modeling the hyperaccretion phase even more
challenging.

\subsection{Massive star collapse in the early universe}

Black hole formation from massive stars in the early universe is
interesting primarily for explaining the ``seeds'' from which
supermassive black holes grew, but also as sources of electromagnetic
and gravitational wave signals.

1D simulations of Population III stars indicate that stars with mass
less than around $260 M_{\odot}$ end their lives in  pair-instability
supernovae, while more massive stars collapse directly to black
holes~\cite{Fryer:2000my,Heger:2002by}.  Nakazato~{\it et
  al.}~\cite{Nakazato:2005ek} study the effects of neutrino emission on
Pop III star collapse using a relativistic Boltzmann transport code.
Effects of rotation are indirectly addressed by the 2D Newtonian
simulations of Ohkubo~{\it et al.}~\cite{Ohkubo:2005pu}, who, in a
manner reminiscent of early collapsar simulations, insert a point mass
by hand into the collapsing star.  Rather than adding rotation and
viscosity, the effects of a disk are modeled on larger scales by
injecting a jet through the inner boundary.  The jet drives an
explosion, and nucleosynthesis outputs are calculated.

Supermassive stars began to be simulated in the 1970s.  At the time,
interest was primarily driven by the prospect of explaining active
galactic nuclei in terms of these objects, which do radiate at their
Eddington luminosity.  These stars are radiation-pressure dominated
and presumed isentropic, so they are nearly $n=3$ polytropes.  (Gas
pressure makes $n$ slightly less than 3, but this deviation decreases
with increasing mass.)  Although they are not compact, the instability
of $>10^5 M_{\odot}$ stars is triggered by general relativity.  These
earliest simulations were 1D, assuming spherical symmetry.
Appenzeller and Fricke~\cite{1972A&A....18...10A,1972A&A....21..285A},
using post-Newtonian gravity but including nuclear reactions,
found prompt collapse to a black hole if $M>10^6M_{\odot}$; for lower
masses nuclear burning of hydrogen explodes the star.  Later
simulations found that nonzero metallicity (albeit high given the
context: $Z\sim 10^{-2}$) catalyzes CNO burning and triggers
explosion~\cite{1986ApJ...307..675F}.  Spherically symmetric full
numerical relativity
simulations of high-mass supermassive stars found prompt collapse to a
black hole~\cite{1979ApJ...234L.177S,Linke:2001mq}.

In fact, this is another system for which we expect rotation to be
extremely important.  As a supermassive star cools and shrinks, it
probably spins up to the mass-shedding limit, thereafter following
a mass-shedding sequence as it cools till it hits a radial instability. 
Equilibrium
sequences of rapidly rotating supermassive stars in general relativity
have been produced by Baumgarte and Shapiro~\cite{Baumgarte:1999nh}
for $n=3$ and by Shibata~{\it et al.}~\cite{Shibata:2016vxy} for
$2.94\le n\le 3$.
Saijo~{\it et al.}~\cite{Saijo:2002qt} evolved an $n=3$ supermassive
star from it's critical point using 3D post-Newtonian physics, finding
that the collapsing star remains axisymmetric (i.e. nonaxisymmetric
modes do not have time to grow even though $T/|W|$ passes the critical
value for bar formation) and that nearly all the mass falls into the
black hole despite its rotation.  Around the same time, 2D numerical
relativity
simulations were carried out by Shibata and Shapiro~\cite{Shibata:2002br}, tracking
the collapse until about
60\% of the mass was inside the apparent horizon.  Because they could
not evolve long past black hole formation, the final disk mass
remained uncertain, but the authors estimated from the angular
momentum distribution that it should be around 10\% of the star's
mass.  

The Illinois group returned to this problem in 2007 with black hole
excision to determine the post-collapse state~\cite{Liu:2007cf}.
Their simulations vindicated the earlier angular-momentum based
predictions of a massive disk (several percent of the total original
rest mass) and a black hole with spin about 70\% of the Kerr limit.
This work also included magnetic fields, which do not affect the
collapse but do affect the disk evolution.  Insertion of a dynamically
unimportant dipole field into the pre-collapse star leads to jet
formation after collapse, with enough Poynting luminosity to
potentially power an ultra-long GRB detectable at high
redshifts~\cite{Sun:2017voo}.  Montero~{\it et
  al.}~\cite{Montero:2011ps} carried out simulations with detailed
microphysics and hydrogen and helium burning, finding that a
metallicity of $10^{-3}$ is needed to cause thermonuclear explosion
rather than black hole formation for mass-shedding supermassive stars
of mass $\sim 5\times 10^5 M_{\odot}$.  Finally, Shibata and collaborators
have included the deviation of $\Gamma$ from $4/3$~\cite{Shibata:2016vzw}
and nuclear reactions~\cite{Uchida:2017qwn}, finding collapse outcomes
similar to the $\Gamma=4/3$ studies~\cite{Liu:2007cf}.

Initial data sets for the above simulations assume the star is able to
maintain uniform rotation and entropy.  Collapse of differentially
rotating supermassive stars has been simulated in 2D by Montero~{\it
  et al.}~\cite{Montero:2011ps} and in 3D by Saijo and
Hawke~\cite{Saijo:2009ub}.  The latter study monitors quasi-periodic
gravitational waves coming from the post-collapse system even after
the hole's quasinormal ringing damps.  When a nearly extremal black
hole is formed, the gravitational wave grows to fairly high amplitude
after collapse for reasons which remain mysterious.  Zink~{\it et
  al.}~\cite{Zink:2005rr} studied the collapse of differentially
rotating toroidal supermassive stars, finding that in this case the
star is subject to strong nonaxisymmetric instabilities.  These lead
to the fragmentation of the star into self-gravitating, collapsing
parts, in some cases leading to the formation of a supermassive \bbh
system~\cite{Reisswig:2013sqa}.

The above all assume that supermassive stars are able to thermally
relax to isentropy, which should be true if they are convective. 
For an alternative scenario, leading to a stellar mass black hole
surrounded by a much more massive envelope, see
Begelman~\cite{2010MNRAS.402..673B,Begelman:2007je}.

\section{Non-vacuum compact binaries}
\label{sec:nsbinaries}
Non-vacuum compact object binaries (that is, binaries made of two
compact objects, at least one of which is not a black hole) inspiral
due to gravitational radiation just like \bbhs.  Tidal
deformation of the star(s) constitute an additional time-varying
quadrupole, subtly affecting the inspiral and associated gravitational
waveform.  For most of the inspiral, this small tidal effect can be
adequately modeled using post-Newtonian theory, and to good approximation
the size and structure of the star(s) only affect the waveform via
their dimensionless tidal deformability parameters
$\Lambda =(2/3)\,k_2\,(c^2\,R/G\,M)^5$,
where $k_2$ is the apsidal constant while $R$ and $M$ are the star's
radius and mass, respectively.  $\Lambda$ can be thought of as a
measure of a star's response to an external tidal field.  Thus, one
may hope to use gravitational waveforms from compact neutron star
binaries to constrain the $\Lambda$, and hence the EoS, of neutron
stars~\cite{Flanagan:2007ix}.  An important application of numerical
relativity, not
discussed in this review, is to test--and if necessary improve--these
models of tidal effects on waveforms during inspiral.  Interested
readers are referred to a sample of the papers on the
topic~\cite{Baiotti:2010xh,Bernuzzi:2014owa,Hinderer:2016eia,
  Hotokezaka:2016bzh,Dietrich:2017aum}.

Material effects become dramatic at the end of inspiral,
which will involve a collision or, more often, a tidal disruption. 
The latter happens when the tidal force on a star from its companion
exceeds the star's own self gravity.  This can be illustrated by
a simple Newtonian order-of-magnitude calculation.  Suppose the
disrupting star has mass $M$, radius $R$ and is at a separation
$d$ from its companion of mass $m$.  (In all cases we consider,
this companion will be more massive and more compact, often a
black hole.)  Then the self-gravitation and tidal accelerations
are $~M/R^2$ and $mR/d^3$, respectively, and they match when
\begin{equation}
\label{disruption}
d/m \sim (R/M) (m/M)^{-2/3}\, .
\end{equation}
Tidal disruption is likely marked by a sharp decrease in the
gravitational wave amplitude as the binary loses its
quadrupolar shape.  The subsequent fate of the disrupted
star's mass must be determined by simulations.  Most attention
has been given to the two observationally important possibilities
of gas forming an accretion disk around the remaining binary
object and gas being ejected from the system.

\subsection{White dwarf--compact object binaries}

{\it Neutron star-white dwarf} mergers and {\it stellar mass black
  hole--white dwarf} mergers are not easily amenable to numerical
  relativity because of
the disparity of length scales between the two objects.  Also, white
dwarfs are not in nuclear statistical equilibrium.  Usually, isotope abundances in a white dwarf
can be considered fixed, but a merger event may trigger nuclear reactions,
which would then provide a new energy reservoir and must be explicitly
tracked.  Because the white dwarf
disruption happens on scales much larger than the neutron star or
low-mass black hole, one might ask if a Newtonian white dwarf plus
point mass treatment is sufficient.  Such calculations have been done
in SPH~\cite{Fryer:1998qs}, indicating that the disrupted white dwarf 
shears into an accretion disk, a possible setup for a long-duration
gamma ray burst.

Paschalidis~{\it et al.}~\cite{Paschalidis:2009zz,Paschalidis:2011ez}
have attempted to use numerical relativity to study white dwarf-neutron star mergers. 
To make simulations feasible, the white dwarf is replaced by a
"pseudo-white dwarf", only ten times bigger than the neutron star rather than 500. 
The merger outcome is a Thorne-Zytkow-like object which will cool to
a hypermassive star and eventually collapse to a black hole.  The
authors acknowledge that simulations with more complete microphysics
are still needed.  1D disk calculations by Metzger and
collaborators~\cite{2012MNRAS.419..827M,Margalit:2016joe} indicate
that, at least for systems with mass ratio not close to one, heating
from nuclear reactions may unbind most accreting matter, and they
judge it unlikely that enough mass accumulates on a neutron star
to trigger collapse in those cases.

For {\it Intermediate mass black hole-white dwarf} tidal disruptions,
the disparity of length scales is removed and tidal disruption happens
in the strong gravity regime.  In this case, we would have in mind
nearly parabolic encounters in dwarf galaxies or globular clusters,
rather than quasicircular inspiral and merger.  Rosswog~{\it et
  al.}~\cite{Rosswog:2008ie} carried out Newtonian SPH simulations,
including nuclear burning, of such events, looking especially at cases
where tidal compression triggers explosive nuclear burning.  Because
the code was Newtonian, the black hole had to be approximated by a
Paczynski-Wiita potential.  The effects of black hole spin could only
be studied with numerical relativity simulations, which were carried out by Haas~{\it et
  al.}~\cite{Haas:2012bk}, although without nuclear reaction effects.
Both sets of simulations predict a residual accretion disk and
accompanying soft X-ray flare lasting about a year.

\subsection{Black hole--neutron star binaries}

Because of their potential as gravitational wave sources and short
gamma ray burst progenitors, most non-vacuum numerical relativity work has focused on
\bhns and \bns mergers.  In addition, these mergers are of
interest as possible sources
of short-duration gamma ray bursts, r-process nucleosynthesis,
and kilonovae.  (For reviews of multimessenger astronomy,
see~\cite{Berger:2014,Metzger:2017wot}.)
In this section, we review \bhns
merger simulations, a key application of numerical
relativity dynamical
black hole-handling technology.  For a fuller treatment, see the
Living Review by Shibata and Taniguchi~\cite{Shibata2011}.

\subsubsection{Expectations before numerical relativity simulations}

Black hole-neutron star binaries were historically the last to be
simulated in numerical relativity, but simple arguments and Newtonian simulations gave
some idea what to expect.

If a neutron star disrupts inside the ISCO of its companion
black hole, no massive disk or ejecta is expected.  What's more, the
gravitational wave in these cases should be nearly indistinguishable
from that of a  \bbh system with the same masses.
Defining $d_{\rm ISCO}=\kappa M_{\rm BH}$ and using
Eq.~(\ref{disruption}), we conclude that tidal disruption is likely for
binaries with
\begin{equation}
  \frac{R_{\rm NS}}{M_{\rm NS}} >
  \kappa \left(\frac{M_{\rm BH}}{M_{\rm NS}}\right)^{2/3}
\end{equation}
That is, disruption is favored by low neutron-star compaction $C\equiv M_{\rm
  NS}/R_{\rm NS}$, low mass ratio $q\equiv M_{\rm BH}/M_{\rm NS}$, and
high dimensionless black-hole spin $\chi$ (to reduce $\kappa$).

What happens when the neutron star fills its Roche lobe and mass
transfer begins?  The question was first addressed in Newtonian
simulations by Lee and
Kluzniak~\cite{Kluzniak:1997cm,Lee:2000uz,Lee:2001ae}, with the black
hole
treated as a point mass and the neutron star treated as a polytrope.  These
simulations found that mass transfer is stable for stiff EoS but
unstable for soft EoS~\cite{Lee:2000uz,Lee:2001ae}, a difference that
carried over to rival nuclear theory-based EoS as studied by
Janka~{\it et al.}~\cite{Janka:1999qu} and Rosswog~{\it et
  al.}~\cite{Rosswog:2004zx}.  In the case of unstable mass transfer,
the neutron star is destroyed in a single mass-transfer event.  Stable mass
transfer, on the other hand, is episodic, yielding an unmistakably
different gravitational wave signal.   Replacing the Newtonian point
mass with a Paczynski-Wiita potential makes mass transfer less stable,
so that tidal disruption happens in one pass even for stiff realistic
EoS~\cite{Rosswog:2005mf,Ruffert:2010pw}.  Newtonian simulations also
found massive ejection of unbound matter during mergers.  SPH
simulations around Kerr black holes supported expectations that
prograde black hole spin is favorable to disk
formation~\cite{Rantsiou:2007ct}.

\subsubsection{Inspiral and merger in numerical relativity:  parameter space exploration
and gravitational waves}

The first general relativisitic simulations of black hole systems came soon
after the moving puncture revolution.  A head-on collision with a neutron
star falling into a black hole was successfully modeled (using excision)
by Loffler~{\it et al}~\cite{PhysRevD.74.104018}, and soon after Shibata
and Uryu carried out binary
merger simulations starting from roughly circular orbit (using moving
punctures)~\cite{Shibata:2006bs}.  Simulations by the UIUC, SXS, and
LSU/BYU/LIU groups quickly
followed~\cite{Etienne:2007jg,Duez:2008rb,Chawla:2010sw}.  The former
two of groups used the NOKBSSN formalism with moving punctures; the
latter two used the generalized harmonic formulation with explicit
excision.  These early simulations used simple, polytropic EoS, and in
some cases improper treatment of low-density material led to
underestimates in the disk and ejecta masses.  They all found complete
neutron-star disruption in a single mass transfer event.  When tidal disruption
occurs outside the ISCO, the neutron star deforms into a tidal stream,
with inner material streaming toward the black hole and outer material
streaming outward.  Of the matter falling toward the black hole, most
will fall into the black hole.  Soon after flow into the black hole
commences, material with sufficient angular momentum wraps around the
black hole, causing the tidal stream to crash into itself.  The
resulting shock heats the gas, which begins setting into an accretion
disk very close (tens of km) to the black hole.  Of the nuclear matter
expanding outward, some is bound and eventually falls back onto the
disk, while the rest (the ``dynamical ejecta'') is unbound and escapes
permanently.

Using all the
numerical relativity results available at the time, Foucart devised an analytic fit to
the post-merger disk mass (defined as the rest mass of bound material
outside the black hole 10\,ms after merger; recall that matter is
continuously falling onto the disk and into the horizon) as a function
of $C$, $q$, and $\chi$ which confirms the expectation
that, all else being equal, lower $C$, lower $q$, or higher $\chi$
increases disk mass~\cite{Foucart:2012nc}.  Using their own set of
simulations, Kawaguchi~{\it et al.}~\cite{Kawaguchi:2016ana} devised a
similar analytic fitting formula for the mass and asymptotic speed of
unbound ejecta.

\begin{figure}
 \includegraphics[width=80mm,clip]{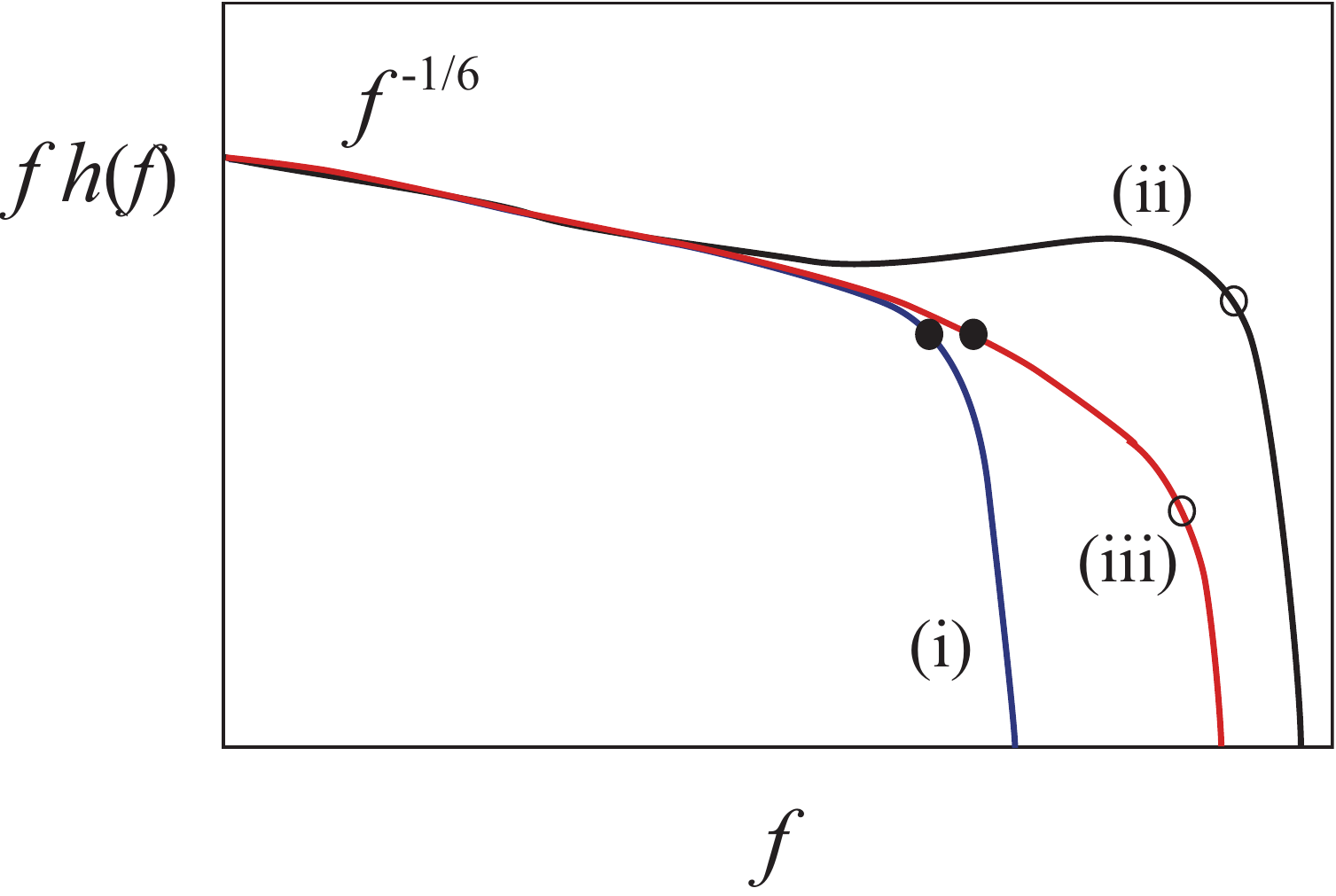} \caption{A schematic
 figure of three types of gravitational-wave spectra from black hole-neutron
 star mergers. Spectrum (i) is
 for the case in which tidal disruption occurs far outside the ISCO, and
 spectrum (ii) is for the case in which tidal disruption does not
 occur. Spectrum (iii) is for the case in which tidal disruption occurs
 and the quasinormal mode (QNM) of the black hole is also excited. 
 The filled and open circles denote $f_{\rm
 tidal}$, the frequency at neutron star tidal disruption, and $f_{\rm QNM}$,
 respectively.  Reproduced with permission
 from~\cite{Kyutoku:2011vz}. } \label{fig:bhnswaves}
\end{figure}

Using numerical relativity simulations, Shibata~{\it et al.}~\cite{Shibata:2009cn} found
gravitational waves from \bhns mergers to always fall into one of three categories,
illustrated in Figure~\ref{fig:bhnswaves}. 
When the neutron star falls into the black hole before being disrupted, the wave is
similar to \bbh waves.  When the neutron star disrupts well outside the ISCO,
the waveform cuts off at this point during the inspiral, and no
merger or ringdown wave is seen.  Disruption close to the ISCO gives a
case with intermediate features:  inspiral and merger waves but
reduced ringdown wave.  Information about the neutron star is
contained in the gravitational wave cutoff frequency.  Shibata~{\it et
  al.} emphasize that this cutoff frequency is not identical to the
gravitational wave frequency at tidal disruption, but rather is
somewhat higher.  Presumably, this is because the star persists for a
time as a clump of matter as it inspirals past the tidal disruption
radius.

BH-polytrope simulations also found a strong dependence of the
post-merger disk mass on the black-hole spin.  Etienne~{\it et
  al.}~\cite{Etienne:2008re} found that much more massive disks could
be formed for neutron stars disrupted by black holes with prograde
spin, which is perhaps to be expected, since such black holes have
smaller ISCOs.  An extreme case--mass ratio of 3 and prograde black
hole spin at 97\% of the Kerr limit--was simulated by Lovelace~{\it et
  al.}~\cite{Lovelace:2013vma}; not even half of the rest mass is
promptly accreted in this case.  Retrograde spin, on the other hand,
makes disruption outside the ISCO less likely and disk masses lower.
The black hole spin orientation has been varied by Foucart~{\it et
  al.}~\cite{Foucart:2010eq,Foucart:2012vn} and by Kawaguchi~{\it et
  al.}~\cite{Kawaguchi:2015bwa}, with the general findings that large
spin misalignments remove the increase in disk mass seen in prograde
spins and lead to disks initially misaligned with the black hole spin.
With high enough prograde black-hole spin, Foucart~{\it et
  al.}~\cite{Foucart:2012vn}  were able to observe tidal disruption
even in systems with mass ratios in the 5--7
range~\cite{Foucart:2012vn} where astrophysical \bhns systems are
thought most likely to lie~\cite{belczynski:08}.

Most contemporary \bhns simulations use more realistic EoS.  Because
the matter only heats up as the gravitational wave signal turns off, waveform studies
have sensibly concentrated on piecewise-polytropic parameter studies
of EoS effects~\cite{Kyutoku:2010zd,Kyutoku:2011vz,Kawaguchi:2015bwa}.
In a truly impressive effort, Kyutoku and collaborators carried out
134 merger simulations, varying both EoS and binary
parameters~\cite{Lackey:2013axa,Kyutoku:2010zd,Kyutoku:2011vz}.  The
equation of state was modeled as a two-piece piecewise polytrope,
which, since the low-density EoS is known, has two free parameters.
They choose to systematically vary the high-density $\Gamma$ and
$P_1$, the pressure at a fiducial density $\rho_{\rm fidu}=10^{14.7}$g
cm${}^{-3}$.  The reason for using $P_1$ rather than the transition
density is because $P_1$ is found to closely correlate with the
neutron star radius and tidal deformability, making it a good
candidate for an equation of state parameter that can be measured by
gravitational wave signatures such as the cutoff frequency.  These
simulations have been used to calibrate analytic \bhns waveform
models, valid in the range $2<q<5$, covering the full inspiral and
merger~\cite{Lackey:2013axa,Pannarale:2013uoa,Pannarale:2015jka,Kumar:2016zlj}.
Fisher matrix and Bayesian analysis of the analytic model shows that
LIGO detections can hope to significantly constrain the tidal
deformability and $P_1$, especially given dozens of realistic
detections~\cite{Lackey:2013axa,Kumar:2016zlj}.

Using piecewise polytrope EoS, East~{\it et al.}~\cite{East:2011xa}
simulated eccentric \bhns encounters.  Here could finally be seen
cases of episodic mass transfer in general relativity, as well as instances of the
``zoom-whirl'' phenomenon observed in \bbh simulations.  Such events
may well occur at interesting rates in dense stellar environments such
as globular clusters~\cite{2010ApJ...720..953L}.  If so, the richer
dynamical possibilities of eccentric merger deserve more attention.

\subsubsection{Post-merger in numerical relativity:  neutrinos, ejecta, and MHD}

Post-merger evolution requires finite-temperature EoS, which have been
employed in simulations by the SXS
collaboration~\cite{Duez:2009yy,Deaton:2013sla,Foucart:2014nda,Foucart:2015vpa}
and by Kyutoku~{\it al.}~\cite{Kyutoku:2017voj}.
These simulations fail to find episodic mass transfer even for stiff
EoS.  It is found that, when $\beta$ equilibrium violation is allowed
during tidal disruption (on these timescales, the lepton number
advects), tidal streams are narrower, and ejecta velocities lower than
$\beta$-equilibrium EoS would
predict~\cite{Foucart:2014nda,Foucart:2016vxd}.  Lepton number
equilibrium in the post-merger disk is re-established about 10\,ms
after merger under the action of intense neutrino emission
($L_{\nu}\sim 10^{53}$erg~s${}^{-1}$, preferentially in electron
antineutrinos), during which time the disk's average $Y_e$ rises to
about 0.1.  (For stiff EoS, $Y_e$ might be as high as
0.2~\cite{Kyutoku:2017voj}.)  As the disk cools, the equilibrium
$Y_e$ decreases and a re-neutronization is seen in the disk.

Whether the neutrino emission
is sufficient to power a gamma-ray burst remains uncertain. 
Just~{\it et al.}~\cite{Just:2015dba} simulate black hole tori using an
energy-dependent M1 neutrino transport scheme, and find that,
at least for favorable cases, neutrino-antineutrino annihilation
can power a relativistic outflow sufficient for a low-energy
short duration GRB.  (\bhns mergers are more promising in this regard
than \bns mergers, because the latter have more baryon loading from
dynamical ejecta on the poles.)

After some early false negatives, tidal ejection of unbound matter was
robustly identified in numerical relativity black hole-neutron star
mergers~\cite{Foucart:2012vn,Kyutoku:2013wxa,Kyutoku:2015gda}.  In
relativistic codes, unbound matter is usually identified as that with
specific orbital energy $e= -u_t-1 > 0$.  The mass of dynamical ejecta
for cases with tidal disruption is often large $\sim 10^{-2}$--$10^{-1}
M_{\odot}$ and highly asymmetric--concentrated on one side in the
orbital plane--with sufficient momentum to impart a kick of $10^2$km
s${}^{-1}$ on the remnant black hole in some
cases~\cite{Kyutoku:2013wxa,Foucart:2012vn}.   Additional matter is
ejected into weakly bound orbit and will fall back onto the central
remnant later.  This fallback material was studied using Newtonian
SPH by Rosswog~\cite{Rosswog:2006rh} and then in numerical relativity
by Chawla~{\it et al.}~\cite{Chawla:2010sw}.  From the
distribution of fallback times, these studies predict a late-time fallback
accretion rate following a $t^{-5/3}$ power law.

The possible importance of black hole-neutron star dynamical ejecta
for r-process nucleosynthesis was pointed out by Lattimer and Schramm
in 1976~\cite{1976ApJ...210..549L}.  Newtonian simulation confirm that
\bhns mergers produce large ejecta masses of neutron-rich material that
should undergo r-process nucleosynthesis and produce an optical/near-IR
transient~\cite{Rosswog:2005su}.  Indeed, so efficient are these
mergers in producing r-process elements that Bauswein~{\it et
  al.}~\cite{Bauswein:2014vfa}, use their own ejecta predictions from
SPH conformally flat general relativistic simulations and the galactic abundance of
r-process material to constrain the rate of black hole-neutron star
mergers.

The SXS simulations confirm expectations that ejecta is
very neutron rich, so as it decompresses it is expected to undergo
r-process nucleosynthesis and produce the second and third r-process
peaks.  This is indeed what Roberts~{\it et al.}~\cite{Roberts:2016igt}
find tracking nuclear reactions in the ejecta, although a weak first
peak can be seeded by neutrino irradiation by the central black
hole-disk system.  Higher $Y_e$ outflow may be provided by winds from
the accretion disk.  Recently, Fernandez~{\it et
  al.}~\cite{Fernandez:2016sbf} have produced models including both
effects.  Using outgoing ejecta and disk profiles from the SXS merger
simulations as initial data, the disk was evolved to late times using
2D Newtonian hydrodynamics with an alpha viscosity to model angular
momentum transport.  As the disk evolves, neutrino cooling decreases
to the point of insignificance, and by $\sim 200$\,ms the disk has
reached an advective state (viscous heating balanced by advection of
hot material inward rather than by radiative cooling) with strong
convection and mass outflow.  In most cases, the dynamical ejecta
dominates, but if the disk outflow and dynamical ejecta should happen
to have comparable masses, a solar-like distribution of r-process
elements would follow.

Radioactive decay of r-process ejection might power a detectable
signal, most likely in the near infrared, called a kilonova or
macronova~\cite{Li:1998bw}.  Tanaka~{\it et al.}~\cite{Tanaka:2013ixa}
use ejecta from numerical relativity black hole-neutron star merger simulations as input
to a (photon) radiation transfer code to predict light curves.  The
simulations did not include nucleosynthesis, but assumed the solar
r-process pattern.  They find that black hole-neutron star kilonovae
can often appear as bright or brighter than neutron star-neutron star
kilonovae, because the former can produce more ejecta, and that the
former will tend to be bluer than the latter.  The same group applied
these models to the purported kilonova associated with GRB 130603B,
showing the observed near-infrared excess is consistent with either a
soft EoS \bns merger or a stiff EoS black hole-neutron
star merger~\cite{Hotokezaka:2013kza}.  Kawaguchi~{\it et
  al.}~\cite{Kawaguchi:2016ana} returned to black hole-neutron star
kilonovae with analytic models for the heating and radiative diffusion
but a large suite of merger simulations.  Lanthanide-free disk wind
might create a bluer signal~\cite{Kasen:2014toa}, but in the models
studied by Fernandez~{\it et al.}, this was all obscured by the
dynamical ejecta~\cite{Fernandez:2016sbf}.  

The most dramatic, and least-understood, post-merger processes are
magnetically driven.  Early simulations with confined poloidal fields
found that the field in the post-merger disk quickly wound into a
toroidally dominated configuration, with no observable
jets~\cite{Chawla:2010sw,Etienne:2011ea} even when the MRI could be
resolved~\cite{Etienne:2012tmf}.  Later simulations by
Paschalidis~{\it et al.}~\cite{Paschalidis:2014qra} found that jets can
more easily emerge from a field initially extending outside the
neutron star.
Such magnetospheric fields might also trigger observable signals that
preceed a GRB from the
merger~\cite{McWilliams:2011zi,Paschalidis:2013jsa}.  In fact, even
confined seed fields may be more promising than they originally
seemed.  Extremely high-resolution studies ($\Delta x\approx 100$m) by
Kiuchi~{\it et al.}~\cite{Kiuchi:2015qua} find winds driven from inner
disk heating, whose strength increases with resolution, with
convergence not yet achieved, can pin magnetic flux to the black hole with
associated Blandford-Znajek jets.  Post-merger disk evolution turns out to be a
difficult, multi-scale problem, and crucial properties like the rates
of wind outflow, magnetic energy outflow, and neutrino annihilation
energy deposit remain poorly constrained.  For the time being, the
long-term evolution of the disk is being investigated using Newtonian
alpha-viscosity models
(e.g.~\cite{Fernandez:2013tya,Fernandez:2014bra}).

\subsection{Neutron star-neutron star mergers}

\subsubsection{Pre-breakthrough simulations}

The story of \bns merger simulations differs from that of other major
numerical relativity problems in that many \bnss evolve well past merger
without encountering black hole formation, so full numerical
relativity merger
simulations began earlier for \bns systems, and the ability to evolve
spacetimes with black holes had a less dramatic effect.  Neutron
star--neutron star binary
simulations have recently received a thorough review by Baiotti and
Rezzolla~\cite{Baiotti:2016qnr}.  They are also the subject of a
Living Review by Faber and Rasio~\cite{Faber2012}.  Techniques for
constructing initial data are described in the review by
Tichy~\cite{Tichy:2016vmv}.  Readers can find
in these sources a more detailed presentation of the large subject
of compact \bnss.

Simulations of \bns mergers were first undertaken in Newtonian
physics with mostly simple polytropic equations of state.  Nakamura
and Oohara in a series of papers performed the first \bns merger
simulations using finite
differencing~\cite{1989PThPh..82..535O,1990PThPh..83..906O,1991PThPh..86...73N,1992PThPh..88.1079S}.
SPH simulations were performed by Rasio and
Shapiro~\cite{1992ApJ...401..226R,1994ApJ...432..242R}.  Further
simulations of both types
followed~\cite{1994PhRvD..50.6247Z,1996PhRvD..54.7261Z,1997ApJ...490..311N}.
Gravitational waves had to be studied in the quadrupole approximation,
and the possibility of black hole formation could not be addressed.
However, these Newtonian simulations showed some features that would
be confirmed by numerical relativity simulations:  the stars merge into a massive
remnant rotating rapidly and differentially, and for stiff EoS the
remnant is subject to bar mode deformations, leading to a sustained
post-merger gravitational wave signal.

Post-Newtonian simulations were the next logical step.  The first
such simulation was carried out using smoothed particle hydrodynamics
by Ayal~{\it et al.}~\cite{Ayal:1999wn}. 
Unfortunately the 1 post-Newtonian-order terms are not always small compared to Newtonian terms,
indicating that truncating at this level is not a valid approximation,
and post-Newtonian studies sometimes resorted to artificially reducing
the post-Newtonian terms
(e.g.~\cite{Faber:2002cg}).  The next advance was to the conformally
flat approximation to general relativity.  Here surprises seemed to
arise when Wilson, Mathews, and
Marronetti~\cite{1995PhRvL..75.4161W,1996PhRvD..54.1317W} reported the
neutron stars in their simulations collapsing individually to black
holes before merging.  However, they used an EoS with fairly low
neutron star maximum mass and were found to have an error in one of
their equations~\cite{Flanagan:1998zt}.  Pre-merger collapse is no
longer considered likely, but the conformal flatness approximation has
turned out to be a useful and reliable tool for \bns modeling
(e.g.~\cite{Oechslin:2001km,Faber:2003sb,Oechslin:2006uk}).

Parallel to efforts toward incorporating relativity were efforts to
include realistic microphysics in Newtonian simulations.  Grid-based
simulations by Ruffert~{\it et
  al.}~\cite{1996A&A...311..532R,1997A&A...319..122R,2001A&A...380..544R}
and SPH simulations by Rosswog~{\it et
  al.}~\cite{Rosswog:2001fh,2003MNRAS.342..673R,2003MNRAS.345.1077R}
began the use of finite-temperature equations of state and inclusion
of neutrino effects (in a leakage approximation) for \bnss.
These simulations highlighted the potential of \bns mergers as GRB
central engines.  The study of \bns mergers with finite-temperature
EoS was continued in conformally flat gravity by Oechslin~{\it et
  al.}~\cite{Oechslin:2006uk}.  Neutrino absorption effects were
studied in Newtonian physics using flux-limited diffusion by
Dessart~{\it et al.}~\cite{Dessart:2008zd} and through ray tracing and
phenomenological extensions to leakage by Perego~{\it et
  al.}~\cite{Perego:2014fma}.  Using these very different methods, both
groups find strong neutrino-driven winds ejected from the merged
remnant, winds that could play important roles in generating r-process
elements and kilonovae and in baryon loading the environment of a
potential GRB.

Newtonian simulations were also able to investigate the ejecta and
its potential for r-process nucleosynthesis before the first numerical
relativity
simulations~\cite{Rosswog:1998hy,1999ApJ...525L.121F,Rosswog:2000nj},
and later numerical relativity simulations have found results
reasonably close to the earlier Newtonian studies.

Neutron star--neutron star mergers in full numerical relativity were first carried out by Shibata and Uryu at the turn
of the century~\cite{Shibata:1999hn,Shibata:1999wm,Shibata:2003ga}.
These initial simulations modeled the neutron stars as $\Gamma=2$
polytropes.  Their most significant discovery was that the remnant
does collapse to a black hole, but only if its mass exceeds a certain
threshold.  Less massive systems form dynamically stable
differentially rotating neutron star remnants (which in some cases are
hypermassive).  Other groups added AMR~\cite{Anderson:2007kz} and
high-resolution shock-capturing techniques~\cite{Baiotti:2008ra}.
Binary polytrope simulations with other sophisticated numerical
relativity codes
followed (BAM~\cite{Bernuzzi:2011aq}, WhiskyTHC~\cite{Radice:2013hxh},
SpEC~\cite{Haas:2016cop}).  All found similar results.  As the next
step in microphysical realism, Shibata and collaborators tried cold
nuclear-theory based EoS (augmented with a Gamma-law thermal component
to capture shock heating)~\cite{Shibata:2005ss}, confirming the
possibility of nonaxisymmetric post-merger structure.  Since then, the
piecewise-polytrope parameterization has often been used to
systematically vary the cold EoS, or even just as a cheap way to
approximate a given cold
EoS~\cite{Hotokezaka:2011dh,Hotokezaka:2012ze,Read:2013zra,Takami:2014zpa}.

With the introduction of moving punctures,
Baiotti~{\it et al.}~\cite{Baiotti:2008ra}
Kiuchi~{\it et al.}~\cite{Kiuchi:2009jt} revisited \bns mergers with
various EoS, following high-mass cases beyond black hole
formation to measure the post-collapse black hole and disk properties. 
For the APR EoS, equal-mass binaries with prompt black-hole formation leave very
little torus mass ($\sim 10^{-4} M_{\odot}$), but about
$10^{-2}M_{\odot}$ disks remain for binaries with mass ratio around 0.8.
Rezzolla~{\it et al.}~\cite{Rezzolla:2010fd,Giacomazzo:2012zt} have produced
analytic fitting
formulae relating post-collapse torus mass to the pre-merger binary
parameters.

\subsubsection{Post-merger evolution:  neutrinos, ejecta, MHD}

Now the main challenge was not metric evolution but microphysics. 
After two neutron stars merge, thermal, neutrino, and magnetic effects
become important.  Numerical relativity simulations with finite-temperature EoS and
neutrino leakage were carried out by Sekiguchi~{\it et
  al.}~\cite{Sekiguchi:2011zd,Sekiguchi:2011mc} for the stiff Shen EoS
and a hyperonic EoS.  A survey of \bns mergers with a large number of
finite-temperature EoS has since been performed (in conformally flat
general relativity) by Bauswein~{\it et al.}~\cite{Bauswein:2013jpa}.  Kastaun and
Galeazzi~\cite{Kastaun:2014fna} carried out a further set of mergers
with finite-temperature EoS (LS220 and SHT) and studied the structure
of the hypermassive remnants in detail.  Contrary to the widespread
presumption that hypermassive stars ``cheat'' the mass-shedding limit
by having rapidly rotating cores with strong centrifugal support and
more slowly rotating envelopes, the authors find that their
hypermassive remnants have slowly rotating (pressure-supported) cores
and extended, quasi-Keplerian envelopes.  Simulations of low-mass
\bns mergers (i.e. with non-hypermassive remnants) by Kastaun~{\it et
  al.}~\cite{Kastaun:2016yaf} and Foucart~{\it et
  al.}~\cite{Foucart:2015gaa} show similar features.  The composition
of low-density regions and outflows is strongly affected by neutrino
absorption.  To capture these effects, numerical relativity simulations with
energy-integrated M1 neutrino transport have been carried out by
Wanajo~{\it et al.}~\cite{Wanajo:2014wha}, Foucart~{\it et
  al.}~\cite{Foucart:2015gaa,Foucart:2016rxm}, and Sekiguchi~{\it et
  al.}~\cite{Sekiguchi:2016bjd}.

A number of studies have focused
particularly on the dynamical
ejecta~\cite{Hotokezaka:2012ze,Radice:2016dwd,Sekiguchi:2016bjd,Foucart:2016rxm},
which, like black hole-neutron star ejecta, is potentially important
for r-process nucleosynthesis and kilonovae, although there are two
important differences.  Both come from the fact that dynamical ejecta
comes not only from the tidal tail, but also from the collision
interface.  Much of this material is polar, leading to the first major
difference from black hole-neutron star ejecta: the distribution of
ejecta is much more
isotropic~\cite{Hotokezaka:2013iia,2013ApJ...773...78B}.  It has even
been suggested that the ejecta forms a cocoon around the central
object that can collimate the GRB outflow~\cite{Nagakura:2014hza}.
Additional ejecta is released in the following milliseconds as the
hot remnant neutron star settles.  These extra sources of early-time
ejecta lead to a second difference from the \bhns case, namely that
ejecta can be hotter and more neutrino-processed.  Due to
$n+e^+\Rightarrow \overline{\nu}_e+p$ and $\nu_e+n\Rightarrow p+e^-$
reactions, a portion of ejecta with $Y_e\sim 0.3-0.4$ can be created. 
The wide range of ejecta $Y_e$ can produce all three r-process
peaks without the need for a subsequent disk
wind~\cite{Wanajo:2014wha}.  Numerical relativity has also been used to study systems
with pre-merger neutron
spin~\cite{Bernuzzi:2013rza,Kastaun:2013mv,Kastaun:2014fna,East:2015vix,Dietrich:2016lyp}
and orbital
eccentricity~\cite{East:2012ww,East:2015vix,Paschalidis:2015mla,Radice:2016dwd}.

MHD simulations of \bns mergers have struggled with the difficulty of
resolving the MRI in high-density regions and, even more challenging,
the growth of the magnetic field in Kelvin-Helmholtz vortices.  Early
studies~\cite{Liu:2008xy,Anderson:2008zp,Giacomazzo:2009mp}
demonstrated the ability of numerical relativity MHD codes to follow
\bns mergers but
could not resolve these effects, although they could resolve the MRI
in a post-collapse torus~\cite{Rezzolla:2011da} and showed the
possibility of jet formation from the black hole-torus system (leading
perhaps to a short duration gamma ray burst).  A series of
unprecedentedly high-resolution (as low as $\delta x = 17.5$m) by
Kiuchi~{\it et al.}~\cite{Kiuchi:2014hja,Kiuchi:2015sga} succeeded in
resolving these effects well enough to demonstrate amplification of
the average field by a factor of $10^3$, but even this is taken as
a lower limit. 

Although this small-scale amplification cannot be adequately resolved,
numerical relativity MHD studies continue to study large-scale processes
during merger.  For example, a few studies
investigate the difference in merger scenarios between ideal and
resistive MHD~\cite{Palenzuela:2013hu,Ponce:2014sza,Dionysopoulou:2015tda}.  Most recently, a series of papers beginning with
Endrizzi~{\it et
  al.}~\cite{Endrizzi:2016kkf,Kawamura:2016nmk,Ciolfi:2017uak} has
surveyed binary properties, EoS, and seed field effects on magnetized
\bnss.  Among their findings is a characteristic
large-scale field structure appearing in many cases.  For cases with
black hole formation, these simulations did not observe polar magnetic
jets, but Ruiz~{\it et al.}~\cite{Ruiz:2016rai} does see them for
stronger seed fields (allowing better MRI resolution) and longer
integration times.

\subsubsection{General relativistic effects:  prompt collapse threshold and post-merger gravitational waves}

Having outlined the development of \bns merger simulations, we turn
to their findings concerning distinctly relativistic effects.  The
threshold mass $M_{\rm th}$ for prompt (i.e. on a dynamical timescale)
collapse after merger to a black hole is found to be 30\% to 70\%
above $M_{\rm TOV max}$, depending on the EoS.  Bauswein~{\it et al.}
find that $M_{\rm th}=(2.43 - 3.38C_{\rm TOV max})M_{\rm TOV max}$ to
reasonable accuracy for all EoS studied, where $C_{\rm TOV max}$ is
the compaction of the TOV maximum mass
configuration~\cite{Bauswein:2013jpa}.  For \bns systems with mass
below $M_{\rm th}$, the massive (perhaps hypermassive) differentially
rotating neutron star remnant is dynamically stable. 

\begin{figure}
\centering
\includegraphics[width=0.95\columnwidth]{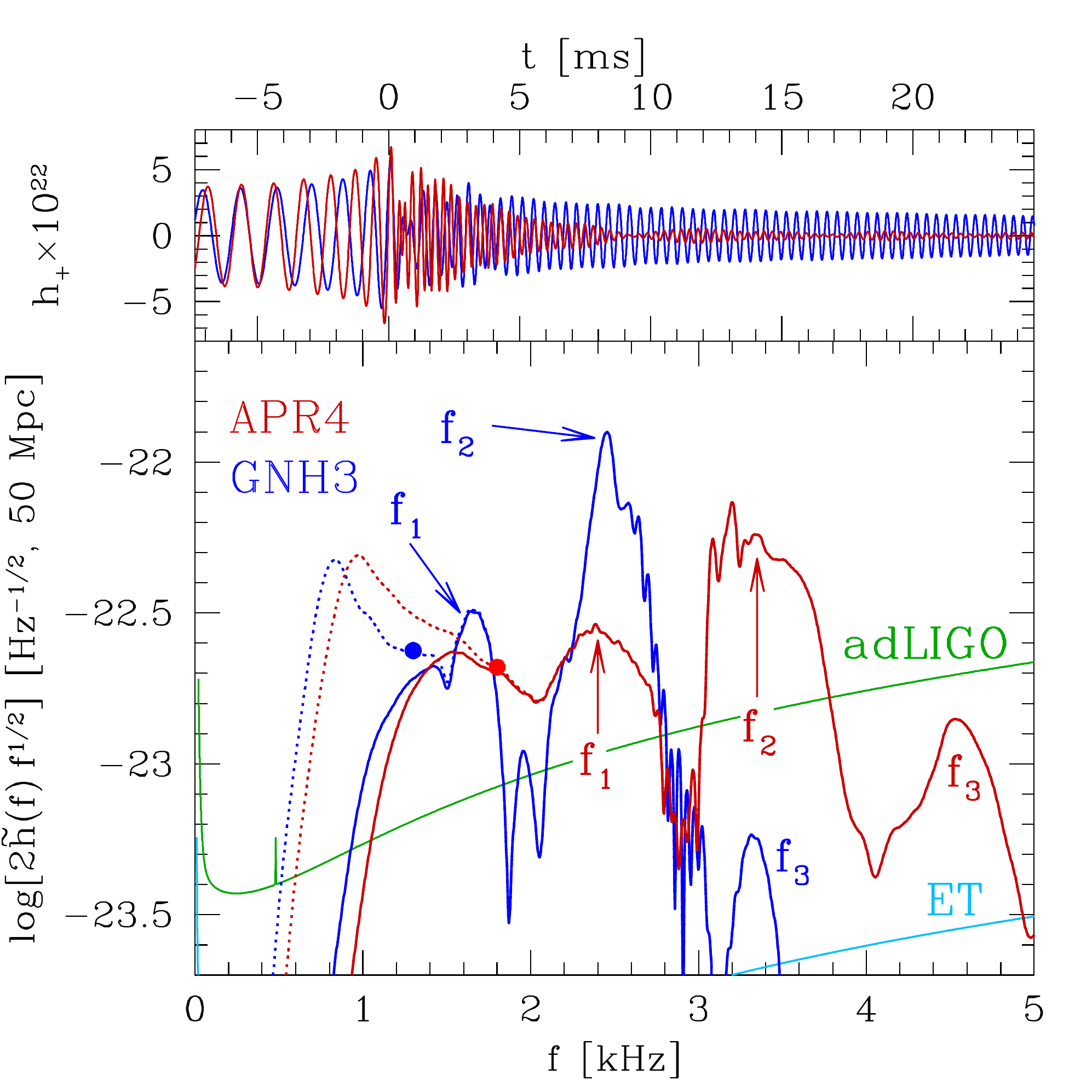}
\caption{\label{fig:nsnswaves} Gravitational waveforms from a \bns merger for two possible nuclear equations of state. 
Top panel: evolution of $h_+$ for
  representative binaries with the APR4 and GNH3 EoSs (dark-red and blue
  lines, respectively) for sources at a polar distance of 50
  Mpc. Bottom panel: spectral density $2 \tilde{h}(f) f^{1/2}$
  windowed after the merger for the two EoSs and sensitivity curves of
  Advanced LIGO (green line) and ET (light-blue line); the dotted lines
  show the power in the inspiral, while the circles mark the contact
  frequency.  Reproduced with permission from~\cite{Takami:2014zpa}.}
\end{figure}

The remnant is
formed with a strong quadrupolar distortion, emitting strong--and,
what's even better,
EoS-sensitive--gravitational waves~\cite{Bauswein:2011tp,Stergioulas:2011gd,Hotokezaka:2013iia,Takami:2014zpa}.
Recall that the remnant is differentially rotating, so the overall
rotating quadrupole in the matter profile is an $m=2$ density mode,
not solid body rotation.  In fact, the mode angular frequency is close
to, but slightly higher than, the maximum angular frequency in the
star, for reasons that remain
unclear~\cite{Kastaun:2014fna,Foucart:2015gaa,Endrizzi:2016kkf}.  This
lack of a corotation radius rules out the possibility of a corotation
shear instability in the $m=2$ mode.  The $\ell=2$, $m=1$ mode has
lower frequency and in fact does grow from the corotation
instability~\cite{East:2015vix,Radice:2016gym,Lehner:2016wjg}.
As shown in Fig.~\ref{fig:nsnswaves},  the post-merger gravitational
wave spectra show
a few distinct sharp peaks.  The strongest peak, at frequency often
called f2 in the literature, comes from the fundamental $m=2$ mode
described above, having a gravitational wave frequency between 2--3 kHz.  For a wide
range of EoS, f2 is an EoS-independent function of $R_{\rm max}$, the
radius of a nonrotating neutron star of the mass $M_{\rm TOV
  max}$~\cite{Bauswein:2011tp}, or to the radius at some other
fiducial mass~\cite{Bauswein:2012ya,Hotokezaka:2013iia}.  Simulations also
consistently find a weaker signal at a lower frequency f1, usually in
the range 1.2--2.5 kHz, which has been explained in terms of
oscillations in the distance between the two cores still distinct for
a short time after merger~\cite{Baiotti:2008ra}, as a spiral density
wave~\cite{Bauswein:2015yca}, or as a coupling between radial and
quadrupolar modes~\cite{Stergioulas:2011gd}.  For nonlinear
perturbations of this sort, these interpretations may not be mutually
exclusive.  For a wide range of EoS, f1 seems to obey a universal
relation to the average compaction (or alternatively, the tidal deformability
$\Lambda$) of the premerger neutron
stars~\cite{Takami:2014zpa,Takami:2014tva,Rezzolla:2016nxn}. 
Measuring these frequencies would
significantly constrain the neutron star EoS.  Unfortunately, this
would only be possible for close mergers ($<40$ Mpc) with Advanced
LIGO or with a next generation gravitational wave observatory.

One worry about the above studies is that they ignore magnetic
field-related stresses.  Simulations by Palenzuela~{\it et
al.}~\cite{Palenzuela:2015dqa} with hot EoS, neutrino cooling,
and MHD included find that the large-scale magnetic fields
(those resolved in global merger simulations) are too weak
to affect the post-merger waveform during the first 10\,ms,
 even when Kelvin-Helmholtz amplification is included in an
approximate way via subgrid modeling.  As we shall shortly
see, the effects of subgrid-scale MHD turbulence may be a
different story.

\subsubsection{Longer term evolution of remnants:  subgrid scale modeling}
\label{sec:subgrid}
The subsequent evolution of the remnant depends on secular processes
that drive the star from one equilibrium to another.  During the first
tens of milliseconds, hydrodynamic torques redistribute angular
momentum outward while gravitational radiation drains the star's total
angular momentum~\cite{Hotokezaka:2013iia}.  On longer timescales of
$\sim 10^1$--$10^2$\,ms, magnetic processes, namely magnetic winding and
turbulent motions triggered by the MRI, also redistribute angular
momentum outward~\cite{Duez:2005cj}.  If the core depends on
rotational support, any of these might trigger collapse to a black
hole-torus system.  However, simulations with finite-temperature
nuclear EoS tend to find thermally supported hypermassive remnants, so
collapse of hypermassive remnants may be delayed until the neutrino
cooling timescale, which would be of order
seconds~\cite{Sekiguchi:2011zd,Paschalidis:2012ff}.  Winds driven by
magnetic fields~\cite{Siegel:2014ita} or
neutrinos~\cite{Perego:2014fma} may carry off a small amount of mass
($10^{-3}$--$10^{-2} M_{\odot}$) and some angular momentum during that
time.  Remnants that are merely supramassive can survive beyond this
time and collapse much later from angular momentum loss due to pulsar
spindown.

Direct modeling of this evolution is, for the time being, out of
reach, due to the multi-scale nature of the problem.  We have seen
how small-scale growth of the MRI and Kelvin-Helmholtz instability
currently frustrate numerical convergence, to which we add a general
observation that in any high Reynolds-number, turbulent system one
must resolve a certain inertial range to accurately estimate mean
stresses.  Such difficulties are not distinctly relativistic; it is
a triumph of sorts that numerical relativity now stumbles against the same challenges
inherent in turbulence and dynamo modeling that would confront us
in Newtonian physics.  

Several numerical relativity groups have attempted to capture
unresolved transport processes using subgrid models, i.e. by
evolving the fluid at large scales while adding
contributions to $T_{\mu\nu}$ meant to represent averaged Reynolds and
Maxwell stresses from unresolved velocity and magnetic field
fluctuations.  
One simple choice is to model these transport processes as a
viscosity.  One then adds a viscous stress term $T^{\rm
  visc}{}_{\mu\nu}=-\eta\sigma_{\mu\nu}$, where $\sigma_{\mu\nu}$ is
the shear tensor associated with the 4-velocity $u^{\alpha}$, and
$\eta(\rho,T)$ sets the strength of the viscosity.  From
$T^{\alpha\beta}{}_{;\beta}=0$, one obtains the relativistic
Navier-Stokes equations.   This was done in 2004 by Duez~{\it et
  al.}~\cite{Duez:2004nf}, who tracked the secular evolution of a set
of hypermassive $\Gamma=2$ polytropes in 2D (axisymmetry) starting from
an initial differential rotation with the angular velocity about three
times higher at the center than at the equator.  For high mass
cases, the core undergoes dynamical collapse when it loses sufficient
angular momentum, leaving a black hole surrounded by a massive torus.
For certain cases, viscous heating provides enough support to avert
collapse, or rather to delay it for the cooling timescale.

A disadvantage of the Navier-Stokes equations is that it results in a
parabolic system that violates causality.  A decade after Duez~{\it et
  al.}'s work, Shibata~{\it et al.}~\cite{Shibata:2017jyf} returned to
this problem,
using a version of the Israel-Stewart formalism for relativistic
viscosity~\cite{Israel:1979wp}, which introduces an evolution equation
for $T^{\rm visc}{}_{\mu\nu}$ and respects causality.  Like in the earlier
study, they evolve hypermassive stars in 2D.  A notable finding is
that, for high viscosities (roughly $\alpha \sim \eta\Omega/P >
10^{-2}$, with $\Omega$ the angular frequency and $P$ the pressure),
outflows driven by viscous heating may be a major source of expelled
matter, with outflow masses comparable to that of the dynamical
ejecta.

Both of the above simulations use artificial initial data.  Recently,
Radice~\cite{Radice:2017zta} has performed \bns merger simulations using
a subgrid turbulence model very similar to a viscosity. 
Because the cores of \bns remnants are slowly rotating,
transport effects spin up the core but spin down the inner envelope, so
that collapse can be delayed or accelerated, depending on the strength of
the effective viscosity.  Meanwhile, Shibata and Kiuchi~\cite{Shibata:2017xht}
find that the effect of viscosity (with a reasonable $\alpha\sim 10^{-2}$)
on nonaxisymmetric deformations is dramatic, with these and their
corresponding gravitational wave signals damping on a viscous
timescale of around 5\,ms.

Subscale effects can also affect the large-scale magnetic field
(e.g. the ``alpha effect'' in dynamo theory).  Giacomazzo~{\it et
  al.}~\cite{Giacomazzo:2014qba} have taken a step to incorporate
these effects in numerical relativity, adding a subgrid EMF to the induction equation. 
The added term is designed to grow the magnetic field to equipartition
with the turbulent kinetic energy (meaning at least the largest eddies
should be present on the grid), as predicted by local simulations
of small-scale dynamo action~\cite{Zrake:2013mra}.  In simulations
with this added term, Giacomazzo~{\it et al.} find that the field can
quickly be amplified to magnetar levels.

Simulations with subgrid terms depend on the reliability of the subgrid
model, an assumption that cannot easily be relaxed, but they do allow
long-term evolutions including effects that would otherwise be inaccessible.

\section{Comparison to observations}
\label{ligo-compare}

\subsection{Gravitational wave astronomy}
To date, the LIGO and Virgo collaborations have announced five
confirmed black hole binary merger observations,
GW150914~\cite{Abbott:2016blz}, GW151226~\cite{Abbott:2016nmj},
GW170104~\cite{Abbott:2017vtc}, GW170608~\cite{Abbott:2017gyy}, and
GW170814~\cite{Abbott:2017oio}, as well as a potential sixth,
LVT151012~\cite{TheLIGOScientific:2016pea}. The observed progenitor
black-hole masses ranged from $7M_\odot$ to $36M_\odot$, the mass
ratios ranged from $0.53$ to $0.83$, and the
final merged black hole masses ranged from $18M_\odot$ to $62
M_\odot$. 

There has been a wealth of new information gleaned from these events.
Perhaps most importantly, a major prediction of general relativity in
the strong-field regime was confirmed: black holes, or at least extremely compact
objects much more massive than the maximum neutron star mass, exist,
form binaries, and merge through the emission of gravitational waves~\cite{Abbott:2016blz,
TheLIGOScientific:2016wfe}. As mentioned above, the observed merger waveforms are consistent with
the predictions of numerical relativity~\cite{Abbott:2016apu,
Lovelace:2016uwp, Lange:2017wki, Healy:2017abq}.
To date, there have been several published tests of general relativity
using the observed waveforms~\cite{TheLIGOScientific:2016src,
TheLIGOScientific:2016pea, Abbott:2017vtc, Abbott:2017oio,
Abbott:2017gyy}.
One such test consists of comparing the observed phase evolution of
the waveform with post-Newtonian predictions.
Another test consists of comparing the inferred parameters of the
merged binary based on the inspiral and merger 
parts of the waveform separately. A difference in the inferred
parameters would then imply that the binary did not evolve according
to the predictions of general relativity.
In all cases, to the precision that current detectors can measure
these effects, the data were consistent with the predictions of
general relativity.

\subsection{GW170817:  The age of multimessenger astronomy begins}

\begin{figure}
\centering
\includegraphics[width=0.95\columnwidth]{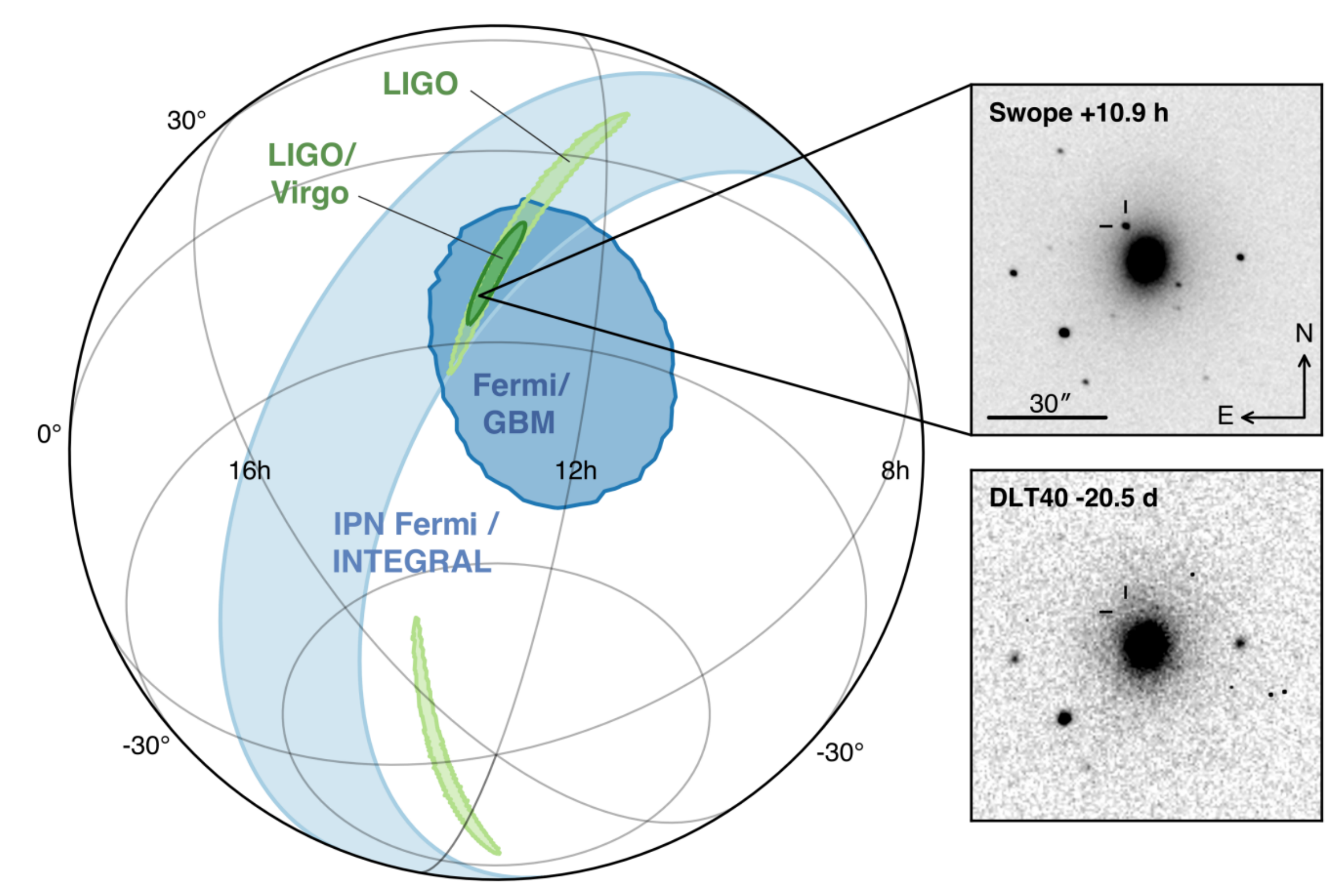}
\label{fig:gw170817}
\caption{
  A reproduction of Fig.~1 of Ref~\cite{GBM:2017lvd} courtesy of the
  authors. The figure shows the
  localization of the gravitational-wave, gamma-ray, and optical signals
  from \bns merger GW170817.  The left panel shows 90\%
  credible regions in the sky from gravitational wave and gamma ray detections.
  On the right are optical images after (top) and before (bottom) the merger. 
  }
\end{figure}

On August  17,  2017, LIGO-Virgo made the first gravitational wave
detection of a late-inspiral \bns
system, labeled GW170817~\cite{TheLIGOScientific:2017qsa}  The identification
as an \bns system could be made from the masses of the binary components,
which from the waveform were estimated to be in the range
1.17-–1.60$M_{\odot}$, with a total mass of $M_{\rm total}$ = 2.73--2.78$M_{\odot}$
and mass ratio in the range 0.7--1~\footnote{The exact allowed range of masses depends on
whether one allows the possibility of large binary component spins.  If spins are assumed to
be not larger than those observed Galactic binary neutron stars, the component masses are
inferred to lie in the range 1.16--1.6$M_{\odot}$.  Allowing for large spins, the range expands
to 1.00--1.89$M_{\odot}$~\cite{Abbott:2018wiz}.  The data itself does not, at the time of writing,
exclude the possibility that one component was a black hole, but known black hole formation
scenarios would not produce black holes of such low mass.}.  As with the first \bbh
detection, nature was unexpectedly kind, supplying an \bns merger at a
quite close luminosity distance of 40Mpc.  Tidal effects were not seen in
the waveform, leading to a maximum tidal deformability of $\Lambda<800$ at
90\% confidence.  A short gamma ray burst,
GRB 170817A, was detected by the Fermi Gamma-ray Burst Monitor at a time
1.7\,s after the GW170817 merger time in a region of the sky consistent
with LIGO-Virgo's 31 deg${}^2$ localization.  A bright optical/infrared/UV
transient was identified about half a day later and labeled AT2017gfo,
followed by X-ray and radio signals in the coming weeks~\cite{GBM:2017lvd}.

The GRB was unusually dim (isotropic) luminosity $L\sim 10^{47}$erg s${}^{-1}$,
perhaps due to some combination of being seen off-axis and various forms
of interaction between the jet and enveloping matter ejected during merger.
Comparisons to numerical relativity results have mostly focused on the optical and infrared
AT2017gfo signal, which strongly resembles an anticipated kilonova signal. 
Recall that, for the $v\sim 0.1c$ outflows expected from \bns mergers,
neutron-rich $Y_e<0.25$ outflow will synthesize lanthanide
elements, have high opacity, and is expected to peak after a week in the
near infrared.  Less neutron-rich outflow (perhaps made so by neutrino
processing) will have lower opacity and is expected to peak after about
a day in optical wavelengths.  AT2017gfo showed signs of both signals,
an early UV-blue signal with a near-IR tail~\cite{Soares-Santos:2017lru},
which can naturally be explained if outflows of both kinds are present,
and the high opacity material doesn't completely occult the low opacity
material.  Kilonova models can accommodate the observed model with two
components to the ejecta:  a $M=0.01M_{\odot}$, $v=0.3c$ lanthanide-poor
ejecta for the ``blue'' component and a $M=0.04M_{\odot}$, $v=0.1c$
lanthanide-rich ejecta for the ``red'' component~\cite{Cowperthwaite:2017dyu,
  Nicholl:2017ahq,Chornock:2017sdf}.

These numbers can be rather directly compared to numerical predictions.
A first conclusion is that dynamical ejecta from tidal forces and the
collision shock are inadequate.  For the range of realistic EoS, the
total dynamical ejecta mass does not exceed about
0.02$M_{\odot}$~\cite{Shibata:2017xdx,Radice:2017lry}.  More ejecta can
be produced by outflows from the stellar remnant or surrounding accretion
disk.  Numerical relativity has already told us something interesting:  the merged object
must be of a kind to give such outflows.

Next, numerical relativity provides another crucial piece of information.  If the remnant
collapses promptly, it will leave a black hole and very low-mass disk, lower
than the remaining mass that needs to be ejected.  (A warning is in order
here.  If the binary was very asymmetric, of the order $q\approx 0.7$, it
may be possible to get massive disks even in a high-mass, prompt-collapse
scenario.  The inferences below mostly assume that this was not the case.)
Therefore, it is surmised
that the remnant did not promptly collapse.  This means the binary's mass,
which we know, is below the threshold mass for prompt collapse, which is
loosely connected to the neutron star maximum mass.  Alternatively, the
need to avoid prompt collapse and small disk led
Bauswein~{\it et al}~\cite{Bauswein:2017vtn}
to set a {\it lower} limit on the radius of a 1.6$M_{\odot}$ neutron star of
10.6\,km and
Radice~{\it et al.}~\cite{Radice:2017lry} to place a {\it lower} limit on
the tidal deformability of $\Lambda>400$,
both based on numerical relativity \bns simulations with a variety of realistic EoS.

Next, we may ask whether the remnant was above or below the
supramassive-hypermassive cutoff mass.  (See Section~\ref{sec:stars} above.) 
Arguments have been made to the effect that the remnant
must have suffered delayed collapse, meaning that it was above this
mass.  First, a long-lived magnetar would have released energy via
dipole radiation at $L\sim 10^{50}$erg s${}^{-1}$ ($B$/$10^{15}$G)${}^2$,
which would accelerate the ejecta to $v\approx c$ and produce bright
emissions not matching observations~\cite{Margalit:2017dij}.  Second,
the collapse to a black hole with disk may be needed to explain the
GRB.  The supramassive mass limit is related to the TOV mass limit; the
former is about 20\% larger than the latter.  (Note, though, that
thermal effects effectively alter the EoS and hence the supramassive limit,
an additional source of uncertainty.)  Reasoning of this sort has been
used in several papers~\cite{Margalit:2017dij} to set an upper limit to
the neutron star maximum mass in the range
2.16--2.28$M_{\odot}$~\cite{Margalit:2017dij,Ruiz:2017due}.

Finally, there may be clues in the presence of the blue kilonova. 
Shibata~{\it et al.}~\cite{Shibata:2017xdx} compare merger simulations
with the stiff DD2 EoS with those of the soft SFHo EoS, combining 3D
merger simulations with 2D viscous simulations of the subsequent
$\sim$s of secular evolution, with neutrino transport included in
both.  SFHo has a lower maximum mass and so
predicts prompt collapse, while DD2 results in a long-lived stellar remnant.
Viscosity drives ejecta from both the hypermassive remnant (for DD2) and
the disk.  However, a long-lived remnant seems to be necessary to provide
the neutrino irradiation needed for a high-$Y_e$ outflow component that
produces the blue kilonova.  Once again, the kilonova combined with
numerical relativity constrains the EoS generally in the direction of excluding soft EoS.
However, the EoS-related parameter to which the \bns outcome is most
sensitive is the maximum mass.

\section{conclusion}
The birth of gravitational wave astronomy presents a remarkable story
of long-term planning and investment, with numerical relativity being
only one of the fields built up largely in anticipation of discoveries
known to be decades away.  The investment of time, money, and careers
has now been vindicated.  Although it is more broadly useful, the
majority of effort in numerical relativity has been devoted where it was needed by
LIGO-Virgo, to the three types of compact object binaries.  For each
type, numerical relativity has established some definite results that are beyond the
reach of Newtonian physics or perturbation theory:  the possibility of
super-kicks in \bbh mergers and the threshold mass for prompt collapse
in \bns merger, to name just two.  The study of these systems is not yet
finished.  Of the three system types, the simulation of \bbhs
is the most mature, but here also the binary parameter space
is most intimidating, and the application of numerical relativity to devising templates
for the full 7D space is ongoing work.  Neutron star--neutron star and \bhns simulations are
less accurate, and their ability to capture the multi-scale magnetic
and neutrino effects of the post-merger evolution might not even be
qualitatively adequate.

However, even where numerical relativity simulations of \bhns and
\bns mergers are
inadequate, there is no longer anything distinctively relativistic
about the problems.  Newtonian simulations of \bns mergers are no
more advanced; they face all the same difficulties.  In fact,
Newtonian simulations of these systems are becoming less common.
Since it is not much more difficult, why not just work in general
relativity?  With the advent of open-source, publicly available
numerical relativity
codes such as the Einstein Toolkit~\cite{Loffler:2011ay,Moesta:2013dna},
the barrier to an
interested astrophysicist doing numerical relativity work has never been lower.  This is
for the best.  One could say that the goal of numerical relativity all along has been to
abolish itself as a distinct subfield, to make solving the Einsteins
equations as routine as solving Poisson's equation, and thus to
dissolve into computational astrophysics.

And yet, numerical relativity will also remain as a tool for addressing questions in
gravitational physics.  Numerical experiments are used to investigate
features of black hole physics such as cosmic censorship and
the generic structure of spacetime
singularities~\cite{Berger2002}.  Perhaps more important, as
gravitational wave detections grow in number and accuracy, numerical
relativity will aid
in testing alternative theories of gravity.  Ultimately, to test
general relativity (and--dare we hope?--supersede it), we will need to
compare general relativity predictions, say of \bbh mergers, with more general
possibilities, to identify
the signatures of new physics that cannot be reproduced in general
relativity by
tinkering with parameters in the vast 7D \bbh parameter space.  A
systematic generalization of general relativity (analogous to the
parameterized post-Newtonian formalism) does not exist, and any generalization will
expand an already prohibitive parameter space.  Nevertheless,
exploratory simulations using scalar-tensor~\cite{Healy:2011ef} and
Chern-Simons~\cite{Okounkova:2017yby} theories of gravity have
already been carried out.  Neutron star
systems are ``messier'', with EoS uncertainties mixing with gravitational
uncertainties, but at least in scalar-tensor theories, they also
allow distinct signatures if the neutron star undergoes spontaneous
scalarization~\cite{Damour:1996ke}, so these systems may be useful
gravity test probes as well.

When all else is done, numerical relativity results leave us with a problem of their
own, the problem of their interpretation.  The asymptotic
gravitational wave output is gauge invariant, but the dynamics of the
interior is encoded in tensor functions on grids in an evolved
coordinate system.  It is actually remarkable that the movies
numerical relativists produce of our mergers look so qualitatively
reasonable, an artifact of gauges designed to minimize unnecessary
coordinate dynamics.  When one wants to know something concrete about
the strong-field region, the difficulty of disentangling coordinate
effects becomes acute.  Even to prove that the result of a \bbh
merger settles to a Kerr spacetime is a surprisingly
intricate affair~\cite{Campanelli:2008dv}.  Apparent horizons are
foliation-dependent.  Event horizons are not, but even they will be
visualized on some arbitrary coordinate system.  Ray tracing can be
used to reconstruct what a nearby observer would actually see from a
\bbh merger~\cite{Bohn:2014xxa}.  To try to give some intuition for
the physics of
the merger, Nichols~{\it et al.}~\cite{Nichols:2011pu} have proposed plotting
(gauge-dependent) field lines to illustrate the local tidal stretches
and twists.  The challenge of making sense of a background-independent
field theory extends more widely in theoretical physics, but it
confronts us in a particularly concrete form in numerical relativity.

\acknowledgments 
We thank John Baker, Andreas Bauswein, Manuela Campanelli,
Koutarou Kyutoku, Carlos
Lousto, Richard O'Shaughnessy, Frans Pretorius, Luciano Rezzolla,
and Masaru Shibata for
careful reading of this manuscript. We thank the referees and
the editorial board member
for their many helpful suggestions.   MD gratefully acknowledge the NSF
for financial support from Grant PHY-1806207.  YZ  gratefully
acknowledge the NSF for financial support from Grants No.~PHY-1607520,
No.~PHY-1707946, No.~OAC-1811228, No.~ACI-1550436, No.~OCI-1515969,
No.~AST-1516150, No.~ACI-1516125, and No.~AST-1028087.

\bibliographystyle{apsrev4-1}
\bibliography{references}
\end{document}